\documentclass[a4paper,11pt]{article}
\pdfoutput=1
\usepackage{jheppub}

\usepackage[dvipsnames]{xcolor}
\usepackage[utf8]{inputenc}
\usepackage[T5]{fontenc}
\DeclareTextSymbolDefault{\ohorn}{T5}
\usepackage[english]{babel}
\usepackage{mathtools}
\usepackage{amsfonts}
\usepackage{amsthm}
\usepackage{physics}
\usepackage{comment}
\usepackage{float}
\usepackage{subcaption}
\usepackage{tikz}
\usepackage{tkz-euclide}
\usetikzlibrary{decorations.markings}
\usetikzlibrary{patterns}
\usepackage{tikz-feynman}
\usepackage{simpler-wick}

\usepackage{hyperref}
\hypersetup{urlcolor=PineGreen, citecolor=Maroon, linkcolor=NavyBlue}
\usepackage{url}

\renewcommand{\a}{\alpha}
\renewcommand{\b}{\beta}
\newcommand{\g}{\gamma}
\renewcommand{\d}{\delta}
\newcommand{\eps}{\varepsilon}
\renewcommand{\th}{\theta}
\renewcommand{\l}{\lambda}
\newcommand{\s}{\sigma}
\renewcommand{\t}{\tau}
\newcommand{\D}{\Delta}
\newcommand{\w}{\omega}
\newcommand{\W}{\Omega}
\newcommand{\G}{\Gamma}

\newcommand{\DeclareAutoPairedDelimiter}[3]{%
  \expandafter\DeclarePairedDelimiter\csname Auto\string#1\endcsname{#2}{#3}%
  \begingroup\edef\x{\endgroup
    \noexpand\DeclareRobustCommand{\noexpand#1}{%
      \expandafter\noexpand\csname Auto\string#1\endcsname*}}%
  \x}
\DeclareAutoPairedDelimiter{\p}{(}{)}

\usepackage{bbm}
\newcommand{\one}{\mathbbm{1}} 
\newcommand{\bb}[1]{\mathbb{#1}} 
\renewcommand{\cal}[1]{\mathcal{#1}} 
\usepackage{bm} 
\renewcommand{\it}[1]{\textit{#1}} 
\renewcommand{\bf}[1]{\textbf{#1}} 
\newcommand{\up}[1]{\mathrm{#1}} 

\newcommand{\N}{\bb{N}} 
\newcommand{\Z}{\bb{Z}} 
\newcommand{\R}{\bb{R}} 
\newcommand{\C}{\bb{C}} 

\usepackage{cancel} 
\newcommand{\f}[2]{\frac{#1}{#2}} 
\newcommand{\too}{\longrightarrow} 
\newcommand{\llrr}{\longleftrightarrow}

\renewcommand{\tilde}[1]{\widetilde{#1}}
\DeclarePairedDelimiterX{\avg}[1]{\langle}{\rangle}{#1} 

\newcommand{\pd}{\partial}

\renewcommand{\L}{\mathcal{L}}
\renewcommand{\O}{\mathcal{O}}

\newtheorem{theorem}{Theorem}
\theoremstyle{plain}

\title{\bf{Heavy States in 3d Gravity and 2d CFT}}

\author{David Grabovsky}
\affiliation{Department of Physics, University of California at Santa Barbara \\ Santa Barbara, CA 93106, U.S.A.}
\emailAdd{davidgrabovsky@physics.ucsb.edu}

\abstract{
We discuss correlators of light fields in heavy states in $\mathrm{AdS}_3$ gravity and holographic 2d CFTs. In the bulk, the propagator of free fields in AdS backgrounds containing a conical defect or a BTZ black hole can be obtained by solving a wave equation, as well as by the method of images. On the boundary, these geometries are sourced by heavy operator insertions, and the propagator is dual to a heavy-light (HHLL) correlator. By matching its expansion in Virasoro blocks to our bulk results, we determine the OPE coefficients of all contributing states in both the $s$ and $t$ channels. In the $s$ channel, these states are excitations of the light field on top of the heavy state, and their OPE coefficients are the amplitudes to create them. The $t$-channel OPE is dominated by the Virasoro vacuum block, but there is also an infinite family of light two-particle states that contribute to the correlator. The OPE coefficients that couple these states to heavy operators represent their expectation values in heavy backgrounds. We determine them exactly, derive their asymptotic form at large twist, and discuss their behavior near and above the BTZ threshold, where they become thermal one-point functions.
}


\begin{document}
\maketitle
\flushbottom

\section{Introduction and Overview}

How can we understand high-energy states in quantum gravity? Our intuition from general relativity suggests that the answer is universal: at high energies, essentially any localized object will form a black hole. At the semiclassical level, black holes are characterized by a universal Hawking temperature, and in AdS/CFT \cite{Witten:1998qj,Gubser:1998bc,Banks:1998dd} they are dual to thermal states of the boundary theory. One way to understand their properties is to add light probe fields to the system and ask how they behave in the presence of a heavy object. We can often answer such questions precisely by studying the correlation functions of light fields in heavy states, especially in holographic settings.

There are two ways to understand how heavy objects affect light ones in the bulk. From one perspective, the light state and the heavy state come together, bind gravitationally, and form two-particle composite states that can be thought of as orbiting configurations of the light field around the heavy object. From another perspective, the light and heavy states each propagate independently but exchange gravitons and other virtual particles, like light-light or heavy-heavy composites, to mediate the interactions between them. Both of these viewpoints are illustrated schematically below:
\begin{align}
\label{eqn:binding-exchange}
\begin{gathered}
\scalebox{0.9}{
\begin{tikzpicture}
    \begin{feynman}
    \vertex (a);
    \vertex [above left = 0.1cm and 1.5cm of a] (a1) {L};
    \vertex [above right = 0.1cm and 1.5cm of a] (a2) {H};
    \vertex [above = 0.1cm of a] (c1);
    \vertex [below = 3cm of a] (b);
    \vertex [below left = 0.1cm and 1.5cm of b] (b1) {L};
    \vertex [below right = 0.1cm and 1.5cm of b] (b2) {H};
    \vertex [below = 0.1cm of b] (c2);
    \diagram* {
    (a1) -- [thick] (a) -- [thick] (a2),
    (a) -- [align=left, scalar, edge label = {bound \\[-4pt] states}, thick] (b),
    (b1) -- [thick] (b) -- [thick] (b2),
    };
    \end{feynman}
\end{tikzpicture}}
\end{gathered}
\qquad \text{\scalebox{1.3}{$=$}} \qquad
\begin{gathered}
\scalebox{0.9}{
\begin{tikzpicture}
    \begin{feynman}
    \vertex (a);
    \vertex [left = 0.1cm of a] (a1);
    \vertex [above left = 1.5cm and 0.1cm of a] (b) {L};
    \vertex [below  left = 1.5cm and 0.1cm of a] (d) {L};
    \vertex [right = 3cm of a] (f);
    \vertex [right = 0.1cm of f] (g);
    \vertex [above right = 1.5cm and 0.1cm of f] (c) {H};
    \vertex [below right = 1.5cm and 0.1cm of f] (e) {H};
    \diagram* {
        (b) -- [thick] (a) -- [thick] (d),
        (a) -- [align=center, scalar, edge label = {virtual \\[-4pt] exchanges}, thick] (f),
        (c) -- [thick] (f) -- [thick] (e),
    };
    \end{feynman}
\end{tikzpicture}}
\end{gathered}
\end{align}

In the dual conformal field theory (CFT), the ``binding'' and ``exchange'' philosophies correspond to the $s$- and $t$-channel expansions of a holographic correlator, respectively, and are related by crossing symmetry. The requirement of crossing symmetry is highly constraining because the data that enters a calculation in one channel can often be used to bootstrap what must happen in the other channel \cite{Belavin:1984vu}. In this way, one can gain access to detailed information about CFT observables and their bulk duals. In particular, (\ref{eqn:binding-exchange}) can teach us about the amplitudes to create the composite states described above, as well as their expectation values in heavy states.

In this paper, we will make this precise within the framework of $\up{AdS}_3/\up{CFT}_2$, a model where many calculations that are essentially intractable in higher dimensions simplify and often yield exact results. Below, we review how the relative simplicity of 3d gravity \cite{Deser:1983tn,Deser:1983nh} and the powerful tools available in 2d CFT \cite{Belavin:1984vu} improve our analytical control.

\subsection{3d Gravity with Defects}

In $(2+1)$-dimensional gravity, there are no local propagating bulk degrees of freedom \cite{Deser:1983tn}. As a consequence, every solution to pure 3d gravity with a negative cosmological constant is locally isometric to the empty anti-de Sitter (AdS) spacetime. The only solutions are $\up{AdS}_3$ itself and certain quotients of $\up{AdS}_3$, the simplest of which is the BTZ black hole \cite{Banados:1992wn,Banados:1992gq}. BTZ black holes of any size can exist, but their masses must lie above a critical threshold mass $M_* = \f{1}{8G}$ separating them from the pure $\up{AdS}_3$ vacuum.\footnote{We work throughout in natural units $c = \hbar = 1$, and we set the AdS radius $\ell_{\up{AdS}} = 1$. We also take the energy of the pure $\up{AdS}_3$ vacuum to be zero, instead of $-1$ as is more commonly done.} What happens if the mass goes below the BTZ threshold?

In pure gravity, there are no such geometries. But if the theory is coupled to pointlike sources, then it admits conical defect solutions describing a massive particle in $\up{AdS}_3$ with mass below the BTZ threshold \cite{Deser:1983nh}. An important class of defects is the $\up{AdS}_3/\Z_N$ orbifolds, which have been important in string theory \cite{Lunin:2001jy,Balasubramanian:2005qu,Giusto:2004ip,Giusto:2020mup} and in connection with gravitational path integrals \cite{Lewkowycz:2013nqa,Benjamin:2020mfz}. But the parameter $N$ characterizing the defect's strength does not need to be an integer, and most conical defect solutions are not quotients of $\up{AdS}_3$ in an obvious way.\footnote{There are also conical excess solutions, but we do not consider them here. Their masses lie below the AdS vacuum, and they are dual to CFT states with negative conformal dimensions. A theory containing such states cannot be unitary; nevertheless, it would be interesting to extend our results to this regime.} In fact, the metric of conical AdS is identical to that of BTZ, but with the horizon radius analytically continued so that the mass falls below the BTZ threshold. This suggests that we should be able to probe the physics of black holes from below the threshold by treating heavy and thermal states on the same footing.

We pursue this treatment by studying the propagation of free fields in heavy AdS backgrounds. We allow the fields to be massive, but we require them to be light enough that they do not backreact on the geometry. Both below and above the BTZ threshold, we construct the propagator for free fields by solving the Klein--Gordon equation and summing over the spectrum of normal or quasinormal modes, which play the role of the orbiting bound states on the LHS of (\ref{eqn:binding-exchange}). We also obtain the propagator in a different way: using the method of images for quotients of $\up{AdS}_3$. A sum over images is always available in BTZ, but only makes sense below the threshold in $\up{AdS}_3/\Z_N$. In both cases, it may be viewed as a saddle-point sum over the lengths of all geodesics between two points in the quotient geometry. Although it is not immediately obvious, this representation of the propagator corresponds to the ``exchange'' calculation on the RHS of (\ref{eqn:binding-exchange}).

\subsection{HHLL Correlators in 2d CFT} 

Semiclassical $\up{AdS}_3$ gravity has a dual description in terms of a holographic 2d CFT living at the asymptotic AdS boundary.\footnote{The existence of such a CFT has not been shown, and the appropriate holographic dual may instead involve an ensemble of theories \cite{Chandra:2022bqq}. However, since every member of this ensemble must share the same semiclassical properties, we may refer to ``the'' boundary theory when discussing its large-$c$ behavior.} Holographic CFTs have a very large central charge $c = \f{3}{2G} \gg 1$ \cite{Brown:1986nw} corresponding to the large-$N$ limit of gauge theories. We work at leading order in the $1/c$ expansion, where many aspects of the CFT become universal \cite{Hartman:2014oaa,Fitzpatrick:2014vua,Fitzpatrick:2015zha,Collier:2019weq}.

In the CFT, the bulk field is dual to a light primary operator $\O_{\up{L}}$ with conformal dimension $\Delta_{\up{L}} = \Delta$, where ``light'' means that $\f{\Delta}{c} \too 0$ as $c \too \infty$. Since the bulk field is free, its CFT dual is a generalized free field: this encodes the notion of an effective field theory where fluctuations on top of the heavy state are essentially free. Meanwhile, a conical defect is created by a heavy CFT operator $\O_{\up{H}}$ whose conformal dimension $\Delta_{\up{H}} = O(c)$ lies below the BTZ threshold.\footnote{Some authors use ``heavy'' to refer to conformal dimensions that lie strictly above the BTZ threshold as $c \too \infty$. We relax this condition slightly and refer to both conical defects and BTZ black holes as heavy.} The defects we consider are static and non-rotating, so in the CFT they are represented by scalar primaries. By the operator-state correspondence, they can also be thought of as highly excited pure states $\ket{\O_{\up{H}}}$. By contrast, BTZ black holes are dual to thermal states in the CFT. But the Eigenstate Thermalization Hypothesis (ETH) \cite{Srednicki:1994mfb} tells us that up to exponentially small corrections, it does not matter whether we consider a thermal state or a pure state of the same average energy. Therefore both above and below the threshold, we may describe heavy states by CFT primary insertions.

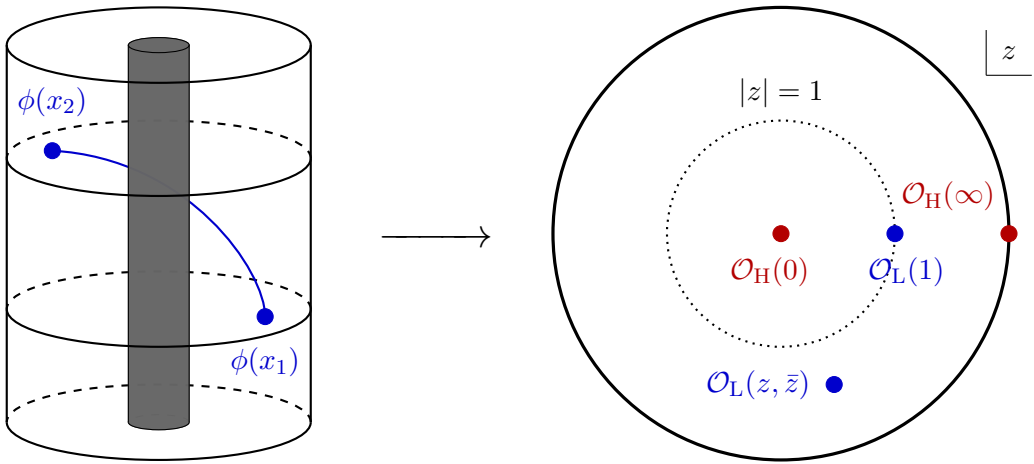
\begin{figure}[t]
\begin{align*}
\begin{gathered}
\begin{tikzpicture}
\draw [dashed, thick] (2,1.5) arc(0:180:2 and .5);
\draw [dashed, thick] (2,3.5) arc(0:180:2 and .5);
\draw [dashed, thick] (2,0) arc(0:180:2 and .5);
\filldraw [blue!80!black] (1.4,1.4) circle (3pt);
\filldraw [blue!80!black] (-1.4,3.6) circle (3pt);
\draw[thick, blue!80!black] (1.4,1.4) .. controls (1.4,2) and (.4,3.5) .. (-1.4, 3.6);
\draw [dashed] (.4,0) arc(0:180:.4 and .1);
\begin{scope}[opacity=.98,transparency group]
\filldraw[black!60] (0,0) ellipse (.4 and .1);
\filldraw[black!60] (0,5) ellipse (.4 and .1);
\filldraw[black!60] (-.4,0) rectangle (.4,5);
\end{scope}
\draw (0,5) ellipse (.4 and .1);
\draw (-.4,0) arc(180:360:.4 and .1);
\draw (-.4,0) -- (-.4,5);
\draw (.4,0) -- (.4,5);
\draw [thick] (0,5) ellipse (2.0 and .5);
\draw [thick] (-2,0) -- (-2, 5);
\draw [thick] (2,0) -- (2, 5);
\draw [thick] (-2,0) arc(180:360:2 and .5);
\draw [thick] (-2,1.5) arc(180:360:2 and .5);
\draw [thick] (-2,3.5) arc(180:360:2 and .5);
\node [blue!80!black] (a) at (1.4,.8) {$\phi(x_1)$};
\node [blue!80!black] (a) at (-1.4,4.25) {$\phi(x_2)$};
\end{tikzpicture}
\end{gathered}
\qquad\; \text{\scalebox{1.4}{$\xrightarrow{\hspace{24pt}}$}} \qquad
\begin{gathered}
\begin{tikzpicture}
\draw[thick, dotted] (0,0) circle (1.5); 
\draw[very thick] (0,0) circle (3); 
\filldraw [red!70!black] (0,0) circle (3pt); 
\node [red!70!black] (b1) at (-.15,-.5) {$\O_{\up{H}}(0)$}; 
\filldraw [blue!80!black] (1.5,0) circle (3pt); 
\node [blue!80!black] (b2) at (1.65,-.5) {$\O_{\up{L}}(1)$}; 
\filldraw [blue!80!black] (.7,-2) circle (3pt); 
\node [blue!80!black] (b3) at (-.3,-2) {$\O_{\up{L}}(z, \bar{z})$}; 
\filldraw [red!70!black] (3,0) circle (3pt); 
\node [red!70!black] (b4) at (2.2,.5) {$\O_{\up{H}}(\infty)$}; 
\draw[-] (2.7,2.1) -- (2.7, 2.7); 
\draw[-] (2.7,2.1) -- (3.3,2.1); 
\node (a) at (3,2.4) {$z$}; 
\node [label = {[label distance=40]90:$|z| = 1$}] {};
\end{tikzpicture}
\end{gathered}
\end{align*}
\caption{The configuration of fields defining the HHLL correlator is shown on the left in AdS (where we have a nontrivial background geometry) and on the right in the CFT (where, in radial quantization, the background is created by two heavy insertions).}
\label{fig:HHLL-cartoon}
\end{figure}

With heavy AdS backgrounds viewed in this way, our bulk propagator becomes a heavy-heavy-light-light (HHLL) CFT correlator of the form $\avg{\O_{\up{H}} \O_{\up{L}} \O_{\up{L}} \O_{\up{H}}}$. This situation is illustrated in Figure~\ref{fig:HHLL-cartoon}, both in AdS (left) and in radial quantization (right). CFT correlators are computed by using the operator product expansion (OPE) to sum over intermediate states exchanged by the external operators, as in (\ref{eqn:binding-exchange}). The organization of CFT states into irreducible representations of the Virasoro algebra turns the OPE into an expansion in functions called Virasoro conformal blocks, which have figured prominently in the conformal bootstrap program \cite{Heemskerk:2009pn,El-Showk:2011yvt,Fitzpatrick:2012yx,Komargodski:2012ek,Jackson:2014nla} and the study of holographic entanglement \cite{Ryu:2006ef,Headrick:2010zt,Hartman:2013mia,Asplund:2014coa}. Fortunately, the semiclassical HHLL Virasoro blocks are known \cite{Fitzpatrick:2015zha}, leaving only the questions of which states are exchanged and how much they each contribute.

The main contributions to the OPE come from double-trace, or double-twist, operators built from $\O_{\up{L}}$ and $\O_{\up{H}}$. In the $s$ channel, these are heavy-light composite states that describe excitations of the light field in the bulk and correspond to the Fourier modes one obtains by solving the wave equation in AdS. The associated OPE coefficients, which we deduce by matching the OPE to our bulk result, are interpreted as the amplitudes of these excitations. In the $t$ channel, the dominant contribution comes from the Virasoro vacuum block, which encodes gravitational interactions between $\O_{\up{L}}$ and $\O_{\up{H}}$. The vacuum block gives the same contribution to the correlator as the minimal geodesic in the sum over images \cite{Hartman:2013mia} and fully captures the singular structure of the OPE limit where the two light operators are brought together \cite{Berenstein:2022ico}. The non-minimal geodesics making up the rest of the correlator are then responsible for the regular terms in the OPE, and should have a CFT interpretation in terms of non-vacuum exchanges. Indeed, there is an infinite family of light-light double-twist states which provides exactly these contributions. 

The OPE coefficients of these primaries are identified as their expectation values in heavy states, and much of the technical work in this paper is devoted to computing them explicitly by matching the $t$-channel OPE to the sum over images. Our main results, comprising Sections \ref{sec:expectation-values}--\ref{sec:thermal-one-point-functions}, include explicit expressions for the expectation values of low-lying operators, asymptotic formul{\ae} for the HHLL Virasoro blocks and OPE coefficients at large twist, and a discussion of their behavior near and above the BTZ threshold.

\subsection{Main Results and Outline}

By carrying out the analysis described above, we aim to achieve four goals:

\begin{enumerate}
\item \bf{Understand the emergence of thermal behavior.} By studying conical defects of any mass (not just $\up{AdS}_3/\Z_N$), we fill the gaps between the integers and illustrate how physics below the BTZ threshold smoothly connects to the physics in the black hole regime. In particular, we describe how to analytically continue the sum over images for non-integer $N$. The expectation values we compute analytically continue, above the threshold, to thermal one-point functions in BTZ or thermal AdS.
\item \bf{Give a unified treatment from a Lorentzian perspective.} By treating both defects and black holes as heavy operator insertions on the Lorentzian cylinder, we avoid the need to switch to Euclidean signature above the BTZ threshold. (Of course, the ETH tells us that it makes no difference whether we work on the torus.)
\item \bf{Move beyond vacuum dominance.} By focusing on non-vacuum $t$-channel exchanges, we point out that there is more to learn about CFT correlators beyond their perturbative expansions in singular powers. These effects are inaccessible to large-spin asymptotics in the $s$ channel, but are required by crossing symmetry.
\item \bf{Compute double-trace expectation values.} In higher dimensions, it is of great interest---but even greater difficulty---to understand the correlators of double-trace operators in black hole backgrounds. Now, at least in a simplified toy model, we have the answers. The expectation values we compute do not factorize at large $N$, in the sense that (for example) $\avg{\O_{\up{L}}^2} \neq \avg{\O_{\up{L}}}\avg{\O_{\up{L}}}$. Instead, the effective classical background created by $\O_{\up{H}}$ modifies the expectation values of $\avg{\O_{\up{L}}^2}$ directly. 
\end{enumerate}
This paper is organized into two broad parts. In the first part, we compute the HHLL correlator in four different ways: twice in the bulk, first by solving the wave equation (Section~\ref{sec:correlators-wave-equation} and Appendix~\ref{sec:appendix-wave-equations}) and then using the method of images (Section~\ref{sec:correlators-method-of-images}); and twice on the boundary, using the $s$-channel and $t$-channel OPEs (Section~\ref{sec:correlators-2d-cft}). These computations are the bulk and boundary avatars of the LHS and RHS of (\ref{eqn:binding-exchange}), respectively. To some extent these preliminary sections review old material, although some details of the construction and structure of the correlator had not been discussed in the literature before. 

In the second part, we calculate the expectation values of light double-trace operators (Section~\ref{sec:expectation-values}) and analyze their properties. We derive an asymptotic formula for the expectation values at large twist (Section~\ref{sec:asymptotics-large-twist}), study their behavior as a function of the mass of the heavy state (Section~\ref{sec:crossing-btz-threshold} and Appendix~\ref{sec:appendix-light-defects}), and discuss their interpretation as thermal one-point functions (Section~\ref{sec:thermal-one-point-functions}). We conclude and discuss future directions in Section~\ref{sec:conclusions-future-work}.

\section{Summing Over Field Modes}
\label{sec:correlators-wave-equation}

The background geometry created by a static, heavy object in $\up{AdS}_3$ can be written in global coordinates $(t,\th,r)$, where $\th \sim \th + 2\pi$, as follows:
\begin{align}
\label{eqn:heavy-metric}
\dd s^2 &= -f(r) \dd t^2 + \f{\dd r^2}{f(r)} + r^2 \dd\th^2, \qquad
f(r) = \begin{cases}
\displaystyle r^2 + \f{1}{N^2} &\text{below the threshold}, \\[11pt]
r^2 - r_0^2 &\text{above the threshold}.
\end{cases}
\end{align}
Below the BTZ threshold, the parameter $N \geq 1$ describes the strength of a conical defect. Integer values of $N$ correspond to the cyclic orbifolds $\up{AdS}_3/\Z_N$, and the limit $N \too \infty$ corresponds to the BTZ threshold. Above the threshold, a horizon of radius $r_0 \in \R_+$ forms, and the state becomes thermal with a temperature given by $r_0 = \f{2\pi}{\b} = 2\pi T_{\up{H}}$. The geometries above and below the threshold are related by the analytic continuation $N \llrr -\f{i}{r_0}$, so conical defects can be viewed as black holes with imaginary temperature, and conversely BTZ can be thought of as a conical geometry with an imaginary angular deficit. To unify our descriptions of the two regimes, we write $f(r) = r^2 + \a^2$, where
\begin{align}
\label{eqn:analytic-continuation}
\a = \f{1}{N} \llrr ir_0 = \f{2\pi i}{\b}, \qquad \up{i.e.} \qquad \a \equiv \begin{cases}
\f{1}{N} \in [0,1] &\text{below the threshold} \\
ir_0 \in i\R_+ &\text{above the threshold}.
\end{cases}
\end{align}

Consider a free massive scalar field $\phi$ propagating in this background. We decompose the solutions of its equation of motion, the Klein--Gordon equation, into Fourier modes:
\begin{align}
\label{eqn:EOM}
\p{\Box - m^2} \phi_{\w,\ell} = 0, \qquad \phi_{\w,\ell}(t,\th,r) = e^{-i\w t} e^{i\ell\th} R_{\w,\ell}(r), \qquad \ell \in \Z.
\end{align}
On this ansatz, (\ref{eqn:EOM}) reduces to a differential equation for the radial modes:
\begin{align}
\label{eqn:radial-equation}
\qty[\f{1}{r} \dv{}{r} \p{r f(r) \dv{}{r}} + \f{\w^2}{f(r)} - \f{\ell^2}{r^2} - m^2] R_{\w,\ell}(r) = 0.
\end{align}
The solutions to the radial equation can be expressed in terms of hypergeometric functions. With appropriate boundary conditions, these mode solutions can be used to construct the propagator. We review how this is done in Appendix~\ref{sec:appendix-wave-equations}, which includes a detailed treatment of the wave equation, its solutions in various backgrounds, the relevant boundary conditions, and the resulting propagators. Below, we summarize the essential results.

\subsection{Correlators Below the Threshold}
\label{sec:correlators-below-threshold}

In conical AdS, we require the radial modes to be regular at the origin and normalizable, i.e. decaying at infinity. Those solutions $R_{\w,\ell}(r)$ which are regular at $r=0$ behave near the AdS boundary as linear combinations of normalizable and non-normalizable modes:
\begin{align}
\label{eqn:AdS-bdy-asymptotics}
R_{\w,\ell}(r) \sim \cal{N}_{\w,\ell} \p{A(\w,\ell) r^{\Delta - 2} + B(\w,\ell) r^{-\Delta}} \qquad \up{as}\quad r \too \infty.
\end{align}
Here $\cal{N}_{\w,\ell}$ is a normalization constant, $\Delta$ is the conformal dimension of the CFT operator dual to $\phi$, and $A(\w,\ell)$ and $B(\w,\ell)$ are certain ratios of Gamma functions. To ensure the normalizability of the modes, we must have $A(\w,\ell) = 0$. This condition is satisfied on a discrete set of normal frequencies $\w_{n\ell}$ derived in Appendix~\ref{sec:appendix-below-threshold} and given by
\begin{align}
\label{eqn:normal-frequencies}
\w_{n\ell} = |\ell| + \f{1}{N} \big(\Delta + 2n\big) \in \R_+, \qquad \ell \in \Z, \;\, n \in \N.
\end{align}
The resulting normal modes can be made orthonormal with respect to the Klein--Gordon inner product, and any solution to (\ref{eqn:EOM}) can be expanded in normal modes:
\begin{align}
\phi(x) = \sum_{n=0} \sum_{\ell \in \Z} \qty[\phi_{n\ell}(x) a_{n\ell} + \phi_{n\ell}^*(x) a_{n\ell}^{\dagger}], \qquad \phi_{n\ell}(t,\th,r) = e^{-i\w_{n\ell} t} e^{i\ell\th} R_{\w_{n\ell},\ell}(r).
\end{align}

We quantize the field by promoting $a_{n\ell}$ and $a_{n\ell}^{\dagger}$ to operators obeying the canonical commutation relations $[a_{n\ell}, a_{n'\ell'}^{\dagger}] = \d_{nn'} \d_{\ell\ell'}$. We choose the Hartle--Hawking vacuum $\ket{0}$ as the one annihilated by all of the lowering operators $a_{n\ell}$, and finally we construct the bulk-to-bulk propagator $G(x,x')$ as a sum over the normal modes:
\begin{align}
\label{eqn:bulk-to-bulk-modes}
G(x,x') = \mel{0}{\phi(x) \phi(x')}{0} = \sum_{n\ell} e^{-i\w_{n\ell} (t-t')} e^{i\ell(\th-\th')} R_{\w_{n\ell},\ell}(r) R_{\w_{n\ell},\ell}^*(r').
\end{align}
As we send $r \too \infty$, the boundary values of the field modes are rescaled by $r^{\Delta}$ \cite{Gubser:1998bc}:
\begin{equation}
\label{eqn:field-bdy-values}
\begin{aligned}
\phi_{n\ell}(t,\th) &= \lim_{r \to \infty} r^{\Delta} \phi_{n\ell}(t,\th,r) = \cal{C}_{n\ell}^{\up{HL}} e^{-i\w_{n\ell} t} e^{i\ell\th}, \qquad \cal{C}_{n\ell}^{\up{HL}} \equiv \lim_{r \to \infty} r^{\Delta} R_{\w_{n\ell},\ell}(r).
\end{aligned}
\end{equation}
The coefficients $\cal{C}_{n\ell}^{\up{HL}} = \cal{N}_{\w_{n\ell},\ell} B(\w_{n\ell},\ell)$, given by (\ref{eqn:HL-coefficients-conical}), are computed in Appendix~\ref{sec:appendix-below-threshold} by normalizing the modes $\phi_{n\ell}(x)$. These coefficients represent the amplitudes to produce an excitation of the scalar on the boundary with energy $\w_{n\ell}$ and angular momentum $\ell$; we will see later that they are also OPE coefficients for the CFT operators dual to $\phi_{n\ell}$. 

We obtain the boundary-to-boundary propagator by sending $r,r' \too \infty$ in (\ref{eqn:bulk-to-bulk-modes}) with $t' = \th' = 0$ and using (\ref{eqn:field-bdy-values}). The result is a sum of Fourier modes on the boundary:
\begin{equation}
\label{eqn:conical-propagator-modes}
\begin{aligned}
G(t,\th) = \lim_{r,r' \to \infty} \Big( r^{\Delta} r'^{\Delta} G(x,x') \Big) = \sum_{n\ell} \p{\cal{C}_{n\ell}^{\up{HL}}}^2 e^{-i\w_{n\ell} t} e^{i\ell\th} = \sum_{n\ell} \cal{G}(\w_{n\ell},\ell) e^{-i\w_{n\ell} t} e^{i\ell\th}.
\end{aligned}
\end{equation}
Here we have introduced the momentum-space boundary propagator $\cal{G}(\w, \ell)$ as the Fourier inverse of $G(t,\th)$. In conical AdS, it is defined only at the normal frequencies.

\subsection{Correlators Above the Threshold}
\label{sec:correlators-above-threshold}

The story in BTZ is similar, but there are some important differences. For instance, normal modes are not well defined because waves in black hole backgrounds can fall through the horizon. To probe the black hole's thermal response to perturbations, we need to impose boundary conditions suitable for the retarded Green's function $G_{\up{R}}(x,x')$. The modes should be purely ingoing at the horizon and decaying at infinity:
\begin{align}
\label{eqn:ingoing-conditions}
R_{\w,\ell}(r) \sim 
\begin{cases}
(r - r_0)^{-\f{i\w}{2r_0}}, &r \too r_0, \\
r^{-\Delta}, &r \too \infty
\end{cases}
\end{align}
The ingoing modes behave like (\ref{eqn:AdS-bdy-asymptotics}) near the boundary, and the condition $A(\w,\ell) = 0$ (which ensures that $R_{\w,\ell}(r)$ decays at infinity) is satisfied on a discrete set of complex frequencies $\tilde{\w}_{n\ell}$ in the lower half plane associated to the quasinormal modes (QNMs):
\begin{align}
\label{eqn:quasinormal-frequencies}
\tilde{\w}_{n\ell} = -\ell - ir_0 \big( \Delta + 2n \big) \in \bb{H}_-, \qquad \ell \in \Z, \;\; n \in \N.
\end{align}
The real and imaginary parts of $\tilde{\w}_{n\ell}$ represent the energy and decay rate of the QNMs, respectively. $\up{Im}(\tilde{\w}_{n\ell}) < 0$ ensures that the black hole is stable to small perturbations. The QNMs have a characteristic lifetime $\t_n = \f{1}{r_0 \p{\Delta + 2n}}$, independent of their angular momentum. The longest-lived modes ($n=0$) dominate the retarded propagator at late times, leading to an exponential decay $G_{\up{R}}(t, \th) \sim e^{-r_0 \Delta t}$ in the semiclassical limit.

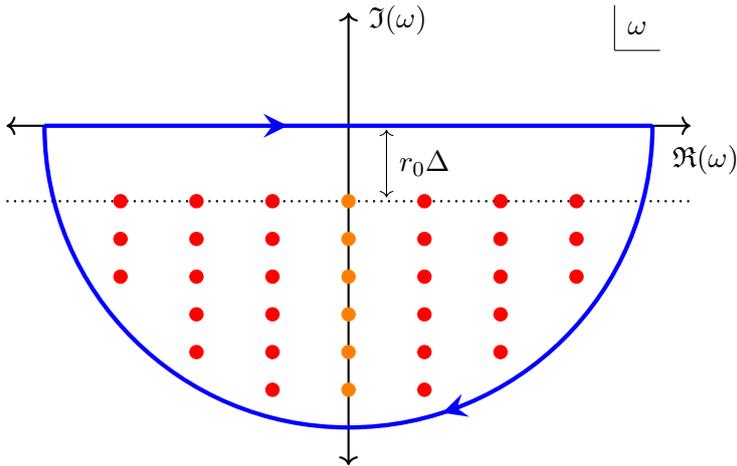
\begin{figure}[t]
\centering
\begin{tikzpicture}
[decoration={markings, 
  mark= at position 0.4 with {\arrow[scale=1.5,>=stealth]{>}}}] 
\draw[<->, thick] (-4.5,0) -- (4.5,0); 
\draw[<->, thick] (0,-4.5) -- (0,1.5); 
\draw[dotted, thick] (-4.5,-1) -- (4.5,-1); 
\foreach \l in {-2,-1,1,2}{\foreach \n in {0,...,4}{\filldraw[red] (\l, -.5*\n-1) circle (2.5pt);}} 
\filldraw[red] (-1, -3.5) circle (2.5pt);
\filldraw[red] (1, -3.5) circle (2.5pt);
\foreach \l in {-3,3}{\foreach \n in {0,...,2}{\filldraw[red] (\l, -.5*\n-1) circle (2.5pt);}} 
\foreach \n in {0,...,5}{\filldraw[orange] (0, -.5*\n-1) circle (2.5pt);} 
\draw[postaction={decorate}, ultra thick, blue] (-4,0) -- (4,0); 
\draw[postaction={decorate}, ultra thick, blue] (4,0) arc (0:-180:4); 
\draw[<->] (.5,-.95) -- (.5, -.05); 
\node (a) at (1, -.5) {$r_0 \Delta$}; 
\draw[-] (3.5,1) -- (3.5, 1.6); 
\draw[-] (3.5,1) -- (4.1,1); 
\node (a) at (3.8,1.3) {$\w$}; 
\node (b) at (4.7, -.45) {$\mathfrak{R}(\w)$};
\node (c) at (.65,1.4) {$\mathfrak{I}(\w)$};
\end{tikzpicture}
\caption{The poles of the momentum-space propagator in BTZ lie in the lower half-plane and are located at the QNM frequencies $\tilde{\w}_{\ell} = -\ell + ir_0 \p{\Delta + 2n}$, where $\ell \in \Z$ and $n \in \N$. They are all simple poles (red), except for the double poles at $\ell = 0$ (orange). The contour used to perform the integral (\ref{eqn:BTZ-propagator-integral}) encloses all of these poles and is shown in blue.}
\label{fig:QNM-poles}
\end{figure}

The QNMs are neither normalizable nor complete \cite{Warnick:2013hba}, so we cannot naïvely sum over them as we did below the BTZ threshold. Instead, we identify $A(\w,\ell)$ and $B(\w,\ell)$ in (\ref{eqn:AdS-bdy-asymptotics}) as the source $J$ and linear response $\avg{\O}_J$, respectively, for small perturbations of the boundary theory by an operator dual to the bulk field \cite{Klebanov:1999tb,Skenderis:2008dg}. In momentum space, the linear response is the retarded Green's function multiplied by the source, so we have
\begin{align}
\label{eqn:btz-propagator-momentum}
\avg{\O(\w,\ell)}_J = \cal{G}_{\up{R}}(\w,\ell) J(\w,\ell) \implies
\cal{G}_{\up{R}}(\w,\ell) = \f{B(\w,\ell)}{A(\w,\ell)}.
\end{align}
To obtain the position-space propagator from (\ref{eqn:btz-propagator-momentum}), we perform a Fourier transform. The spectrum of normal modes in conical AdS becomes dense near the BTZ threshold, so an integral over all frequencies is understood to replace the discrete sum in (\ref{eqn:conical-propagator-modes}):
\begin{align}
\label{eqn:BTZ-propagator-integral}
G_{\up{R}}(t,\th) = \sum_{\ell \in \Z} \int_{-\infty}^{\infty} \f{\dd\w}{2\pi}\, \cal{G}_{\up{R}}(\w,\ell) e^{-i\w t} e^{i\ell\th}.
\end{align}
The momentum-space retarded Green's function $\cal{G}_{\up{R}}(\w,\ell)$ is a meromorphic function of $\w$ whose only singularities are simple poles at the QNM frequencies $\tilde{\w}_{n\ell}$.\footnote{There is one important exception. At $\ell = 0$, the QNM singularities with $\ell > 0$ and $\ell < 0$ coalesce into double poles on the negative imaginary axis. These poles still have nontrivial residues; see below.} The integral in (\ref{eqn:BTZ-propagator-integral}) may then be performed by closing the contour at infinity in the lower half-plane and picking up all of the residues of $\cal{G}_{\up{R}}(\w,\ell)$, as shown in Figure~\ref{fig:QNM-poles}:
\begin{align}
\label{eqn:BTZ-propagator-QNM}
G_{\up{R}}(t,\th) = \sum_{n\ell} \big(\tilde{\cal{C}}_{n\ell}^{\up{HL}}\big)^2 e^{-i\tilde{\w}_{n\ell} t} e^{i\ell\th}, \qquad \textbf{}\big(\tilde{\cal{C}}_{n\ell}^{\up{HL}}\big)^2 \equiv i\, \up{Res}\big[\cal{G}_{\up{R}}(\w,\ell),\, \tilde{\w}_{n\ell}\big].
\end{align}
Both $\cal{G}_{\up{R}}(\w,\ell)$ and its residues are computed in Appendix~\ref{sec:appendix-above-threshold}; we discuss them below.

\subsection{Analytic Continuation to BTZ}
\label{sec:crossing-threshold}

The formal similarity between the conical and BTZ propagators (\ref{eqn:conical-propagator-modes}) and (\ref{eqn:BTZ-propagator-QNM}) is striking, especially given that their derivations are somewhat different. It is natural to expect that the two should be related by an analytic continuation in $\a$, perhaps in the sense that taking $N \too -\f{i}{r_0}$ will convert (\ref{eqn:conical-propagator-modes}) into (\ref{eqn:BTZ-propagator-QNM}) by sending
\begin{align}
\w_{n\ell} \stackrel{?}{\too} \tilde{\w}_{n\ell} \qquad \up{and} \qquad \p{\cal{C}_{n\ell}^{\up{HL}}}^2 \stackrel{?}{\too} \big(\tilde{\cal{C}}_{n\ell}^{\up{HL}}\big)^2.
\end{align}
This naïve expectation turns out to be almost, but not quite, correct. As we show in Appendix~\ref{sec:appendix-above-threshold}, it is actually the momentum-space Green's function, \emph{not} the coefficients $\cal{C}_{n\ell}^{\up{HL}}$, which analytically continues across the BTZ threshold via $N \too -\f{i}{r_0}$. In terms of the parameter $\a$, the Green's function both above and below the threshold is \cite{Son:2002sd}
\begin{align}
\label{eqn:momentum-space-G}
\cal{G}(\w,\ell) \propto \f{\G\Big(\f{1}{2} \big(\D + \f{\w + \ell}{\a}\big){}_{\!} \Big)\, \G\Big(\f{1}{2} \big(\D + \f{\w - \ell}{\a}\big){}_{\!} \Big)}{\G\Big(\f{1}{2} \big(2-\D + \f{\w + \ell}{\a}\big){}_{\!} \Big) \G\Big(\f{1}{2} \big(2-\D + \f{\w - \ell}{\a}\big){}_{\!} \Big)}.
\end{align}

What about the amplitudes $\cal{C}_{n\ell}^{\up{HL}}$ and the residues $\tilde{\cal{C}}_{n\ell}^{\up{HL}}$? They end up being nearly identical, but they differ by an important factor that appears only in the BTZ regime:
\begin{subequations}
\label{eqn:HL-coefficients}
\begin{align}
\label{eqn:HL-coefficients-conical}
\p{\cal{C}_{n\ell}^{\up{HL}}}^2 &= \f{N^{1-2\Delta}}{\G(\Delta)^2} \f{\G(\Delta + n + N|\ell|) \G(\Delta + n)}{\G(1 + n + N|\ell|)\, \G(1+n)}, \\
\label{eqn:HL-coefficients-BTZ}
\big( \tilde{\cal{C}}_{n\ell}^{\up{HL}} \big)^2 &= \f{r_0^{2\Delta - 1}}{\G(\Delta)^2} \f{\G(\Delta + n - i|\ell|/r_0) \G(\Delta + n)}{\G(1 + n - i|\ell|/r_0) \G(1+n)} f_{n\ell}.
\end{align}
\end{subequations}
The BTZ factor $f_{n\ell}$ is given by
\begin{align}
\label{eqn:fnl-factor}
f_{n\ell} &=\begin{cases}
\displaystyle \f{\sinh\p{\pi|\ell|/r_0 + i\pi\Delta}}{\sinh\p{\pi|\ell|/r_0}}, & \ell \neq 0, \\
\displaystyle 2\cos\p{\pi \Delta} - \f{2}{\pi} \sin\p{\pi\Delta} \big(\psi(1+n) - \psi(\Delta + n)\big), & \ell = 0,
\end{cases}
\end{align}
where $\psi(z) = \f{\G'(z)}{\G(z)}$ is the digamma function. This factor arises from the use of Euler's reflection formula $\G(z) \G(1-z) = \f{\pi}{\sin(\pi z)}$ in evaluating the QNM residues (\ref{eqn:BTZ-propagator-QNM}). If we had naïvely taken $N \too -\f{i}{r_0}$ in (\ref{eqn:HL-coefficients-conical}), we would have gotten (\ref{eqn:HL-coefficients-BTZ}) wrong.

One way to understand the apparent asymmetry between these coefficients is to view $f_{n\ell}$ as a thermal factor that becomes trivial at zero temperature and influences the high-temperature behavior of the correlator. As a function of the inverse temperature,
\begin{align}
\label{eqn:thermal-factor}
f_{n\ell}(\b) = \f{\sinh\p{\b|\ell|/2 + i\pi\Delta}}{\sinh\p{\b|\ell|/2}}, \qquad
\big|f_{n\ell}(\b)\big|^2 = \cos^2\p{\pi\Delta} + \coth^2\p{\f{\b|\ell|}{2}} \sin^2\p{\pi\Delta}
\end{align}
for $\ell \neq 0$. Up to an overall phase $e^{-i\pi\Delta}$ that can be absorbed into the normalization of the propagator, $f_{n\ell}(\b)$ quickly approaches 1 at low temperatures and stays frozen at 1 below the BTZ threshold: this explains its apparent absence in conical AdS.

\subsection{Asymptotics at Large Spin}
\label{sec:asymptotics-large-l}

With this understanding, we can write down the propagator both above and below the threshold in a unified manner. We adopt complex coordinates $z = e^{\t + i\th}$ and $\bar{z} = e^{\t - i\th}$ with $\t = it$ on the AdS boundary cylinder, so that (\ref{eqn:conical-propagator-modes}) and (\ref{eqn:BTZ-propagator-QNM}) become\footnote{\label{fn:conf-transf}To see this, observe that the modes $\phi_{n\ell}(t,\th) = e^{-i\w_{n\ell} t} e^{i\ell\th}$ become $\phi_{n\ell}(z, \bar{z}) = |z|^{-\Delta} |z|^{\a(\Delta + 2n)} z^{\ell}$ in complex coordinates. The leading factor $|z|^{-\Delta}$ is the cost of the conformal transformation from the plane to the cylinder. It is understood in the definition of $\cal{C}_{n\ell}^{\up{HL}}$ that $|f_{n\ell}| \equiv 1$ below the BTZ threshold.}
\begin{equation}
\label{eqn:mode-sum}
\begin{aligned}
G(z, \bar{z}) &= \abs{z}^{(\a - 1) \Delta} \sum_{n\ell} \p{\cal{C}_{n\ell}^{\up{HL}}}^2 \abs{z}^{2\a n} z^{\ell}, \\
\p{\cal{C}_{n\ell}^{\up{HL}}}^2 &= \f{\a^{2\Delta-1}}{\G(\Delta)^2} \f{\G(\Delta + n + |\ell|/\a) \G(\Delta + n)}{\G(1 + n + |\ell|/\a)\, \G(1+n)} f_{n\ell}(\b).
\end{aligned}
\end{equation}

\begin{figure}[t]
\centering
\includegraphics[width=\textwidth]{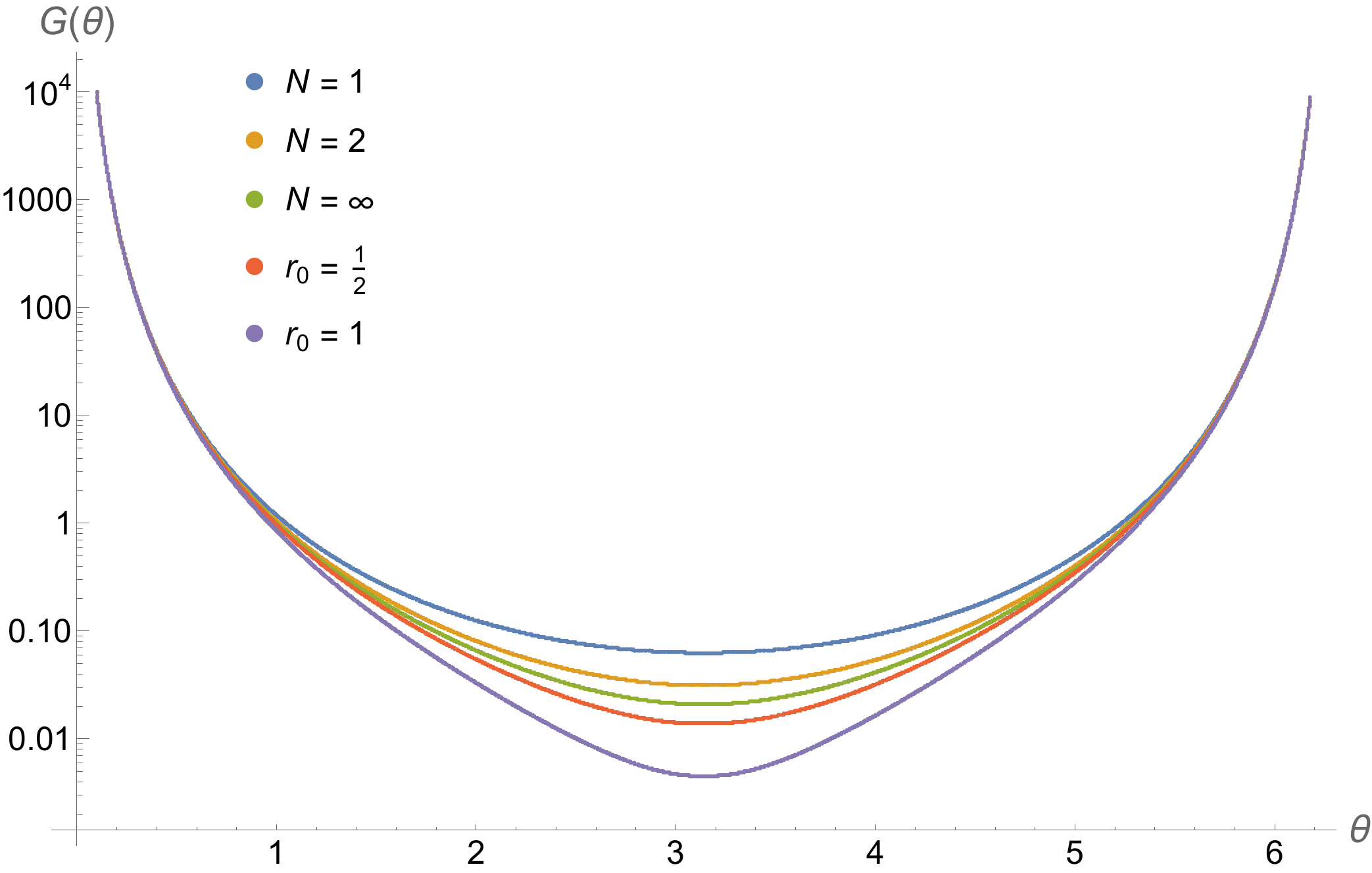}
\caption{The equal-time propagator $G(\th)$ is shown for $\Delta = 2$ and several values of the mass, both below and above the BTZ threshold. The UV divergence at $\th = 0$ is universal, and the propagator is periodic in $\th$. Long-range correlations are suppressed as the mass increases, and for larger black holes $G(\th)$ decays exponentially in $\th$ away from $\th = 0$.}
\label{fig:propagator}
\end{figure}

The power series (\ref{eqn:mode-sum}) is analytic in $|z| < 1$, but it diverges at $z, \bar{z} = 1$. In Lorentzian signature, these singularities occur when $e^{i(t \pm \th)} = 1$, i.e. on the boundary light cones $t \pm \th = 2\pi m$, where $m \in \Z$ is the number of times the light cones wrap the cylinder. To see this more clearly, it is useful to study the correlator at equal times. Setting $t=0$ puts $|z| = 1$ and makes $G(z, \bar{z})$ a function only of the angular separation $\th$ between the field insertions on the boundary. In this case the sum over $n$ can be performed, and we find
\begin{align}
\label{eqn:equal-time}
G(\th) = \f{\a^{2\Delta - 1}}{2\G(2\Delta) \cos\p{\pi \Delta}} \sum_{\ell \in \Z} \p{\f{\G(\Delta + |\ell|/\a)}{\G(1 + |\ell|/\a - \Delta)}} f_{0\ell}(\b) e^{i\ell\th}.
\end{align}
Figure~\ref{fig:propagator} illustrates the behavior of $G(\th)$ as $\a$ varies across the BTZ threshold. The $z, \bar{z}=1$ singularity occurs at vanishing separation $\th = 0$; this is the standard UV divergence that arises when two CFT operators are brought close together. To diagnose the strength of the singularity, we examine the behavior of the Fourier coefficients of (\ref{eqn:equal-time}) at large $|\ell|$. The ratio of Gamma functions grows like $\big( |\ell|/\a \big)^{2\Delta - 1}$; meanwhile, $|f_{n\ell}|^2$ approaches 1 nonperturbatively quickly as $|\ell| \too \infty$.\footnote{By this we mean that the Taylor series of $|f_{n\ell}|^2$ at $|\ell| = \infty$, i.e. in powers of $\f{1}{|\ell|/\a}$, is identically 1.} We substitute $f_{n\ell} \sim 1$ and the asymptotic series for the ratio of Gamma functions \cite{Erdelyi:1951aa} into (\ref{eqn:equal-time}), resum the resulting Fourier series term by term, and expand in powers of $\th$ to extract the divergence structure:
\begin{equation}
\begin{aligned}
\label{eqn:equal-time-powerseries}
G(\th) &\sim \f{\a^{2\Delta - 1}}{2\G(2\Delta) \cos\p{\pi \Delta}} \sum_{\ell \in \Z}\p{\big(|\ell|/\a\big)^{2\Delta - 1} + 
O\p{\big(|\ell|/\a\big)^{2\Delta - 3}}} e^{i\ell\th} \\ &= \th^{-2\Delta} \p{1 + \f{\Delta \a^2}{12} \th^2 + \f{\Delta(5\Delta + 1)\a^4}{1440} \th^4 + O\p{\th^6}}.
\end{aligned}
\end{equation}

This analysis shows that $f_{n\ell}$ is not felt at large $|\ell|$ and cannot contribute to the singular part of the correlator. (Intuitively, the large-$|\ell|$ modes cannot tell the difference between conical AdS and BTZ because only modes with small $|\ell|$ probe deep into the bulk.) It does, however, affect the first few terms of the series, which contribute mainly to the part of $G(z,\bar{z})$ that is regular at $z, \bar{z} = 1$. We will show in Section~\ref{sec:expectation-values} that the regular part of $G(z, \bar{z})$ contains important information about the expectation values of certain two-particle states in heavy backgrounds, and that it encodes certain OPE data in the dual CFT.

\section{The Method of Images}
\label{sec:correlators-method-of-images}

In spacetimes that can be realized as quotients of $\up{AdS}_3$ by the action of some group, we can obtain the bulk and boundary propagators by passing to the pure AdS covering space and using the method of images. In this section, we review this construction, discuss the singular and regular parts of the propagator, explain the connection between the method of images and the mode sums of Section~\ref{sec:correlators-wave-equation}, and relate the images to bulk geodesics.

\subsection{The Sum Over Images}
\label{sec:sum-over-images}

The bulk-to-bulk propagator in pure $\up{AdS}_3$ is known and can be expressed as a function of the geodesic distance $\s(x,x')$ between two bulk points. It takes the form \cite{Hijano:2015qja}
\begin{align}
\label{eqn:pure-AdS-propagator}
G_{\up{AdS}}(x,x') = \f{e^{-\Delta \s(x,x')}}{e^{-2\s(x,x')} - 1} = \p{\f{\xi}{2}}^{\Delta} {}_2 F_1 \p{\f{\Delta}{2}, \f{\Delta + 1}{2}, \Delta; \xi^2} = \f{\xi^{\Delta} \p{1 + \sqrt{1 - \xi^2}}}{2 \sqrt{1 - \xi^2}},
\end{align}
where $\s(x,x') = \cosh^{-1}(1/\xi)$, and $\xi$ is given in global coordinates by
\begin{align}
\xi \equiv \f{1}{\cosh \s(x,x')} = \f{1}{\sqrt{\p{r^2 + 1} \p{r'^2 + 1}} \cos(t-t') - r r' \cos(\th-\th')}.
\end{align}
On the boundary, (\ref{eqn:pure-AdS-propagator}) reduces to the standard result
\begin{align}
G_{\up{AdS}}(t, \th) = \f{1}{\big(2 \cos t - 2 \cos \th\big)^{\Delta}}.
\end{align}

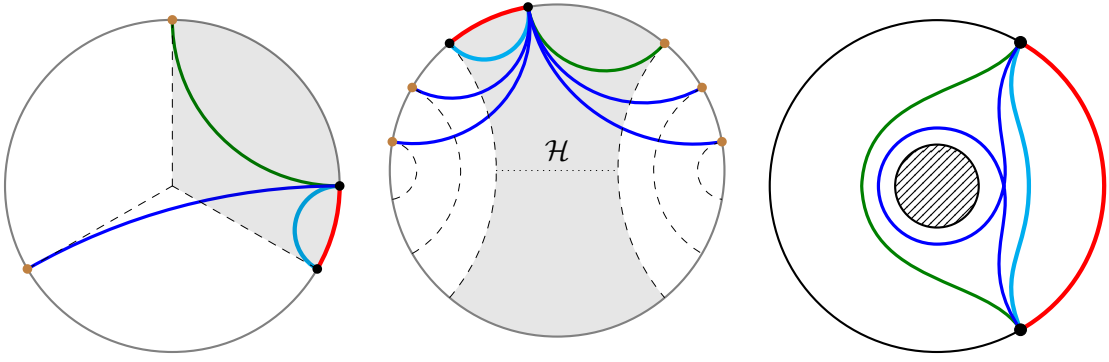
\begin{figure}[t]
\centering
\begin{tikzpicture}[scale=2.2] 

\tkzDefPoint(0,0){O}
\tkzDefPoint(0:1.001){p}
\tkzDefPoint(330:1.001){q0}
\tkzDefPoint(90:1.001){q1}
\tkzDefPoint(210:1.001){q2}
\tkzDrawCircle[thick](O,p)

\tkzDefCircle[orthogonal through = p and q0](O,p)
\tkzGetPoint{z0}
\tkzDefCircle[orthogonal through = p and q1](O,p)
\tkzGetPoint{z1}
\tkzDefCircle[orthogonal through = p and q2](O,p)
\tkzGetPoint{z2}
\tkzDrawArc[cyan, ultra thick](z0,p)(q0)
\tkzDrawArc[green!50!black, very thick](z1,q1)(p)
\tkzDrawArc[blue, very thick](z2,p)(q2)

\tkzDrawSegment[dashed](q0,O)
\tkzDrawSegment[dashed](q1,O)
\tkzDrawSegment[dashed](q2,O)
\tkzFillSector[black, very nearly transparent](O,q0)(q1)
\tkzDrawArc[ultra thick, red](O,q0)(p)
\tkzDrawPoints[ultra thick](p,q0)
\tkzDrawPoints[ultra thick, brown](q1,q2)

\end{tikzpicture}
\begin{tikzpicture}[scale=2.2] 
\clip(0,0) circle (1.1);

\tkzDefPoint(0,0){O}
\tkzDefPoint(100:1.001){p}

\tkzDefPoint(130:1.001){a1}
\tkzDefPoint(150:1.001){a2}
\tkzDefPoint(170:1.001){a3}
\tkzDefPoint(50:1.001){b1}
\tkzDefPoint(30:1.001){b2}
\tkzDefPoint(10:1.001){b3}
\tkzDefPoint(-130:1.001){ap1}
\tkzDefPoint(-150:1.001){ap2}
\tkzDefPoint(-170:1.001){ap3}
\tkzDefPoint(-50:1.001){bp1}
\tkzDefPoint(-30:1.001){bp2}
\tkzDefPoint(-10:1.001){bp3}

\tkzDefCircle[orthogonal through = p and a1](O,p)
\tkzGetPoint{ca1}
\tkzDefCircle[orthogonal through = p and a2](O,p)
\tkzGetPoint{ca2}
\tkzDefCircle[orthogonal through = p and a3](O,p)
\tkzGetPoint{ca3}
\tkzDefCircle[orthogonal through = p and b1](O,p)
\tkzGetPoint{cb1}
\tkzDefCircle[orthogonal through = p and b2](O,p)
\tkzGetPoint{cb2}
\tkzDefCircle[orthogonal through = p and b3](O,p)
\tkzGetPoint{cb3}

\tkzDefCircle[orthogonal through = a1 and ap1](O,p)
\tkzGetPoint{za1}
\tkzDefCircle[orthogonal through = a2 and ap2](O,p)
\tkzGetPoint{za2}
\tkzDefCircle[orthogonal through = a3 and ap3](O,p)
\tkzGetPoint{za3}
\tkzDefCircle[orthogonal through = b1 and bp1](O,p)
\tkzGetPoint{zb1}
\tkzDefCircle[orthogonal through = b2 and bp2](O,p)
\tkzGetPoint{zb2}
\tkzDefCircle[orthogonal through = b3 and bp3](O,p)
\tkzGetPoint{zb3}

\tkzFillCircle[very nearly transparent](O,p)
\tkzFillSector[white](za1,ap1)(a1)
\tkzFillSector[white](zb1,b1)(bp1)
\tkzDrawCircle[thick](O,p)

\tkzDefPoint(1,0){p1}
\tkzInterLC[](O,p1)(za1,a1)
\tkzGetPoints{d1}{d2}
\tkzInterLC[](O,p1)(zb1,b1)
\tkzGetPoints{e1}{e2}
\tkzDrawSegment[dotted](d2,e1)
\tkzLabelSegment[above](d2,e1){$\cal{H}$}

\tkzDrawArc[cyan, ultra thick](ca1,a1)(p)
\tkzDrawArc[blue, very thick](ca2,a2)(p)
\tkzDrawArc[blue, very thick](ca3,a3)(p)
\tkzDrawArc[green!50!black, very thick](cb1,p)(b1)
\tkzDrawArc[blue, very thick](cb2,p)(b2)
\tkzDrawArc[blue, very thick](cb3,p)(b3)

\tkzDrawArc[black, dashed](za1,ap1)(a1)
\tkzDrawArc[black, dashed](za2,ap2)(a2)
\tkzDrawArc[black, dashed](za3,ap3)(a3)
\tkzDrawArc[black, dashed](zb1,b1)(bp1)
\tkzDrawArc[black, dashed](zb2,b2)(bp2)
\tkzDrawArc[black, dashed](zb3,b3)(bp3)

\tkzDrawArc[ultra thick, red](O,p)(a1)
\tkzDrawPoints[ultra thick](p, a1)
\tkzDrawPoints[ultra thick, brown](a2,a3,b1,b2,b3)

\end{tikzpicture}
\;
\begin{tikzpicture}[scale=1.1] 

\draw[thick] (0,0) circle (2); 
\draw[thick, pattern=north east lines] (0,0) circle (.5); 
\draw[ultra thick, red] (-60:2) arc (-60:60:2) {};

\draw[ultra thick, cyan] 
(60:2) .. controls (.7,1) and (1.1,.8) .. (1.1,0)
       .. controls (1.1,-.8) and (.7,-1) .. (-60:2);

\draw[very thick, green!50!black] 
(60:2) .. controls (.6,1.1) and (-.85,1.2) .. (-.9,0)
       .. controls (-.85,-1.2) and (.6,-1.1) .. (-60:2);

\draw[very thick, blue]
(60:2) .. controls (.5,1) and (.9,.5) .. (.8,0)
       .. controls (.7,-.5) and (.5*.7,-.7) .. (0,-.7)
       .. controls (-.55*.7,-.7) and (-.7,-.55*.7) .. (-.7,0)
       .. controls (-.7,.55*.7) and (-.55*.7,.7) .. (0,.7)
       .. controls (.5*.7,.7) and (.7,.5) .. (.8,0)
       .. controls (.9,-.5) and (.5,-1) .. (-60:2);

\filldraw (60:2) circle (2 pt);
\filldraw (-60:2) circle (2 pt);

\end{tikzpicture}
\caption[LoF entry]{The quotient structure of $\up{AdS}_3/\Z_N$ (left) and BTZ (middle) is shown on constant-time slices of $\up{AdS}_3$. The fundamental domain (gray) is embedded in pure AdS, and images of this domain (dashed lines) are identified in the quotient. Two field insertions (black points) are separated by some opening angle (red), and the action of the quotient group rotates one of them to image points (brown) in the AdS cover. As the fields are brought together, the renormalized length of the minimal geodesic (cyan) diverges, while the other geodesics have finite lengths because their endpoints remain well separated. \\ \hspace*{15pt} The winding number of each geodesic is equal to the number of dashed lines it crosses. In BTZ, we also require that the geodesics do not cross the horizon (dotted line). The view from the quotient geometry (right) shows that in addition to $\g_0$, there is always another non-winding geodesic (green) which competes for dominance with $\g_0$ for opening angles near $\th = \pi$. In BTZ and $\up{AdS}_3/\Z_N$ with $N \geq 3$, there are also winding geodesics (blue).}
\label{fig:quotients}
\end{figure}

Below the BTZ threshold, the method of images applies only to the orbifolds $\up{AdS}_3/\Z_N$, i.e. for $N \in \Z_+$. Above the threshold, it applies to black holes of any mass, since every BTZ spacetime is a quotient of $\up{AdS}_3$ by an action of $\Z$. The quotient structure is depicted in Figure~\ref{fig:quotients}. In both cases, we start from the pure AdS propagator (\ref{eqn:pure-AdS-propagator}), pass to the quotient by rescaling the coordinates $(t, \th, r) \mapsto (\a t, \a\th, \f{r}{\a})$, and sum over images that rotate one of the insertion points by an angle $2\pi k \a$, where $k \in \{0, ..., N-1\}$ in conical AdS and $k \in \Z$ in BTZ. Each field insertion also picks up a factor of $\a^{\Delta}$ in response to the coordinate transformation. As a result, the bulk-to-bulk propagator in an $\up{AdS}_3$ quotient is
\begin{align}
\label{eqn:bulk-to-bulk-images}
G(x,x') &= \a^{2\Delta} \sum_{k} \f{e^{-\Delta \s_k(x, x')}}{e^{-2\s_k(x,x')} - 1}, \quad
\s_k(x,x') = \s\p{\a t, \a t';\, \a(\th + 2\pi k), \a\th';\, \f{r}{\a}, \f{r'}{\a}}.
\end{align}
It will also be useful, for future reference, to record the boundary-to-boundary propagator below and above the threshold. Sending $r \too \infty$ in (\ref{eqn:bulk-to-bulk-images}), we obtain
\begin{subequations}
\label{eqn:images}
\begin{align}
\label{eqn:images-conical}
G_{\up{con}}(t,\th) &= \sum_{k=0}^{N-1} \f{N^{-2\Delta}}{\Big[2\cos\p{\f{t}{N}} - 2\cos\p{\f{\th + 2\pi k}{N}} \Big]^{\Delta}}, \\
\label{eqn:images-BTZ}
G_{\up{BTZ}}(t,\th) &= \sum_{k \in \Z} \f{r_0^{2\Delta}}{\Big[2\cosh\!\big(r_0(\th + 2\pi k)\big) - 2\cosh\p{r_0 t} \Big]^{\Delta}}. 
\end{align}
\end{subequations}

The results (\ref{eqn:images}) can be unified using the holomorphic coordinates $(z, \bar{z})$, which become $(z^{\a}, \bar{z}^{\a})$ in the quotient geometry. The images are generated by the action $z^{\a} \mapsto \zeta^k z^{\a}$, where $\zeta = e^{2\pi i \a}$ represents a $2\pi\a$ rotation.\footnote{Below the BTZ threshold, $\zeta = e^{2\pi i /N}$ is a root of unity. Above the threshold, $\zeta = e^{-2\pi r_0}$ is real. In holomorphically factorized expressions like (\ref{eqn:propagator-images}), we take the ``conjugate'' of $\zeta$ to be $e^{+2\pi r_0}$.} The boundary propagator in $\up{AdS}_3$ is the CFT two-point function $G_{\up{AdS}}(z, \bar{z}) = \abs{1-z}^{-2\Delta}$, so we have
\begin{align}
\label{eqn:propagator-images}
G(z, \bar{z}) = \abs{z}^{(\a - 1)\Delta} \sum_k \abs{\f{1 - \zeta^k z^{\a}}{\a}}^{-2\Delta} \qquad \up{with} \quad \zeta = e^{2\pi i \a}
\end{align}
for any value of $\a$. The only remaining difference between the conical and BTZ cases is the range of the sum, reflecting the structure of the quotient ($\Z_N$ vs. $\Z$).

\subsection{Short-Distance Behavior}
\label{sec:singular-structure}

As the two field insertions are brought close together, the propagator diverges. The nature of this singularity is controlled by the direct ($k=0$) image in (\ref{eqn:bulk-to-bulk-images}--\ref{eqn:propagator-images}), which inherits its UV behavior from the pure AdS propagator. All of the other images are regular in the coincident limit, since they correspond to configurations where the field insertions remain well separated (see Figure~\ref{fig:quotients}). This gives us a natural decomposition of the propagator into its regular and singular parts: we have $G(x,x') = G_0(x,x') + G_{\up{reg}}(x,x')$, with
\begin{align}
G_0(x,x') = \a^{2\Delta} \f{e^{-\Delta \s_0(x, x')}}{e^{-2\s_0(x,x')} - 1}, \qquad 
G_{\up{reg}}(x,x') = \a^{2\Delta} \sum_{k \neq 0} \f{e^{-\Delta \s_k(x, x')}}{e^{-2\s_k(x,x')} - 1}.
\end{align}

At short distances, the bulk-to-bulk propagator exhibits a divergence that goes like the inverse of the shortest geodesic length $\s_0(x,x')$ between nearby points:
\begin{equation}
\begin{aligned}
G(x,x') \approx G_0(x,x') = \f{\a^{2\Delta} e^{-\Delta \s_0}}{e^{-2\s_0} - 1} = -\f{\a^{2\Delta}}{2\s_0} + (\up{regular}) \qquad \up{as} \quad x \too x'.
\end{aligned}
\end{equation}
For short radial geodesics, the geodesic length is approximately $\s_0(x,x') \sim \f{r}{\a}$; therefore $G(x,x')$ diverges like $1/r$. We may understand this behavior physically by zooming in on two nearby points with $|r - r'| \ll R_{\up{AdS}} = 1$. At these scales the curvature of AdS becomes negligible, so $G(x,x')$ approaches the flat-space propagator in 3 dimensions---but this is nothing more than the Green's function $1/|r-r'|$ for the flat-space Laplacian.

The short-distance behavior of the boundary propagator is markedly different, since points on the conformal boundary with very small $t-t'$ and $\th - \th'$ are still formally infinitely far apart. The ``coincident'' limit is still controlled by the $k=0$ image, but this time $G(z,\bar{z})$ exhibits a much stronger power-law divergence. From (\ref{eqn:propagator-images}), near $z, \bar{z} = 1$ we have
\begin{align}
\label{eqn:singular-part}
G_0(z, \bar{z}) = |z|^{(\a - 1)\Delta} \abs{\f{1-z^{\a}}{\a}}^{-2\Delta} = \abs{1-z}^{-2\Delta}\Big(1 + O\big(|1-z|\big){}_{\!}\Big).
\end{align}
Meanwhile, the regular part of $G(z,\bar{z})$ remains finite at $z, \bar{z} = 1$:
\begin{align}
\label{eqn:regular-part}
G_{\up{reg}}(z, \bar{z}) = \abs{z}^{(\a - 1)\Delta} \sum_{k \neq 0} \abs{\f{1 - \zeta^k z^{\a}}{\a}}^{-2\Delta} = \sum_{k \neq 0} \abs{\f{2}{\a} \sin(\pi k\a)}^{-2\Delta} + O\big(|1-z|\big).
\end{align}

It can be checked that $G_0(z, \bar{z})$ has the same singular structure as the full mode sum (\ref{eqn:mode-sum}). This is easiest to see explicitly by comparison with the equal-time propagator. Taking $z = e^{i\th}$ and expanding (\ref{eqn:singular-part}) in powers of $\th \ll 1$, we get
\begin{align}
G_0(\th) = \p{\f{2}{\a} \sin\p{\f{\a\th}{2}}}^{-2\Delta} = \th^{-2\Delta} \p{1 + \f{\Delta \a^2}{12} \th^2 + \f{\Delta(5\Delta + 1)\a^4}{1440} \th^4 + O\big(\th^6\big){}_{\!}},
\end{align}
in agreement with (\ref{eqn:equal-time-powerseries}). We see that $G_0(z, \bar{z})$ characterizes the asymptotic behavior of the mode sum at large $|\ell|$, while $G_{\up{reg}}(z, \bar{z})$ is influenced by the small-$|\ell|$ terms in the series, and is sensitive to the thermal factor $f_{n\ell}(\b)$ above the threshold.

\subsection{Analytic Continuation in $N$}
\label{sec:analytic-continuation}

While the BTZ propagator is manifestly analytic in $r_0$, in conical AdS the sum over images is finite and only defined for integer $N$. However, most conical geometries have $N \notin \Z$ and are not quotients of $\up{AdS}_3$, so it is desirable to analytically continue (\ref{eqn:images-conical}) to arbitrary $N$. A partial solution to this problem is to approximate $G(z, \bar{z})$ by its singular term $G_0(z, \bar{z})$, which is analytic in $N$ and dominates near $z, \bar{z} = 1$. But this is a poor approximation away from $z, \bar{z} = 1$, and it does not explain how one might also analytically continue $G_{\up{reg}}(z,\bar{z})$ to non-integer $N$. Below, we explain how to proceed more carefully.

\paragraph{The full correlator.} We begin by applying Newton's binomial formula
\begin{align}
\p{1 - x}^{-\Delta} = \f{1}{\G(\Delta)} \sum_{m=0}^{\infty} \f{\Gamma(\Delta + m)}{\Gamma(1 + m)} x^m, \qquad |x| < 1,
\end{align}
in (\ref{eqn:propagator-images}), taking $x = \zeta^k z^{\a}$. Expanding in $z$ and $\bar{z}$ separately, we obtain
\begin{align}
\label{eqn:images-to-modes}
G(z,\bar{z}) &= \f{\a^{2\Delta}}{\Gamma(\Delta)^2} \abs{z}^{(\a - 1)\Delta} \sum_{k=0}^{N-1} \sum_{m,m'=0}^{\infty} \f{\Gamma(\Delta + m) \Gamma(\Delta + m')}{\Gamma(1 + m) \Gamma(1 + m')} z^{\a m} \bar{z}^{\a m'} \zeta^{k(m-m')}.
\end{align}
Now we can sum over the roots of unity $\zeta^{k(m-m')}$ to get rid of the sum over $k$. When this is done carefully,\footnote{The sum of the roots of unity is a Kronecker-delta comb spaced by multiples of $N$,
\begin{align}
\sum_{k = 0}^{N-1} e^{2\pi i k(m-m')/N} = N \sum_{\ell \in \Z} \d_{m,m' + N\ell}.
\end{align}
Substituting this into (\ref{eqn:images-to-modes}) sets $m' = m + N|\ell|$ and replaces the sum over $m'$ with a sum over $\ell$. The absolute value is necessary because we may have either $\ell \geq 0$ or $\ell < 0$, but $m'$ must be nonnegative.} one obtains precisely the sum over normal modes (\ref{eqn:mode-sum}). We conclude that the method of images knows about the wave equation; and vice versa, the sum over modes analytically continues the sum over images to cases where it does not apply. As we will discuss in Section~\ref{sec:crossing-inversion}, this calculation implements the crossing symmetry of CFT four-point functions and is related to the conformal bootstrap.

\paragraph{Regular and singular parts.} This calculation shows that $G_{\up{reg}}(z, \bar{z})$ may be analytically continued to non-integer $N$ by writing it as the difference between the mode sum (\ref{eqn:mode-sum}) and the singular image (\ref{eqn:singular-part}). This prescription can be somewhat unwieldy, however, because it involves the direct subtraction of two divergent quantities.

We can do better by taking advantage of a symmetry: in $\up{AdS}_3/\Z_N$, the $k$th image in (\ref{eqn:propagator-images}) is equal to the $(N-k)$th image; and in BTZ, the $k$th image is equal to the $-k$th image. Moreover, near $z, \bar{z} = 1$, the dominant contribution to $G_{\up{reg}}(z, \bar{z})$ comes from (twice) the $k=1$ image. All of the others are suppressed because they correspond to geodesics with nontrivial winding around the singularity---see the discussion below. By keeping only one contribution, we get an approximate analytic continuation of $G_{\up{reg}}(z,\bar{z})$:
\begin{align}
\label{eqn:Greg-approximate}
G_{\up{reg}}(z, \bar{z}) \approx 2|z|^{(\a - 1)\Delta} \abs{\f{1 - \zeta z^{\a}}{\a}}^{-2\Delta}, \qquad \a \in i\R_+ \quad \up{or} \quad \a \leq \f{1}{3}.
\end{align}
In conical AdS, the condition $N \geq 3$ is necessary because the $k=1$ and $k = N-1$ images do not really exist for defect geometries with fewer than 3 distinct images. That said, (\ref{eqn:Greg-approximate}) is exact for $N=3$, and is also exact for $N=2$ without the leading factor of 2.

\subsection{Images and Geodesics}
\label{sec:images-geodesics}

The form of the propagator obtained from the method of images has a nice geometric interpretation in terms of the geodesic approximation for heavy correlators. For sufficiently massive fields, $G(z, \bar{z})$ is well approximated by a saddle-point sum over all bulk geodesics between two boundary points \cite{Balasubramanian:1999zv}. There are $N$ such geodesics in $\up{AdS}_3/\Z_N$, and infinitely many in BTZ. Each geodesic $\g_k$ contributes $e^{-m \L_{\up{ren}}[\g_k]}$ to the propagator, where $\L_{\up{ren}}[\g_k]$ is the renormalized geodesic length of $\g_k$. Direct calculations of the geodesic length \cite{Berenstein:2022ico, Martinez:2019nor} show that each of the images in (\ref{eqn:images}) have precisely this form.

In general, the geodesic approximation is valid only for very heavy fields; in particular, one requires that $\Delta = 1 + \sqrt{1 + m^2} \approx m \gg 1$. However, in the present case the method of images is exact for any $\Delta$ and matches the geodesic method, so in fact the assumption $\Delta \gg 1$ is unnecessary. What we do need to assume is that the field is light enough that it does not backreact on the geometry. This means that the CFT operator dual to $\phi$ has a conformal dimension independent of the central charge, which is taken to be very large. Probe operators with dimensions $1 \ll \D \ll c$ are sometimes called ``hefty,'' to distinguish them from heavy operators whose dimensions scale (at least) linearly with $c$.

For nearby boundary points, the singular part dominates the correlator because it corresponds to $\g_0$, which is always the shortest geodesic between nearby two points. The other geodesics either pass around the other side of the conical singularity or horizon, or else have a nontrivial winding number around it. As a result, they are all longer than $\g_0$, and their contributions to the propagator are exponentially suppressed at large $\Delta$. But for larger opening angles, $\g_0$ starts to compete for dominance with another geodesic on the other side of the boundary cylinder. In conical AdS, this is $\g_{N-1}$, while in BTZ, it is $\g_{-1}$: see Figure~\ref{fig:quotients}. This Ryu--Takayanagi phase transition \cite{Hartman:2013mia} occurs at $\th = \pi$ (i.e. $z, \bar{z} = -1$) and cannot be seen solely by looking at the singular part of the correlator.

\section{HHLL Correlators in 2d CFT}
\label{sec:correlators-2d-cft}

We now turn to the boundary description of heavy states in $\up{AdS}_3$ gravity. We consider a holographic 2d CFT with large central charge $c \gg 1$, working at leading order in the $1/c$ expansion. In this setting, the propagator $G(z,\bar{z})$ becomes a heavy-heavy-light-light (HHLL) four-point function, and to compute it we need to understand the spectrum of the theory and the Virasoro blocks that contribute to the OPE. The computation of $G(z,\bar{z})$ is based on techniques developed by Fitzpatrick, Kaplan, and Walters \cite{Fitzpatrick:2014vua, Fitzpatrick:2015zha}, and in this section we use their methods to interpret the results of Sections~\ref{sec:correlators-wave-equation} and~\ref{sec:correlators-method-of-images} holographically. 

\subsection{The Spectrum of Primaries}
\label{sec:spectrum-of-primaries}

At the semiclassical level, the spectrum of a holographic 2d CFT dual to pure $\up{AdS}_3$ gravity is universal: besides the identity operator, it contains a continuum of heavy primaries $\O_{\up{H}}$ whose conformal dimensions lie above the BTZ threshold $\Delta_* = \f{c}{12}$, with a Cardy density of states \cite{Hartman:2014oaa}. These states are dual to black holes,\footnote{This is true in a coarse-grained sense. The spectrum of a holographic CFT is conjectured to be discrete and chaotic, and one should really average over the heavy states in a small energy window or over an ensemble of 2d CFTs \cite{Collier:2019weq, Chandra:2022bqq}. However, all of these subtleties concern effects that are subleading or exponentially suppressed in $c$, and are therefore undetectable in the semiclassical limit.} and their dimensions
\begin{align}
\label{eqn:heavy-dimension}
\Delta_{\up{H}} = M = \f{c}{12} \p{1 - \a^2}, \qquad \a \equiv \sqrt{1 - \f{12 \Delta_{\up{H}}}{c}}
\end{align}
correspond to the ADM masses of the BTZ spacetimes if $\a \in i \R_+$ is identified with the parameter introduced in (\ref{eqn:analytic-continuation}). To this universal part of the spectrum, we can add matter in two ways. Firstly, we will add conical defects by allowing $\a$ to take on real values in $[0,1]$, so that  $\Delta_{\up{H}}$ falls below the threshold and causes $\O_{\up{H}}$ to create a static particle in the bulk with mass (\ref{eqn:heavy-dimension}). And secondly, we will add by hand a light scalar primary $\O_{\up{L}}$ of conformal dimension $\Delta_{\up{L}} = \Delta$ independent of $c$. We take $\O_{\up{L}}$ to be a generalized free field, so that it is dual in the bulk to the free scalar $\phi$ studied in Section~\ref{sec:correlators-wave-equation}. By definition, all correlators of $\O_{\up{L}}$ are determined from its two-point function via Wick's theorem, and its OPE data is universal and governed by mean-field theory (MFT).

We may use both $\O_{\up{L}}$ and $\O_{\up{H}}$ to build composite states. Given any two primaries $\O_1$ and $\O_2$, one can construct an infinite family of double-trace or \emph{double-twist} primaries that must appear in their OPE and have the schematic form $[\O_1 \O_2]_{n\ell} \sim\, :\!\O_1 \Box^n \overset{{}_\leftrightarrow}{\pd} {}^{\ell} \O_2\!: - \,(\up{traces})$. The trace subtraction is necessary to ensure that $[\O_1 \O_2]_{n\ell}$ is a primary; moreover, if $\O_1$ and $\O_2$ are identical, then the spin $\ell$ must be even. These operators should be thought of as two-particle states whose bulk description involves an orbiting configuration of $\O_1$ and $\O_2$, with angular momentum $\ell$ and excitation energy labeled by $n$. Depending on $\O_1$ and $\O_2$, there are three classes of double-twist states to consider, described below. The operator content that will be important to us is summarized in Table~\ref{tab:spectrum}.

\paragraph{LL double-twists.} If both $\O_1$ and $\O_2$ are light, then at large $|\ell|$ the spectrum of $[\O_1 \O_2]_{n\ell}$ approaches the universal MFT spectrum of double-trace operators \cite{Fitzpatrick:2014vua}:
\begin{align}
\label{eqn:double-twist-spectrum}
h_{[\up{LL}]_{n\ell}} = \begin{cases} 
h_1 + h_2 + n + \ell, & \ell \geq 0, \\
h_1 + h_2 + n, & \ell < 0;
\end{cases} \qquad
\bar{h}_{[\up{LL}]_{n\ell}} = \begin{cases} 
\bar{h}_1 + \bar{h}_2 + n, & \ell \geq 0, \\
\bar{h}_1 + \bar{h}_2 + n - \ell, & \ell < 0.
\end{cases}
\end{align}

In dealing with OPEs that sum over the light-light (LL) double-twist spectrum, we will occasionally find it useful to change summation variables from $(n,\ell)$ to $(n, n+|\ell|)$. To accomplish this, we can take advantage of holomorphic factorization to decompose $[\O_1 \O_2]_{n\ell}(z, \bar{z})$ into a product of chiral and anti-chiral primaries labeled by nonnegative integers $m,m' \in \N$ that satisfy $m+m' = 2n + |\ell|$ and $m-m' = \ell$. We write
\begin{align}
\label{eqn:chiral-double-twist}
[\O_1 \O_2]_{n\ell}(z, \bar{z}) = [\O_1 \O_2]_m(z) [\O_1 \O_2]_{m'}(\bar{z}), \qquad n = \min(m,m'), \quad \ell = m-m'.
\end{align}
These primaries have weights $h_m = h_1 + h_2 + m$ and $\bar{h}_{m'} = \bar{h}_1 + \bar{h}_2 + m'$, respectively. We will focus on LL states with $\O_1 = \O_2 = \O_{\up{L}}$. The simplest such primary is the composite field $\O_{\up{L}}^2 = [\O_{\up{L}} \O_{\up{L}}]_{00}$, but the whole family of LL double-twists will be important to us.

\paragraph{HL double-twists.} If $\O_1$ is heavy and $\O_2$ is light, then the twist $\t = \min(h, \bar{h})$ of the states $[\O_{\up{H}} \O_{\up{L}}]_{n\ell}$ must be rescaled by a factor of $\a$ relative to (\ref{eqn:double-twist-spectrum}):
\begin{align}
\label{eqn:double-twist-dimension}
\Delta_{[\up{LL}]_{n\ell}} \equiv \Delta_{n\ell} = \Delta_1 + \Delta_2 + 2n + |\ell|, \qquad
\Delta_{[\up{HL}]_{n\ell}} = \Delta_1 + \a\p{\Delta_2 + 2n} + |\ell|.
\end{align}
The HL double-twists $[\O_{\up{H}} \O_{\up{L}}]_{n\ell}$ are dual to the bulk field modes $\phi_{n\ell}$. This can be seen from their spectrum, which reproduces the energies (\ref{eqn:normal-frequencies}) of the modes on top of the heavy state: $\Delta_{[\up{HL}]_{n\ell}} = \Delta_{\up{H}} + \w_{n\ell}$. (In pure AdS, these modes are descendants of $\O_{\up{L}}$.)

\paragraph{HH double-twists.} Heavy-heavy (HH) double-twist operators are dual to bulk states containing two defects. The range of dimensions these states can have is nontrivial: if the defects are too heavy, it is impossible to combine them without creating a BTZ black hole \cite{Gott:1990zr}. Such high-energy states make for very short-lived, unstable virtual particles, and processes involving them are exponentially suppressed in the central charge. Therefore in the present semiclassical calculation, it is safe to essentially ignore them.

\begin{table}
\centering
\begin{tabular}{ c | c | c | c }
\bf{Primary} & \bf{Dimension} $\Delta$ & \bf{Spin} $\ell$ & \bf{AdS interpretation} \\
$\O$ & $h + \bar{h}$ & $h - \bar{h}$ & (bulk excitations) \\ \hline
$\one$ & $0$ & $0$ & pure AdS vacuum \\
$\O_{\up{L}}$ & $\Delta_{\up{L}} = \Delta$ & $0$ & free scalar in AdS \\
$\O_{\up{H}}$ & $\Delta_{\up{H}} = \f{c}{12} \p{1 - \a^2}$ & $0$ & heavy background \\
$[\O_{\up{L}} \O_{\up{L}}]_{n\ell}$ & $\Delta_{n\ell} = 2\Delta + 2n + |\ell|$ & $\ell \in 2\Z$ & light 2-particle state \\
$[\O_{\up{H}} \O_{\up{L}}]_{n\ell}$ &  $\Delta_{\up{H}} + \a\p{\Delta + 2n} + |\ell|$ & $\ell \in \Z$ & orbiting light mode
\end{tabular}
\caption{The spectrum $(h, \bar{h})$ of the primary operators important for our calculations.}
\label{tab:spectrum}
\end{table}

\subsection{The Structure of the OPE}
\label{sec:structure-of-ope}

We work in radial quantization, using the coordinates $z = e^{\t + i\th}$ and $\bar{z} = e^{\t - i\th}$ (with $\t = it$ the Euclidean time) to map the cylinder at the boundary of $\up{AdS}_3$ to the complex plane. The boundary correlator $G(z,\bar{z})$ is the two-point function of $\O_{\up{L}}$ in the heavy state $\ket{\O_{\up{H}}} = \O_{\up{H}}(0) \ket{0}$, or in other words the normalized HHLL four-point function
\begin{align}
\label{eqn:4-pt-function}
G(z, \bar{z}) = \avg{\O_{\up{H}}(\infty) \O_{\up{L}}(1) \O_{\up{L}}(z, \bar{z}) \O_{\up{H}}(0)}.
\end{align}
(Here the normalization $\O(\infty) = \lim_{y \to \infty} |y|^{2\Delta_{\O}} \O(y, \bar{y})$ is understood implicitly.)

Let us review how to compute correlation functions like $G(z, \bar{z})$ by using the OPE to sum over intermediate states. We start by inserting into (\ref{eqn:4-pt-function}) a complete set of states $\one = \sum_{\psi} \ket{\psi}\!\bra{\psi}$, where the basis runs over all of the primaries $\O_{h, \bar{h}}$ in the spectrum and their descendants. This decomposes $G(z,\bar{z})$ into a sum of primary and descendant three-point functions; by convention, we keep the sum over descendants implicit. The OPE becomes a sum over the weights $(h, \bar{h})$ of all primaries in the theory, and assumes one of two forms depending on whether distinct or identical operators are paired up in (\ref{eqn:4-pt-function}):
\begin{subequations}
\label{eqn:s-t-OPE}
\begin{align}
\label{eqn:s-OPE}
G(z, \bar{z}) &= \avg{\wick[sep=0pt, offset=1.3em]{
\c1 \O_{\up{H}} \c1 \O_{\up{L}} \c2 \O_{\up{L}} \c2 \O_{\up{H}}}} = 
\sum_{\psi} \bra{\O_{\up{H}} \O_{\up{L}}}\ket{\psi}\!\bra{\psi}\ket{\O_{\up{L}} \O_{\up{H}}} = 
\sum_{h, \bar{h}} \big(\cal{C}_{h, \bar{h}}^{\up{HL}}\big)^2 
\cal{F}^{(\up{s})}_{h} \overline{\cal{F}}^{(\up{s})}_{\bar{h}} \\
\label{eqn:t-OPE}
&=
\avg{\wick{\c1 \O_{\up{H}} \c2 \O_{\up{L}} \c2 \O_{\up{L}} \c1 \O_{\up{H}}}} = 
\sum_{\psi} \bra{\O_{\up{L}} \O_{\up{L}}}\ket{\psi}\!\bra{\psi}\ket{\O_{\up{H}} \O_{\up{H}}} =
\sum_{h, \bar{h}} \cal{C}^{\up{LL}}_{h, \bar{h}} \cal{C}^{\up{HH}}_{h, \bar{h}} 
\cal{F}^{(\up{t})}_{h} \overline{\cal{F}}^{(\up{t})}_{\bar{h}}.
\end{align}
\end{subequations}
The holomorphically factorizing functions $\cal{F}_h(z) \overline{\cal{F}}_{\bar{h}}(\bar{z})$ are Virasoro blocks: they encode the contributions to $G(z,\bar{z})$ of a primary and all of its descendants by summing up the descendant three-point functions implicit in (\ref{eqn:s-t-OPE}). Because CFT three-point functions are fixed by conformal symmetry up to normalization, the Virasoro blocks are (in principle) also fixed by symmetry, while the corresponding OPE coefficients $\cal{C}_{h, \bar{h}}$ are left unfixed.

What are the intermediate states that propagate in each OPE channel? Because $\O_{\up{L}}$ is a generalized free field, only the identity and the double-twist primaries can contribute. Their contributions in both OPE channels can be summarized pictorially:
\begin{align}
\label{eqn:OPE-diagram}
\displaystyle \sum_{n\ell}
\scalebox{0.7}{$\abs{
\begin{gathered}
\begin{tikzpicture}
    \begin{feynman}
    \large
    \vertex (a);
    \vertex [above left = 0.1cm and 1.5cm of a] (a1) {\( \O_{\up{L}} \)};
    \vertex [above right = 0.1cm and 1.5cm of a] (a2) {\( \O_{\up{H}} \)};
    \vertex [above = 0.1cm of a] (c1) {$\mathcal{C}_{n\ell}^{\up{HL}}$};
    \vertex [below = 3cm of a] (b);
    \vertex [below left = 0.1cm and 1.5cm of b] (b1) {\( \O_{\up{L}} \)};
    \vertex [below right = 0.1cm and 1.5cm of b] (b2) {\( \O_{\up{H}} \)};
    \vertex [below = 0.1cm of b] (c2){$\mathcal{C}_{n\ell}^{\up{HL}}$};
    \diagram* {
    (a1) -- [thick] (a) -- [thick] (a2),
    (a) -- [scalar, edge label = $\;{[\O_{\up{H}} \O_{\up{L}}]}_{n\ell}$, thick] (b),
    (b1) -- [thick] (b) -- [thick] (b2),
    };
    \end{feynman}
\end{tikzpicture}
\end{gathered}
}$}^{\,2}
&= \scalebox{0.7}{$\abs{
\begin{gathered}
\begin{tikzpicture}
    \begin{feynman}
    \large
    \vertex (a);
    \vertex [left = 0.1cm of a] (a1);
    \vertex [above left = 1.5cm and 0.1cm of a] (b) {\(\O_{\up{L}}\)};
    \vertex [below  left = 1.5cm and 0.1cm of a] (d) {\(\O_{\up{L}}\)};
    \vertex [right = 3cm of a] (f);
    \vertex [right = 0.1cm of f] (g);
    \vertex [above right = 1.5cm and 0.1cm of f] (c) {\(\O_{\up{H}}\)};
    \vertex [below right = 1.5cm and 0.1cm of f] (e) {\(\O_{\up{H}}\)};
    \diagram* {
        (b) -- [thick] (a) -- [thick] (d),
        (a) -- [scalar, edge label = {$\one,\; T(z),\; ...$}, thick] (f),
        (c) -- [thick] (f) -- [thick] (e),
    };
    \end{feynman}
\end{tikzpicture}
\end{gathered}
}$}^{\,2}
+ \displaystyle \sum_{\substack{n\ell \\ \ell\; \up{even}}}
\scalebox{0.7}{$\abs{
\begin{gathered}
\begin{tikzpicture}
    \begin{feynman}
    \large
    \vertex (a);
    \vertex [left = 0.1cm of a] (a1);
    \vertex [above left = 1.5cm and 0.1cm of a] (b) {\(\O_{\up{L}}\)};
    \vertex [below  left = 1.5cm and 0.1cm of a] (d) {\(\O_{\up{L}}\)};
    \vertex [right = 3cm of a] (f);
    \vertex [right = 0.1cm of f] (g);
    \vertex [above right = 1.5cm and 0.1cm of f] (c) {\(\O_{\up{H}}\)};
    \vertex [below right = 1.5cm and 0.1cm of f] (e) {\(\O_{\up{H}}\)};
    \vertex [below right = 0.0cm and 0.0cm of a] (a2){$\mathcal{C}_{n\ell}^{\up{LL}}$};
    \vertex [below left = 0.0cm and 0.0cm of f] (a2){$\mathcal{C}_{n\ell}^{\up{HH}}$};
    \diagram* {
        (b) -- [thick] (a) -- [thick] (d),
        (a) -- [scalar, edge label = {$[\O_{\up{L}} \O_{\up{L}}]_{n\ell}$}, thick] (f),
        (c) -- [thick] (f) -- [thick] (e),
    };
    \end{feynman}
\end{tikzpicture}
\end{gathered}
}$}^{\,2} \\
G(z, \bar{z}) \qquad\quad &= 
\qquad\quad G_0(z, \bar{z}) \qquad\quad\;\, + 
\qquad\qquad\;\; G_{\up{reg}}(z, \bar{z}). \notag
\end{align}

In the $s$ channel, $\O_{\up{H}}$ and $\O_{\up{L}}$ are brought together and combine to form the HL double-twists $[\O_{\up{H}} \O_{\up{L}}]_{n\ell}$. The bulk picture is that of a light field mode $\phi_{n\ell}(t,\th) = \cal{C}_{n\ell}^{\up{HL}} e^{-i\w_{n\ell} t} e^{i\ell\th}$ orbiting a heavy object. We will show soon that the $s$-channel sum over HL double-twists agrees with the sum over these bulk field modes in (\ref{eqn:mode-sum}), and this will allow us to identify the amplitudes $\cal{C}_{n\ell}^{\up{HL}}$ with the OPE coefficients of the same name in (\ref{eqn:OPE-diagram}).

In the $t$ channel, pairwise identical operators are brought together. The dominant OPE contribution comes from the Virasoro vacuum block, which encodes gravitational interactions between $\O_{\up{H}}$ and $\O_{\up{L}}$. As we will see, the vacuum block reproduces the singular image $G_0(z,\bar{z})$. But there is more to $G(z, \bar{z})$ than its singular part, so there must be other states besides the vacuum running in the $t$ channel. In other words, gravitons cannot be the only virtual particles that mediate long-range interactions between $\O_{\up{H}}$ and $\O_{\up{L}}$. There are two natural candidates for additional states that could be exchanged: two $\O_{\up{L}}$ states could fuse to form $[\O_{\up{L}} \O_{\up{L}}]_{n\ell}$, or two $\O_{\up{H}}$ states could fuse to form $[\O_{\up{H}} \O_{\up{H}}]_{n\ell}$. But as discussed above, the latter process is strongly suppressed, so we focus on the former.

The fact that the LL double-twists $[\O_{\up{L}} \O_{\up{L}}]_{n\ell}$ contribute nontrivially to $G(z,\bar{z})$ implies that the corresponding OPE coefficients $\cal{C}_{n\ell}^{\up{LL}}$ and $\cal{C}_{n\ell}^{\up{HH}}$ are nonzero. We will study these coefficients extensively in the remainder of the paper. For now, the bulk interpretation is that whenever heavy states in 3d gravity couple to matter, they interact at long range by exchanging an infinite family of light two-particle states. Another way to say this is that conical defects and BTZ black holes admit infinitely many species of quantum hair.

\subsection{Heavy-Light Virasoro Blocks}
\label{sec:HHLL-virasoro-blocks}

Now we assemble the complete $s$- and $t$-channel OPEs for $G(z,\bar{z})$, including regular terms, and match them to our bulk results to find the OPE coefficients in both channels.

\paragraph{\boldmath The $s$ channel.} The $s$-channel Virasoro blocks $\cal{F}_h^{(\up{s})}(z)$ are sums of three-point functions of the form $\avg{\O_{\up{H}} \O_{\up{L}} \O_h}$, where the internal operator $\O_{h}$ is either a double-twist primary $[\O_{\up{H}} \O_{\up{L}}]_{n\ell}$ or one of its descendants. In the case of primary exchange, we get
\begin{align}
\label{eqn:$s$-channel-primary}
\avg{\O_{h}(\infty) \O_{\up{L}}(z) \O_{\up{H}}(0)} = \cal{C}^{\up{HL}}_h z^{h - h_{\up{H}} - h_{\up{L}}} \implies
\cal{F}_h^{(\up{s})}(z) \supset z^{h - h_{\up{H}} - h_{\up{L}}},
\end{align}
where $\cal{C}^{\up{HL}}_h$ is an OPE coefficient and $\supset$ means ``contains as a term.'' In fact, this is all we need, since all descendant OPE contributions are suppressed at large $c$. To see this, note that descendant three-point functions can be obtained from (\ref{eqn:$s$-channel-primary}) by applying a differential operator to $\O_h$ and then dividing by the norm of the descendant state. The differential operator generates a function of $z$ with a prefactor polynomial in $h - h_{\up{H}} - h_{\up{L}}$, while the norm of the descendant is a polynomial in $c$ and $h$ of the same degree as the level of the descendant. (For instance, $L_{-1} \O_h = \pd \O_h$ has norm $\mel{h}{L_1 L_{-1}}{h} = 2h$.) The HL double-twists have $h = O(c)$ but $h - h_{\up{H}} - h_{\up{L}} = O(1)$, so the contribution of a level-$k$ descendant is $O(c^{-k})$. Thus the $s$-channel blocks of the $[\O_{\up{H}} \O_{\up{L}}]_{n\ell}$ reduce to scaling blocks,
\begin{align}
\label{eqn:$s$-channel-blocks}
\cal{F}_h^{(\up{s})}(z) = z^{h - h_{\up{H}} - h_{\up{L}}} \big(1 + O(1/c) \big), \qquad h = h_{[\up{HL}]_{n\ell}}.
\end{align}

To put together the full correlator, we multiply in the anti-holomorphic part of the block and sum over the double-twist spectrum, as in (\ref{eqn:OPE-diagram}):
\begin{align}
\label{eqn:$s$-channel-OPE}
G(z, \bar{z}) = \sum_{n\ell} \p{\cal{C}_{n\ell}^{\up{HL}}}^2 \cal{F}_{[\up{HL}]_{n\ell}}^{(\up{s})}(z) \overline{\cal{F}}_{[\up{HL}]_{n\ell}}^{(\up{s})}(\bar{z}) = \abs{z}^{(\a - 1)\Delta} \sum_{n\ell} \p{\cal{C}_{n\ell}^{\up{HL}}}^2 \abs{z}^{2\a n} z^{\ell}.
\end{align}
As promised, this OPE takes the same form as the mode sum (\ref{eqn:mode-sum}). Matching the two expressions allows us to identify the OPE coefficients $\cal{C}_{n\ell}^{\up{HL}}$ with the normal mode amplitudes given in (\ref{eqn:HL-coefficients}).\footnote{Above the BTZ threshold, the frequencies $\tilde{\w}_{n\ell}$ and coefficients $\tilde{\cal{C}}_{n\ell}^{\up{HL}}$ are associated to quasinormal modes in the bulk. Since they are complex-valued, they should not be interpreted as the conformal dimensions or OPE data of physical states in a unitary CFT. See \cite{Dodelson:2022eiz, Giusto:2023awo} for a discussion of similar issues.} We conclude that the bulk theory knows how to compute $s$-channel OPEs, and that it also knows the HL double-twist OPE data.

\paragraph{\boldmath The $t$ channel.} The semiclassical HHLL $t$-channel Virasoro blocks were worked out by Fitzpatrick, Kaplan, and Walters \cite{Fitzpatrick:2015zha}. Their key insight was that the presence of heavy operators in the correlator modifies the background geometry in which the light fields propagate. The result is that the HHLL blocks reduce to the all-light (i.e. global) conformal blocks, with the light fields evaluated in a new set of coordinates:
\begin{align}
\label{eqn:$t$-channel-blocks}
\cal{F}_h^{(\up{t})}(z) &= z^{(\a - 1) h_{\up{L}}} \p{\f{w}{\a}}^{h - 2h_{\up{L}}} {}_2 F_1 \p{h,h; 2h; w}, \qquad w(z) = 1 - z^{\a}.
\end{align}
The replacement $z \mapsto z^{\a}$ above is the same as the change of boundary coordinates from pure $\up{AdS}_3$ to a heavy background that shows up in the method of images. From (\ref{eqn:$t$-channel-blocks}), we obtain the vacuum block by setting $(h, \bar{h}) = (0,0)$, and we get the LL double-twist blocks by substituting the MFT spectrum (\ref{eqn:double-twist-spectrum}). The vacuum contribution is
\begin{align}
\label{eqn:vacuum-block}
G_0(z, \bar{z}) = \abs{\cal{F}_0^{(\up{t})}(z)}^2 = \abs{z}^{(\a - 1)\Delta} \abs{\f{1 - z^{\a}}{\a}}^{-2\Delta},
\end{align}
in agreement with the singular image (\ref{eqn:singular-part}), while the contributions of the LL double-twist blocks must sum together to reproduce the regular piece of the correlator:
\begin{align}
\label{eqn:Greg-OPE-LL}
G_{\up{reg}}(z, \bar{z}) &= 
\abs{z}^{(\a - 1)\Delta} \sum_{\substack{n\ell \\ \ell\; \up{even}}} \cal{C}_{n\ell}^{\up{LL}} \cal{C}_{n\ell}^{\up{HH}} \abs{\f{w}{\a}}^{2n} \p{\f{w}{\a}}^{\ell} \Big| {}_2 F_1 (h_{n\ell}, h_{n\ell}; 2h_{n\ell}; w) \Big|^2.
\end{align}
Here we set $h_{n\ell} = h_{[\up{LL}]_{n\ell}}$ and recall that absolute values are shorthand for holomorphically factorized products. The products $\cal{C}_{n\ell}^{\up{LL}} \cal{C}_{n\ell}^{\up{HH}}$ must be even under $\ell \too -\ell$, because the primaries $[\O_{\up{L}} \O_{\up{L}}]_{n, \pm \ell}$ should contribute equally to the OPE $\O_{\up{L}} \times \O_{\up{L}}$ of two scalars.

\subsection{Crossing and Lorentzian Inversion}
\label{sec:crossing-inversion}

Let us make several bootstrap-adjacent comments complementary to our perspective.

\paragraph{The vacuum block.} The Virasoro vacuum block in 2d CFT is nontrivial: it contains the identity, the stress tensor, and an infinite tower of multi-stress tensor descendants. In the bulk, these states excite topological degrees of freedom called boundary gravitons, and their $t$-channel exchange between $\O_{\up{H}}$ and $\O_{\up{L}}$ creates the effective classical background defined by the $w$ coordinate. In the $w$-plane, the vacuum block becomes the two-point function of $\O_{\up{L}}$, which is determined by a geodesic length. This is one way in which the vacuum block knows about bulk geometry. But the vacuum block does not know everything, even at the semiclassical level: it is not single-valued in $z$, and as we have discussed it fails to capture $G_{\up{reg}}(z,\bar{z})$. None of the blocks (\ref{eqn:$t$-channel-blocks}) are single-valued either, but the fact that their sum must be puts strong constraints on the coefficients $\cal{C}_{n\ell}^{\up{LL}} \cal{C}_{n\ell}^{\up{HH}}$.

\paragraph{Crossing symmetry.} Crossing symmetry is the statement that the $s$-channel and $t$-channel OPEs compute the same four-point function $G(z,\bar{z})$. This is formally manifest in (\ref{eqn:s-t-OPE}), and our calculation (\ref{eqn:images-to-modes}) proving the equivalence of the mode sum and the method of images provides an explicit demonstration of crossing. The fact that $G(z,\bar{z})$ is a linear combination of Virasoro blocks in both channels implies that each block in one channel can be expressed as a linear combination of the blocks in the other. The matrix that inverts individual $t$-channel blocks into combinations of $s$-channel blocks is known exactly and is called the Virasoro fusion kernel \cite{Collier:2018exn}. The use of this kernel to bootstrap OPE data gives rise to the Euclidean and Lorentzian inversion formul{\ae} \cite{Simmons-Duffin:2017nub}.

We are essentially working in the opposite direction: by independently computing $G(z,\bar{z})$ using bulk methods, we start with the answer for $G(z,\bar{z})$ as well as all of its contributing Virasoro blocks. By matching the answer to its OPE, we can deduce all of the OPE coefficients and thereby ``derive'' the fusion kernel. 

\paragraph{Lorentzian inversion.} We have seen that the singular structure of (\ref{eqn:vacuum-block}) controls the asymptotic behavior of the OPE (\ref{eqn:$s$-channel-OPE}). This is the statement that applying an inversion formula to the vacuum block produces the dominant contribution to the $s$-channel OPE data. But as emphasized above, the vacuum cannot be the only contribution. Indeed, the LL double-twist blocks were inverted in \cite{Collier:2018exn} (see also \cite{Kusuki:2018wpa}) and shown to give nontrivial subleading contributions to the $s$-channel OPE data. Our ``inversion'' of the full sum over images in (\ref{eqn:images-to-modes}) is another way to see that the OPE coefficients $\cal{C}_{n\ell}^{\up{HL}}$ receive important contributions from both vacuum and non-vacuum exchange.


\section{Expectation Values in Heavy States}
\label{sec:expectation-values}

In this section, we initiate our study of the OPE coefficients $\cal{C}_{n\ell}^{\up{LL}}$ and $\cal{C}_{n\ell}^{\up{HH}}$. Both sets of coefficients are normalization constants for three-point functions involving $[\O_{\up{L}} \O_{\up{L}}]_{n\ell}$:
\begin{align}
\cal{C}_{n\ell}^{\up{LL}} = \avg{\O_{\up{L}} [\O_{\up{L}} \O_{\up{L}}]_{n\ell} \O_{\up{L}}}, \qquad
\cal{C}_{n\ell}^{\up{HH}} = \avg{\O_{\up{H}} [\O_{\up{L}} \O_{\up{L}}]_{n\ell} \O_{\up{H}}}.
\end{align}
The coefficients $\cal{C}_{n\ell}^{\up{LL}}$ represent the amplitudes to fuse two copies of $\O_{\up{L}}$ to create $[\O_{\up{L}} \O_{\up{L}}]_{n\ell}$; they are known in MFT. The coefficients $\cal{C}_{n\ell}^{\up{HH}}$ are more interesting: they are the expectation values of the LL double-twist states in heavy backgrounds. Above the BTZ threshold, they become thermal one-point functions---see Section~\ref{sec:thermal-one-point-functions} for a discussion.

In order to find the OPE coefficients for a correlator whose form is already known, we can Taylor-expand both the correlator and each of the blocks in its OPE, and then match the Taylor coefficients on both sides order by order.\footnote{We could also use an orthogonality relation for Virasoro blocks or conformal partial waves \cite{Simmons-Duffin:2017nub}, which would enable us to obtain the OPE data by integrating each block against the correlator. The results should be the same, and we leave a more detailed exploration of this approach to future work.} We begin by demonstrating how this strategy works in the case of the light double-twist amplitudes $\cal{C}_{n\ell}^{\up{LL}}$ before adapting it to determine the products $\cal{C}_{n\ell}^{\up{LL}} \cal{C}_{n\ell}^{\up{HH}}$, first in general and then at large twist.

\subsection{Universal OPE Data in Mean-Field Theory}
\label{sec:OPE-coefficients-LLLL}

Consider the four-point function of $\O_{\up{L}}$ in MFT. By Wick's theorem, it is given by
\begin{equation}
\label{eqn:LLLL-Wick}
G_{\up{L}}(z, \bar{z}) \equiv \avg{\O_{\up{L}}(\infty) \O_{\up{L}}(1) \O_{\up{L}}(z, \bar{z}) \O_{\up{L}}(0)} = \f{1}{|z|^{2\Delta}} + \f{1}{|1-z|^{2\Delta}} + 1.
\end{equation}
But we can also compute $G_{\up{L}}(z, \bar{z})$ using an OPE. The only primaries that appear in the OPE of $\O_{\up{L}}$ with itself are the identity and the LL double-twists. The Virasoro blocks of these primaries reduce to global conformal blocks in the semiclassical limit, so
\begin{align}
\label{eqn:LLLL-OPE}
G_{\up{L}}(z, \bar{z}) &= \f{1}{|z|^{2\Delta}}\p{1 + \sum_{n\ell} \p{\cal{C}_{n\ell}^{\up{LL}}}^2 \big|k_{h_{n\ell}}(z)\big|^2}, \qquad k_h(z) = z^h {}_2 F_1 \p{h,h; 2h; z},
\end{align}
where as before $h_{n\ell} = h_{[\up{LL}]_{n\ell}}$ with $\ell \in 2\Z$. Now we set (\ref{eqn:LLLL-Wick}) equal to (\ref{eqn:LLLL-OPE}) and strip off the vacuum contribution $1/|z|^{2\Delta}$ from both sides: what remains is the part of $G_{\up{L}}(z, \bar{z})$ that is regular at $z, \bar{z} = 0$. We change summation variables from $(n,\ell)$ to $(m,m')$ as in (\ref{eqn:chiral-double-twist}) and define $p^n_{\ell} = \p{\cal{C}_{n\ell}^{\up{LL}}}^2$. The vacuum-subtracted correlator becomes
\begin{align}
\label{eqn:LLLL-matching}
1 + \f{1}{|1-z|^{2\Delta}} = \f{1}{|z|^{2\Delta}} \sum_{m,m' = 0}^{\infty} \f{1}{2} \qty[1 + (-1)^{m-m'}] p^{\up{min}(m,m')}_{m-m'} k_{\Delta + m}(z) k_{\Delta + m'}(\bar{z}).
\end{align}

Next, we Taylor-expand both sides of (\ref{eqn:LLLL-matching}) in powers of $z, \bar{z}$ around the origin:
\begin{align}
\label{eqn:LLLL-taylor-series}
1 + \sum_{r,r'=0}^{\infty} a^{(r,r')} z^r \bar{z}^{r'} = \sum_{m,m' = 0}^{\infty} \f{1}{2} \qty[1 + (-1)^{m-m'}] p^{\up{min}(m,m')}_{m-m'} \sum_{r,r'=0}^{\infty} b_{m,m'}^{(r,r')} z^r \bar{z}^{r'},
\end{align}
where the Taylor coefficients $a^{(r,r')} = \a^{(r)} \a^{(r')}$ and $b_{m,m'}^{(r,r')} = \b_m^{(r)} \b_{m'}^{(r')}$ are given by
\begin{align}
\label{eqn:ab-coefficients}
\a^{(r)} = \f{\G(\Delta + r)}{\G(\Delta) \G(1+r)}, \qquad
\b_m^{(r)} = \f{\G(\Delta + r)^2 \G(2\Delta + 2m)}{\G(\Delta + m)^2 \G(1 + r-m) \G(2\Delta + r + m)}.
\end{align}
Finally, we match Taylor coefficients at each order $z^r\bar{z}^{r'}$ on both sides of (\ref{eqn:LLLL-taylor-series}): 
\begin{align}
\label{eqn:LLLL-linear-system}
\d_{r,0} \d_{r',0} + a^{(r,r')} = \sum_{m,m' = 0}^{\infty} \f{1}{2} \qty[1 + (-1)^{m-m'}] b_{m,m'}^{(r,r')}\, p^{\up{min}(m,m')}_{m-m'}.
\end{align}
This power-matching condition reflects the fact that multiple LL double-twist conformal blocks contribute to the regular part of the correlator at each order in $z, \bar{z}$. The nontrivial Taylor expansion of each $k_{h_{n\ell}}(z)$ is what causes the products $p^n_{\ell}$ to appear in complicated linear combinations at each order; to disentangle them, we view (\ref{eqn:LLLL-linear-system}) as an infinite linear system, with one equation at each order $(r,r')$, to be solved for the $p^n_{\ell}$.

In the simplest case $(r,r') = (0,0)$, we are looking at the constant term in the Taylor series. From (\ref{eqn:LLLL-matching}), the value of the vacuum-subtracted correlator at $z,\bar{z} = 0$ is 2, and the only block that contributes to the constant term is the one with $m=m'=0$, corresponding to the composite operator $[\O_{\up{L}} \O_{\up{L}}]_{00} = \O_{\up{L}}^2$. Thus we find
\begin{align}
\label{eqn:C00LL}
2 = p^0_0 = \p{\cal{C}_{00}^{\up{LL}}}^2 \implies \cal{C}_{00}^{\up{LL}} = \avg{\O_{\up{L}}\, \O_{\up{L}}^2\, \O_{\up{L}}} = \sqrt{2}.
\end{align}
This OPE coefficient reflects the correct Boltzmann counting one has to do to account for the bosonic exchange statistics that arise when combining two $\O_{\up{L}}$ states to form $\O_{\up{L}}^2$. The higher coefficients $\cal{C}_{n\ell}^{\up{LL}}$ should be understood as combinatorial generalizations of the Boltzmann factor which serve to properly normalize the states $[\O_{\up{L}} \O_{\up{L}}]_{n\ell}$.

The linear system (\ref{eqn:LLLL-linear-system}) is upper-triangular, which means it can be solved recursively by back-substitution. To see how this works, observe that the sums over $m,m'$ are all finite: they truncate when $m$ reaches $r$ or when $m'$ reaches $r'$, due to the Gamma function poles in the denominator of $b^{(r,r')}_{m,m'}$. Therefore only finitely many of the OPE coefficients enter at each order. In fact, the set of equations one gets by truncating the expansion at order $(r,r')$ is sufficient to determine all of the OPE coefficients $\cal{C}_{n\ell}^{\up{LL}}$ with $n \leq \min(r,r')$ and $|\ell| \leq |r - r'|$. One can then increase $r$ and $r'$ until the next ``batch'' of equations closes, solve for the remaining undetermined OPE coefficients, and repeat the process.

Remarkably, this system has a closed-form solution, first written down in \cite{Heemskerk:2009pn}:
\begin{align}
\label{eqn:CnlLL}
\p{\cal{C}_{n\ell}^{\up{LL}}}^2 = \qty[1 + (-1)^{\ell}] c_n c_{n+\ell}, \qquad c_n = \f{\G(\Delta + n)^2 \G(2\Delta + n - 1)}{\G(\Delta)^2 \G(1+n) \G(2\Delta + 2n - 1)}.
\end{align}
This can be deduced either by guessing a pattern from the first few OPE coefficients and checking that (\ref{eqn:CnlLL}) solves (\ref{eqn:LLLL-linear-system}), or by applying the Lorentzian inversion formula.\footnote{We thank Jo\~{a}o Penedones for discussion on this point.}

As an aside, another way to formulate the problem is to ask for the differential operator $\cal{D}_{n\ell} \sim \Box^n \overset{{}_\leftrightarrow}{\pd} {}^{\ell}$ for which $[\O_{\up{L}} \O_{\up{L}}]_{n\ell} = \O_{\up{L}} \cal{D}_{n\ell} \O_{\up{L}}$ is a properly normalized primary state. Then one may obtain the desired OPE coefficients by taking derivatives of (\ref{eqn:LLLL-Wick}):
\begin{align}
\cal{C}_{n\ell}^{\up{LL}} = \avg{\O_{\up{L}}(\infty) [\O_{\up{L}} \O_{\up{L}}]_{n\ell}(1) \O_{\up{L}}(0)} = \cal{D}_{n\ell} \avg{\O_{\up{L}}(\infty) \O_{\up{L}}(1) \O_{\up{L}}(z,\bar{z}) \O_{\up{L}}(0)} \Big|_{z, \bar{z} = 1}
\end{align}
This was the perspective of \cite{Berenstein:2021pxv}, where a recursion relation for $\cal{D}_{n\ell}$ was found and solved, in the case of SL(2), for the normalization coefficients of quasiprimaries.

\subsection{Heavy-Heavy-Light OPE Coefficients}
\label{sec:OPE-coefficients-HHLL}

Let us return to the problem of matching the HHLL correlator $G(z, \bar{z})$ to its $t$-channel OPE. This problem is similar to the one treated above: in both cases, the vacuum-subtracted correlator $G_{\up{reg}}(z, \bar{z})$ is set equal to an infinite sum of LL double-twist Virasoro blocks and then Taylor-expanded to extract the OPE coefficients. In the present case, we have
\begin{align}
\label{eqn:HHLL-matching}
\tilde{G}_{\up{reg}}(w, \overline{w}) = \sum_{k \neq 0} \abs{\f{1 - \zeta^k (1-w)}{\a}}^{-2\Delta} = \f{1}{|w|^{2\Delta}} \sum_{n\ell} \cal{C}_{n\ell}^{\up{LL}} \cal{C}_{n\ell}^{\up{HH}} |\a|^{-(2n + |\ell|)} \big|k_{h_{n\ell}}(w)\big|^2.
\end{align}
Here we have stripped off an unimportant prefactor from (\ref{eqn:regular-part}) and (\ref{eqn:Greg-OPE-LL}) by defining $\tilde{G}_{\up{reg}}(w,\overline{w}) \equiv |z|^{-(\a - 1)\Delta} G_{\up{reg}}(z,\bar{z})$. Taking the same approach as in Section~\ref{sec:OPE-coefficients-LLLL}, we change summation variables from $(n,\ell)$ to $(m,m')$, define $\tilde{p}^n_{\ell} = \cal{C}_{n\ell}^{\up{LL}} \cal{C}_{n\ell}^{\up{HL}}$, expand in powers of $w, \overline{w}$, and equate the Taylor coefficients on both sides order by order:
\begin{align}
\label{eqn:HHLL-taylor-series}
\tilde{G}_{\up{reg}}(w, \overline{w}) = \sum_{r,r'=0}^{\infty} \tilde{a}^{(r,r')} w^r \overline{w}^{r'} = \sum_{m,m'=0}^{\infty} |\a|^{-(m+m')} \tilde{p}^{\up{min}(m,m')}_{m-m'} \sum_{r,r' = 0}^{\infty} b_{m,m'}^{(r,r')} w^r \overline{w}^{r'}.
\end{align}
The coefficients $b_{m,m'}^{(r,r')}$ are the same as the ones in (\ref{eqn:ab-coefficients}), while the new coefficients $\tilde{a}^{(r,r')}$ can be obtained by expanding the images $\p{1 - \zeta^k (1-w)}^{-\Delta}$ in powers of $w$:
\begin{equation}
\label{eqn:a-tilde-coefficients}
\begin{aligned}
\tilde{a}^{(r,r')} &= \f{1}{r!\, r'!} \pd^r \overline{\pd} {}^{r'} \tilde{G}_{\up{reg}}(w, \overline{w}) \eval_{w,\overline{w} = 0} = \sum_{k \neq 0} \tilde{\a}_k^{(r)} \tilde{\a}_{-k}^{(r')}, \\
\tilde{\a}^{(r)}_k &= \f{|\a|^{\Delta} \p{-\zeta^k}^r}{\p{1 - \zeta^k}^{\Delta + r}} \p{\f{\G(\Delta + r)}{\G(\Delta) \G(1+r)}}.
\end{aligned}
\end{equation}

From (\ref{eqn:HHLL-taylor-series}), the linear system that determines the products $\tilde{p}^n_{\ell}$ is
\begin{align}
\label{eqn:HHLL-linear-system}
\tilde{a}^{(r,r')} = \sum_{m,m' = 0}^{\infty} |\a|^{-(m+m')} b_{m,m'}^{(r,r')} \tilde{p}^{\up{min}(m,m')}_{m-m'}.
\end{align}
This matrix can, at least in principle, be inverted to yield OPE coefficients schematically of the form $\cal{C}_{n\ell}^{\up{HH}} \sim (b_{n\ell}^{-1} \cdot \tilde{a})/\cal{C}_{n\ell}^{\up{LL}}$. However, it is unlikely that the system can be solved in closed form in the same way as (\ref{eqn:LLLL-linear-system}), due to the complicated $r$-dependence of $\tilde{a}^{(r,r')}$.

Nevertheless, the first few OPE coefficients can be found by truncating the expansion at low orders in $w, \overline{w}$. For instance, consider $[\O_{\up{L}} \O_{\up{L}}]_{00} = \O_{\up{L}}^2$, the simplest operator with a nontrivial expectation value in a heavy state. Its expectation value can be obtained by looking at the constant term in (\ref{eqn:HHLL-matching}), or from the $(r,r') = (0,0)$ equation of (\ref{eqn:HHLL-linear-system}):
\begin{align}
\label{eqn:CHH00}
\cal{C}_{00}^{\up{HH}} = \avg{\O_{\up{L}}^2}_{\up{H}} = \f{1}{\sqrt{2}} \tilde{G}_{\up{reg}}(0,0) = \f{1}{\sqrt{2}} \sum_{k \neq 0} \abs{\f{2}{\a} \sin(\pi k \a)}^{-2\Delta}.
\end{align}

\begin{figure}[t]
\begin{subfigure}{.5\textwidth}
  \centering
  \includegraphics[width=.95\linewidth]{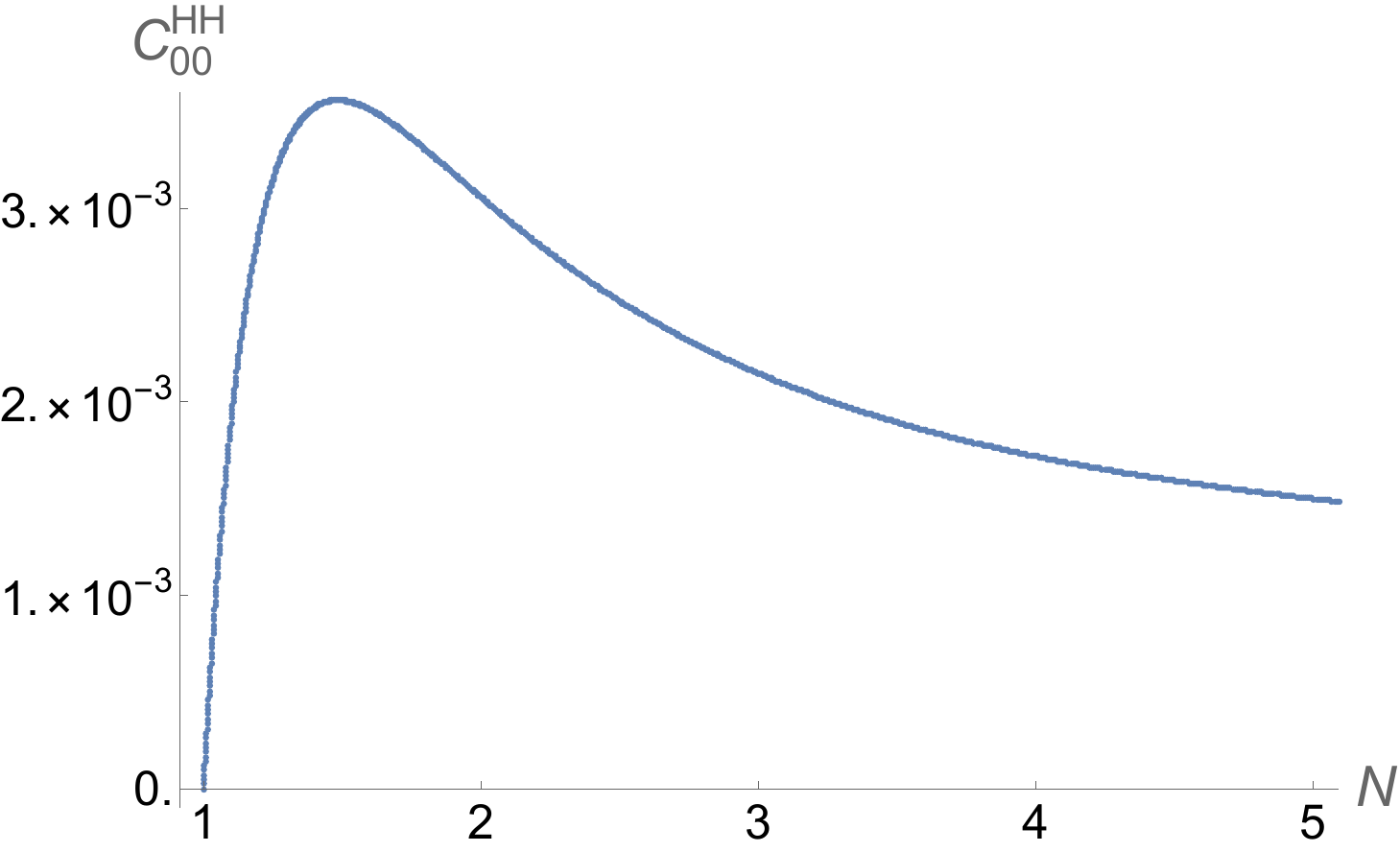}
\end{subfigure}%
\begin{subfigure}{.5\textwidth}
  \centering
  \includegraphics[width=.95\linewidth]{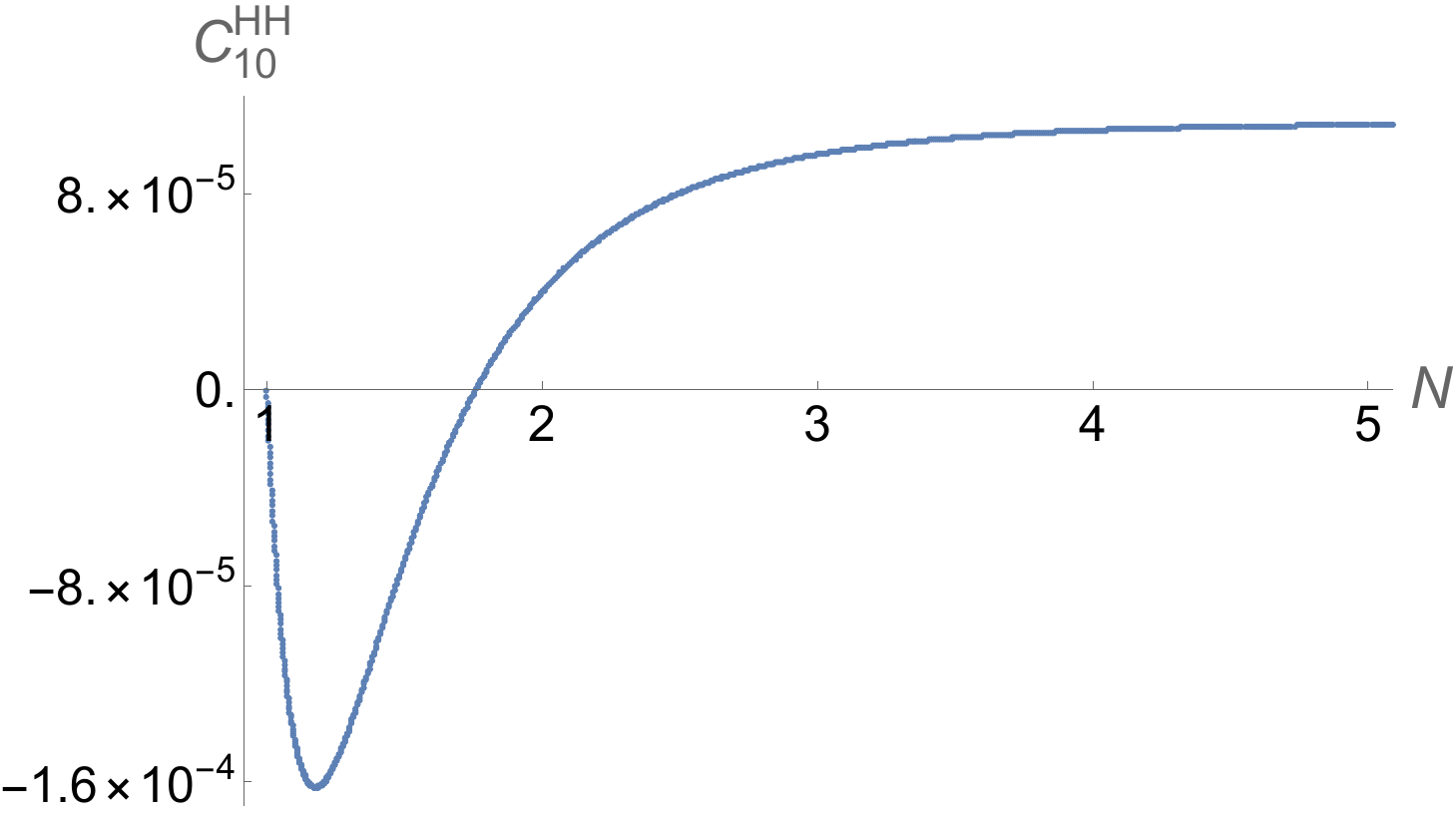}
\end{subfigure} \\[11pt]
\begin{subfigure}{.5\textwidth}
  \centering
  \includegraphics[width=.95\linewidth]{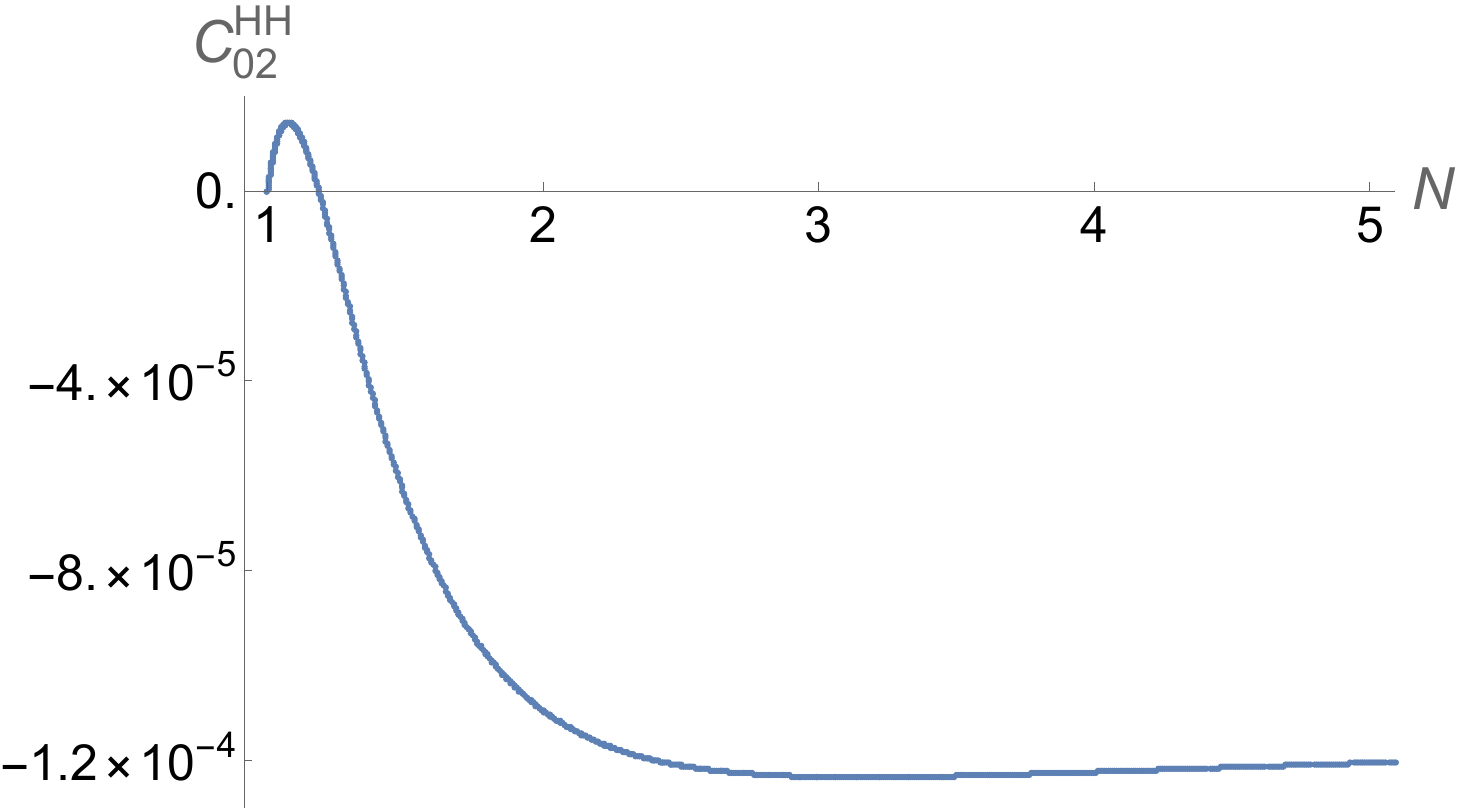}
\end{subfigure}%
\begin{subfigure}{.5\textwidth}
  \centering
  \includegraphics[width=.95\linewidth]{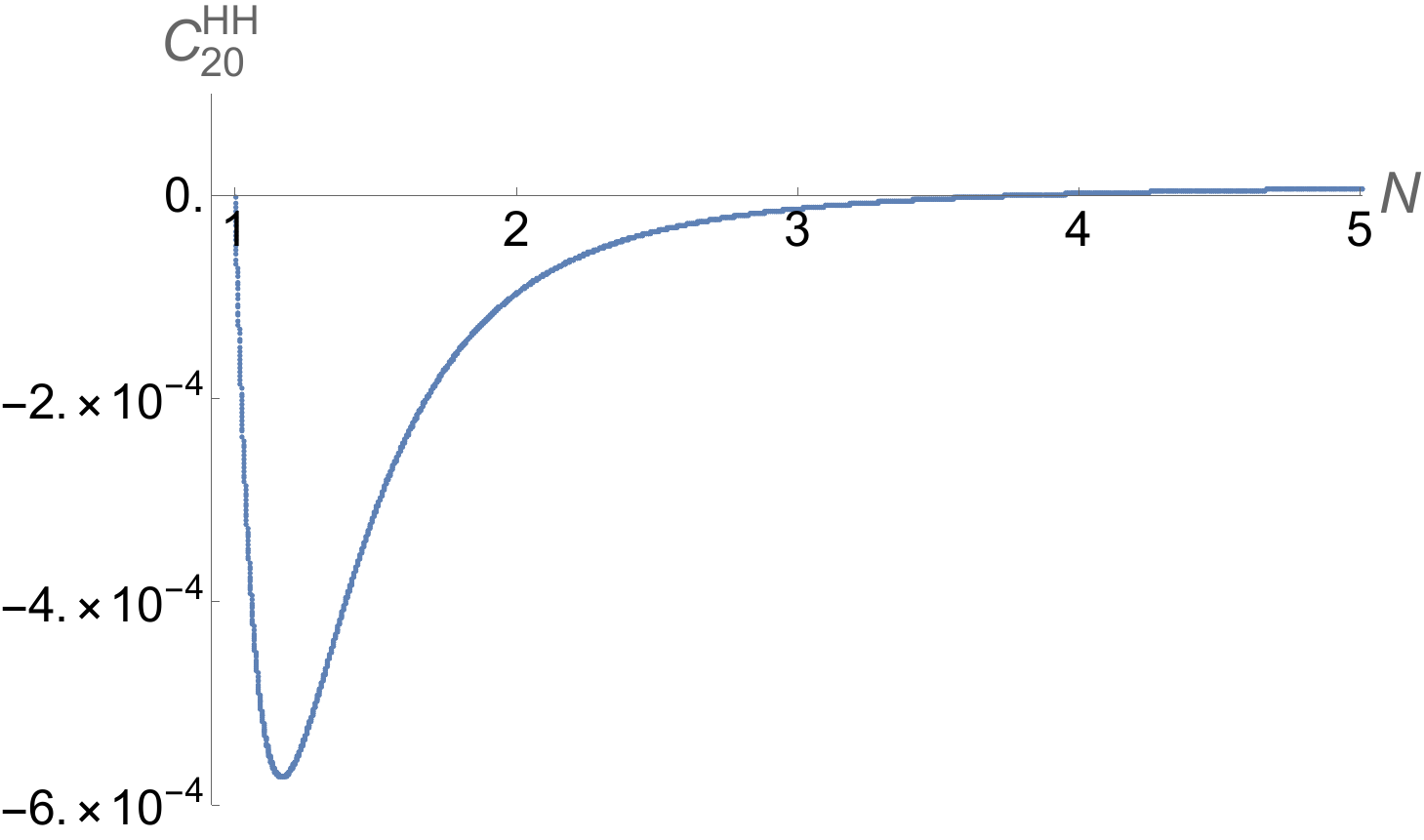}
\end{subfigure}
\caption{The first few low-lying LL double-twist expectation values in conical AdS are plotted numerically as a function of the defect strength. We take $\Delta = 2$ and focus primarily on the regime of light conical defects ($1 \leq N < 3$), where the method of images is not reliable. The expectation values vanish in pure AdS ($N=1$), exhibit nontrivial behavior at finite $N$, and tend to small but nonzero limits as $N \too \infty$.}
\label{fig:numerics}
\end{figure}

The next few OPE coefficients are readily obtained explicitly in terms of $\tilde{G}_{\up{reg}}(w, \overline{w})$ and its derivatives $\tilde{G}_{\up{reg}}^{(r,r')}(w,\overline{w})$, evaluated at $w = \overline{w} = 0$:
\begin{align}
\label{eqn:low-lying}
\cal{C}_{10}^{\up{HH}} &= \f{\a^2}{\cal{C}_{10}^{\up{LL}}} \p{\tilde{G}_{\up{reg}}^{(1,1)}(0,0) - \f{\Delta^2}{4} \tilde{G}_{\up{reg}}(0,0)},  \notag \\
\cal{C}_{02}^{\up{HH}} &= \f{\a^2}{\cal{C}_{02}^{\up{LL}}} \p{\f{1}{2} \tilde{G}_{\up{reg}}^{(2,0)}(0,0) - \f{\Delta(1 + \Delta)^2}{4(1 + 2\Delta)} \tilde{G}_{\up{reg}}(0,0)}, \\
\cal{C}_{20}^{\up{HH}} &= \f{\a^4}{\cal{C}_{20}^{\up{LL}}} \bigg(\f{1}{4}\tilde{G}_{\up{reg}}^{(2,2)}(0,0) - \f{\Delta(1+\Delta)^2}{8(1+2\Delta)} \tilde{G}_{\up{reg}}^{(2,0)}(0,0) - \f{(1 + \Delta)^2}{4} \tilde{G}_{\up{reg}}^{(1,1)}(0,0) + \f{\Delta^2(1 + \Delta)^2}{16} \tilde{G}_{\up{reg}}(0,0) \bigg), \notag
\end{align}
and so on. The main point is that (\ref{eqn:HHLL-linear-system}) provides an algorithm for computing the OPE coefficients $\cal{C}_{n\ell}^{\up{HH}}$ whose input data consists of the universal coefficients (\ref{eqn:ab-coefficients}) and (\ref{eqn:CnlLL}), both derived from the semiclassical Virasoro blocks, and the local behavior of $G_{\up{reg}}(z, \bar{z})$ near $z, \bar{z} = 1$. This data can be obtained either by summing over images, as in (\ref{eqn:a-tilde-coefficients}), or by direct numerical evaluation of the difference $G_{\up{reg}} = G - G_0$, i.e. using
\begin{align}
\label{eqn:numerical-subtraction}
\tilde{G}_{\up{reg}}(w, \overline{w}) \equiv \f{G(z, \bar{z}) - G_0(z, \bar{z})}{|z|^{(\a - 1) \Delta}} = \sum_{n\ell} \p{\cal{C}_{n\ell}^{\up{HL}}}^2 |1-w|^{2n} (1 - w)^{\ell/\a} - \abs{\f{w}{\a}}^{-2\Delta}.
\end{align}
Both terms in (\ref{eqn:numerical-subtraction}) are formally divergent at $w = 0$, but their difference can be computed numerically at very small $|w| \ll 1$. The results of this subtraction are shown in Figure~\ref{fig:numerics}, which plots the expectation values (\ref{eqn:CHH00}--\ref{eqn:low-lying}) in conical AdS as a function of $N$.

\section{Asymptotic Analysis at Large Twist}
\label{sec:asymptotics-large-twist}

For larger values of $n$, it becomes cumbersome to evaluate the $\cal{C}_{n\ell}^{\up{HH}}$ analytically. However, the Virasoro blocks of LL double-twist operators with \emph{very} large $n$ simplify dramatically, and we regain analytical control over the OPE. This will lead to an asymptotic formula for the expectation values of highly excited double-twist operators.

\subsection{Virasoro Blocks with Heavy Exchange}
\label{sec:heavy-exchange}

For preparation, consider the semiclassical $t$-channel blocks (\ref{eqn:$t$-channel-blocks}) at asymptotically large internal dimension, $h \gg 1$.\footnote{Here the large-$c$ limit is taken first, at fixed $h$, before letting $h$ become large. In other words, we take the limits $h \too \infty$ and $c \too \infty$, but with $h/c \too 0$, so that the internal operator is still ``light.''} A classic result of Watson on the asymptotic behavior of hypergeometric functions at large parameters \cite{Luke1969} states that
\begin{align}
\label{eqn:2F1-asymptotics}
k_h(w) = w^h {}_2 F_1(h, h; 2h; w) \sim \f{(4\rho)^h}{\sqrt{1 - \rho^2}} \qquad \up{as}\;\, h \too \infty,
\end{align}
where $\rho$ is the ``radial coordinate'' defined by
\begin{align}
w = \f{4\rho}{\p{1+\rho}^2} \iff 
\rho = \f{1 - \sqrt{1-w}}{1 + \sqrt{1-w}}, \qquad |\rho| < 1.
\end{align}
The transformation $w \mapsto \rho(w)$ maps the cut plane $\C \mathbin{\vcenter{\hbox{$\scriptstyle\setminus{}_{{}_{\!\!\!\!}}\scriptstyle\setminus$}}} [1, \infty)$ to the unit disk. It was first studied in \cite{Pappadopulo:2012jk, Hogervorst:2013sma}, where its use improved the convergence rate of certain OPEs. 

It follows from (\ref{eqn:2F1-asymptotics}) that the blocks (\ref{eqn:$t$-channel-blocks}) have the following asymptotics:
\begin{align}
\label{eqn:large-twist-blocks}
\cal{F}_h^{(\up{t})}(z) \sim z^{(\a - 1) h_{\up{L}}} \f{\p{1 + \rho}^{4h_{\up{L}}}}{\sqrt{1 - \rho^2}} \p{\f{4\rho}{\a}}^{h - 2h_{\up{L}}} \qquad \up{as}\;\, h \too \infty.
\end{align}
Evidently, at large $h$ the semiclassical HHLL blocks reorganize into scaling blocks in the radial variable $\rho$. In this new coordinate, no descendants are excited at all, and the OPE (\ref{eqn:HHLL-matching}) turns into an ordinary power series in $\rho$. This fact will allow us to carry out our power-matching procedure much more directly than in the case of arbitrary $h$.

Since we want both the holomorphic and anti-holomorphic blocks to behave like (\ref{eqn:large-twist-blocks}), it is natural to take the twist $\t = \min(h, \bar{h})$ to be large, rather than either $h$ or $\bar{h}$. In the case of LL double-twist exchange, we have $\t = \Delta + n$, so (\ref{eqn:large-twist-blocks}) is an asymptotic formula either in the hefty regime $1 \ll \Delta \ll c$ or in the highly excited regime $1 \ll n \ll c$.

\subsection{Darboux's Method: Preliminaries}
\label{sec:darboux-preliminaries}

Our next step is to reformulate the matching problem (\ref{eqn:HHLL-matching}) in terms of $\rho, \bar{\rho}$ at large twist. We substitute the asymptotic Virasoro blocks (\ref{eqn:large-twist-blocks}) into the LL double-twist OPE, change variables $(n,\ell) \too (m,m')$ as before, and write the OPE in the following form:
\begin{align}
\label{eqn:F-OPE}
F(\rho, \bar{\rho}) &\equiv \abs{1 - \rho^2} \big|1 + \rho\big|^{-4\Delta} \tilde{G}_{\up{reg}}(w, \overline{w}) \sim \sum_{m,m' = 0}^{\infty} \tilde{p}^{\up{min}(m,m')}_{m-m'} \p{\f{4\rho}{\a}}^{m} \p{\f{4\bar{\rho}}{\a}}^{m'}.
\end{align}
Here $\sim$ indicates that the blocks in the OPE asymptotically approach (\ref{eqn:large-twist-blocks}) at large $m,m'$. From (\ref{eqn:HHLL-matching}), $F(\rho, \bar{\rho})$ also has a representation as a sum over images:
\begin{align}
\label{eqn:F-images}
F(\rho, \bar{\rho}) = \sum_{k \neq 0} f_k(\rho) f_{-k}(\bar{\rho}), \qquad f_k(\rho) = \a^{\Delta} \f{\sqrt{1 - \rho^2}}{\p{1+\rho}^{2\Delta}} \qty[1 - \zeta^k \p{\f{1-\rho}{1+\rho}}^2]^{-\Delta}.
\end{align}
If we can find a formula for the Taylor coefficients of $f_k(\rho)$, then by comparing to the expansion (\ref{eqn:F-OPE}) we can determine the products $\tilde{p}^n_{\ell}$ at large twist. More precisely, if we expand $f_k(\rho)$ in a Taylor series around $\rho, \bar{\rho} = 0$ with coefficients $a_m^{(k)}$, then
\begin{align}
\label{eqn:CnHH-preparation}
\tilde{p}^{\up{min}(m,m')}_{m-m'} \sim \abs{\f{\a}{4}}^{m+m'} \sum_{k \neq 0} a_m^{(k)} a_{m'}^{(-k)} \implies \cal{C}_{n\ell}^{\up{HH}} \sim \f{1}{\cal{C}_{n\ell}^{\up{LL}}} \abs{\f{\a}{4}}^{2n + |\ell|} \sum_{k \neq 0} a_{n+\ell}^{(k)} a_n^{(-k)}.
\end{align}

Thus to determine the expectation values of operators with large twist, it suffices to understand the asymptotic behavior of the Taylor coefficients $a_m^{(k)}$ at large $m$. This can be done using the method of Darboux, which says that the asymptotic Taylor coefficients of an analytic function are controlled by its behavior near the closest of its singularities to the origin. If the function diverges like $f(z) \sim (z - z_0)^{\b}$, then the Taylor coefficients of $f$ are asymptotic to those of $(z - z_0)^{\b}$ itself. The precise statement is as follows \cite{wilf2005}:

\begin{theorem}[Darboux]
\label{thm:darboux}
Let $f(z)$ be analytic in the disk $|z| < R$, and suppose that $f$ has a branch-point singularity at $z = z_0$ on the circle $|z| = R$. Without loss of generality, write $f(z) = (z - z_0)^{\b} g(z)$, where $\b$ is not a positive integer and $g$ is analytic in a neighborhood of $z_0$. Then the Taylor coefficients of $f$ around the origin are asymptotic to
\begin{align}
\label{eqn:darboux}
a_n \sim g(z_0) \f{z_0^{\b - n} \G(n-\b)}{\G(-\b) \G(n+1)} = O(z_0^{-n} n^{-\b - 1}) \qquad \up{as}\;\, n \too \infty.
\end{align}
\end{theorem}

\subsection{Darboux's Method: Results}
\label{sec:darboux-results}

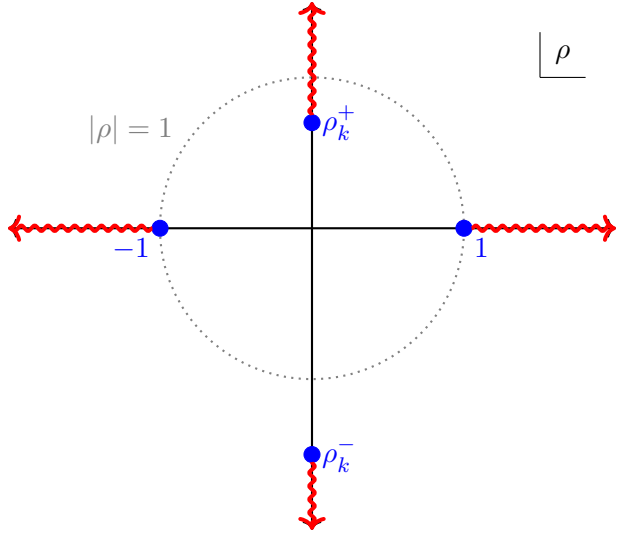
\begin{figure}[t]
\centering
\begin{tikzpicture}
\draw[<->, thick] (-4,0) -- (4,0); 
\draw[<->, thick] (0,-4) -- (0,3); 
\draw[dotted, thick, gray] (0,0) circle (2); 
\filldraw[blue] (2,0) circle (3 pt) node [below right] {$1$}; 
\filldraw[blue] (-2,0) circle (3 pt) node [below left] {$-1$}; 
\filldraw[blue] (0,1.4) circle (3 pt) node [right] {$\rho_k^+$}; 
\filldraw[blue] (0,-3) circle (3 pt) node [right] {$\rho_k^-$}; 
\draw[->, decorate, decoration={snake, segment length = 5pt, amplitude = 1pt}, red, line width = 1.5pt] (2.1,0) -- (4,0); 
\draw[->, decorate, decoration={snake, segment length = 5pt, amplitude = 1pt}, red, line width = 1.5pt] (-2.1,0) -- (-4,0); 
\draw[->, decorate, decoration={snake, segment length = 5pt, amplitude = 1pt}, red, line width = 1.5pt] (0,1.5) -- (0,3); 
\draw[->, decorate, decoration={snake, segment length = 5pt, amplitude = 1pt}, red, line width = 1.5pt] (0,-3.1) -- (0,-4); 
\draw[-] (3,2) -- (3, 2.6); 
\draw[-] (3,2) -- (3.6,2); 
\node (a) at (3.3,2.3) {$\rho$}; 
\node [label = {[label distance=1.8cm]150:$\textcolor{gray}{|\rho| = 1}$}] {};
\end{tikzpicture}
\caption{The singular structure of $f_k(\rho)$, given in (\ref{eqn:F-images}), is shown. One of the two singular points $\rho_k^+$ or $\rho_k^-$ (blue) always lies inside the unit disk. We have chosen the corresponding branch cuts (red) so that $f_k(\rho)$ is analytic in the disk $|\rho| < |\rho_k^+|$.}
\label{fig:branch-cuts}
\end{figure}

As shown in Figure~\ref{fig:branch-cuts}, each image $f_k(\rho)$ has four singular points. Two are at $\rho = \pm 1$, and near these points $f_k(\rho) \sim \sqrt{1 \pm \rho}$. The other two singularities are located at
\begin{align}
\rho_k^+ = i \tan\p{\f{k\pi \a}{2}}, \qquad
\rho_k^- = -i \cot\p{\f{k\pi \a}{2}}.
\end{align}
In conical AdS, $\rho_k^+$ is the same point as $\rho_{N-k}^-$, and $\rho_k^+$ lies inside the unit circle for $k < \f{N}{2}$. Meanwhile, in BTZ we have $\rho_k^+ = \rho_{-k}^-$, and $|\rho_k^+| < 1$ for all $k$. Thus by Theorem~\ref{thm:darboux}, we do not need to worry about the singularities at $\pm 1$, and without loss of generality we may restrict our attention to the behavior of $f_k(\rho)$ near $\rho_k^+$:\footnote{The $\up{AdS}_3/\Z_2$ orbifold is special: it admits only one nontrivial image, and all four singular points of $f_1(\rho)$ lie on the unit circle. The Taylor coefficients are asymptotic to a sum of four terms of the form (\ref{eqn:darboux}), but the two coming from power-law divergences at $\pm i$ dominate over the two coming from the square-root singularities at $\pm 1$. The Taylor coefficients of $F(\rho,\bar{\rho})$ can be computed exactly in this case, and their asymptotic behavior agrees with Darboux's method and is in accordance with the results below.}
\begin{equation}
\begin{aligned}
f_k(\rho) = \p{\f{\a}{4}}^{\Delta} \f{e^{-i\pi k \a \Delta}}{\cos\p{\f{\pi k\a}{2}}} \p{\rho - \rho_k^+}^{-\Delta} \Big[1 + O\p{\rho - \rho_k^+}\Big].
\end{aligned}
\end{equation}

\begin{figure}[t]
\centering
\includegraphics[width=.8\textwidth]{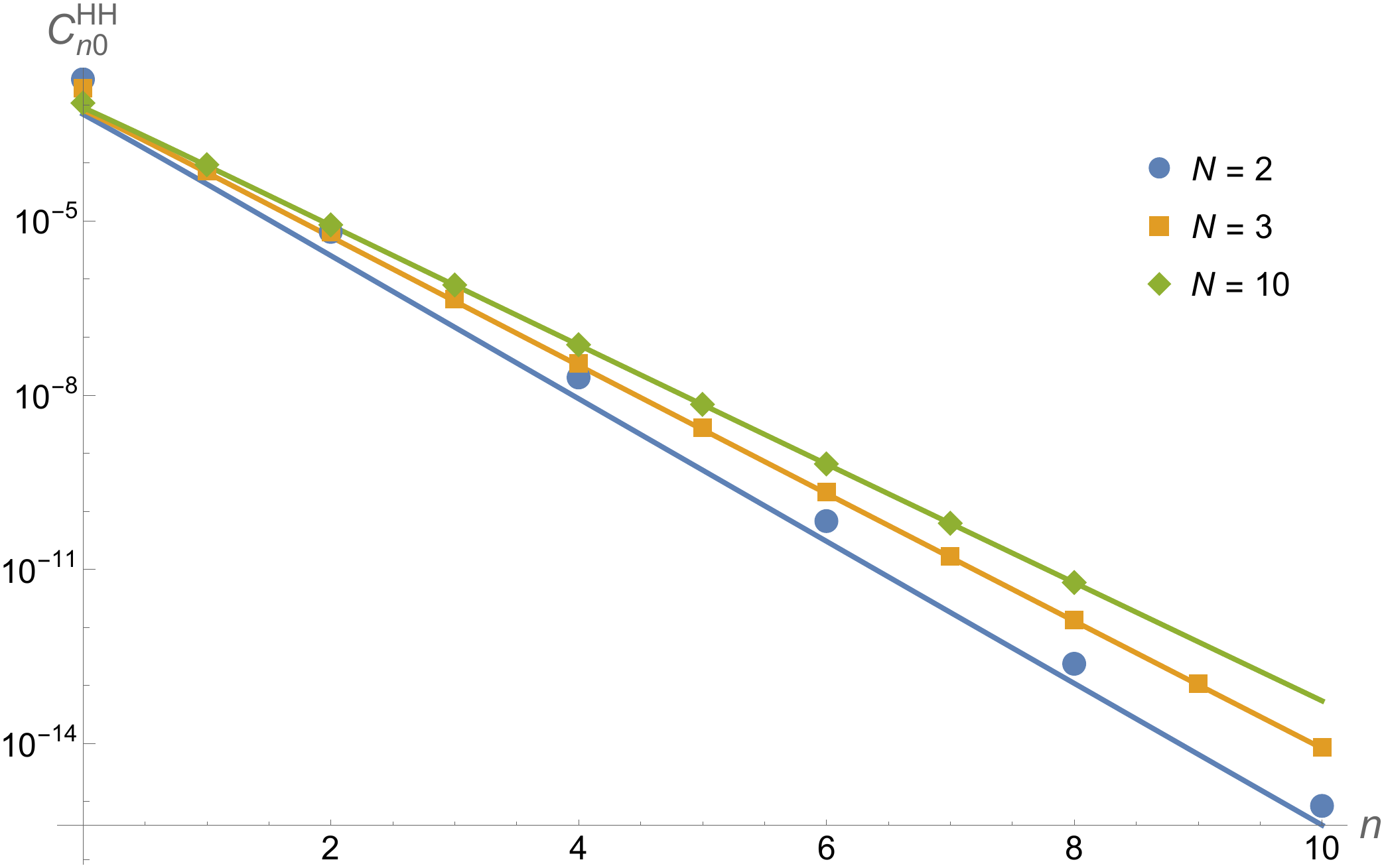}
\caption{The exact scalar expectation values $\cal{C}_{n0}^{\up{HH}}$ (points), as well as the asymptotic predictions of (\ref{eqn:asymptotic-formula}) (lines), are plotted in conical AdS as a function of $n$ for $\Delta = 2$. The asymptotics are already reasonably good in the $\up{AdS}_3/\Z_2$ orbifold, and are essentially indistinguishable from the exact answers for heavier defects.}
\label{fig:asymptotics}
\end{figure}

By Theorem~\ref{thm:darboux}, the Taylor coefficients of $f_k(\rho)$ are asymptotic to
\begin{align}
a_m^{(k)} \sim \p{\f{\a}{4}}^{\Delta} \f{e^{-i\pi k \a \Delta}}{\cos\p{\f{\pi k\a}{2}}} \f{\G(\Delta + m)}{\G(\Delta)\G(1 + m)} \qty[i \tan\p{\f{\pi k \a}{2}}]^{-\Delta - m}
\end{align}
for $1 < k \leq \f{N}{2}$ in conical AdS and for $k \geq 1$ in BTZ. Substituting this into (\ref{eqn:CnHH-preparation}) and using the invariance of $f_k(\rho) f_{-k}(\rho)$ under $k \llrr N-k$ or $k \llrr -k$, we obtain
\begin{align}
\label{eqn:asymptotic-formula}
\cal{C}_{n\ell}^{\up{HH}} \sim \f{i^{\ell}}{\cal{C}_{n\ell}^{\up{LL}}} \p{\f{\G(\Delta + n + \ell) \G(\Delta + n)}{ \G(\Delta)^2 \G(1 + n + \ell) \G(1 + n)}} \sum_{k \neq 0} \f{1}{\cos^2\p{\f{\pi k \a}{2}}} \abs{\f{4}{\a} \tan\p{\f{\pi k \a}{2}}}^{-\Delta_{n\ell}},
\end{align}
where $\Delta_{n\ell} = 2\Delta + 2n + |\ell|$. This is one of our main results: it gives an explicit formula for the expectation values of highly excited composite operators in the presence of a conical defect or a BTZ black hole. It approximates the exact results of (\ref{eqn:low-lying}) surprisingly well, even for modest values of $n$ and $\ell$: see Figure~\ref{fig:asymptotics}. For example, it predicts that
\begin{align}
\cal{C}_{00}^{\up{HH}} \,\;\text{``}\!=\!\text{''}\;\, \f{1}{\sqrt{2}} \sum_{k \neq 0} \f{1}{\cos^2\p{\f{\pi k\a}{2}}} \abs{\f{4}{\a} \tan\p{\f{\pi k\a}{2}}}^{-2\Delta},
\end{align}
which agrees with the exact result (\ref{eqn:CHH00}) near the BTZ threshold, where $|\a| \ll 1$.

The large-$n$ and large-$\Delta$ limits of our results are also worth recording: see Table~\ref{tab:asymptotics}. There we have kept only the $k=1$ contribution to $\cal{C}_{n\ell}^{\up{HH}}$: this term dominates over the others because $\rho_1^+$ is the closest singularity of $F(\rho, \bar{\rho})$ to the origin, and as before this furnishes an approximate analytic continuation of the expectation values to arbitrary $\a$. In both limits, the OPE coefficients are exponentially suppressed in $n$ and $\Delta$, indicating that highly energetic states leave a vanishing imprint on the background geometry.

\begin{table}
\centering
\begin{tabular}{ c | c | c }
{} & $n \too \infty$ & $\Delta \too \infty$ \\ \hline
$\cal{C}_{n\ell}^{\up{LL}}$ & $\displaystyle \f{4\sqrt{2\pi}}{\G(\Delta)^2} 2^{-\Delta_{n\ell}} n^{2\Delta - \f{3}{2}}$ & $\displaystyle \sqrt{\f{2 \p{\Delta/2}^{2n + |\ell|}}{n! (n+\ell)!}}^{{}^{\color{white}{-}}}$ \\[11pt] \hline
\rule{0pt}{1.6\normalbaselineskip}
$\cal{C}_{n\ell}^{\up{HH}}$ & $\displaystyle \f{i^{\ell}}{\cos^2\p{\f{\pi\a}{2}}} \abs{\f{2}{\a} \tan\p{\f{\pi\a}{2}}}^{-\Delta_{n\ell}} \f{1}{2 \sqrt{2\pi n}}$ & $\displaystyle \f{\sqrt{2} i^{\ell}}{\cos^2\p{\f{\pi\a}{2}}} \abs{\f{4}{\a} \tan\p{\f{\pi\a}{2}}}^{-\Delta_{n\ell}} \sqrt{\f{\p{2\Delta}^{2n + |\ell|}}{n! (n+\ell)!}}$ \\[11pt]
\end{tabular}
\caption{Asymptotic behavior of the OPE data $\cal{C}_{n\ell}^{\up{LL}}$ and $\cal{C}_{n\ell}^{\up{HH}}$ in the limits $n \too \infty$ and as $\Delta \too \infty$, based on (\ref{eqn:CnlLL}) and (\ref{eqn:asymptotic-formula}). Here we have used $\Delta_{n\ell} = 2\Delta + 2n + |\ell|$ and kept only the dominant ($k=1$) contribution to the expectation values.}
\label{tab:asymptotics}
\end{table}

\section{Crossing the BTZ Threshold}
\label{sec:crossing-btz-threshold}

\begin{figure}[t]
\centering
\includegraphics[width=\textwidth]{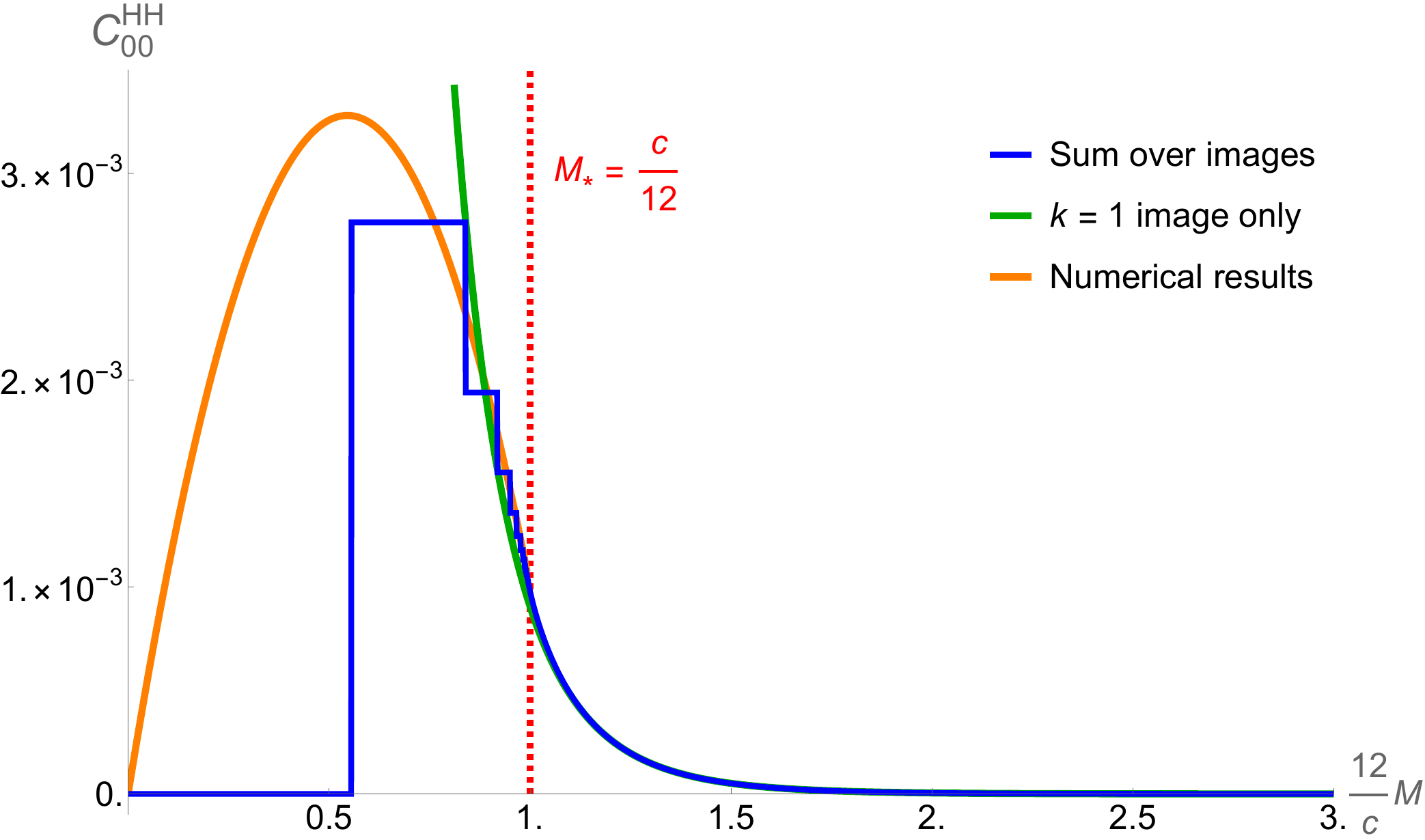}
\caption{The expectation value of $\O_{\up{L}}^2$ is plotted as a function of the mass for $\Delta = 2$, and several methods for computing it are compared. Below the BTZ threshold (shown in red), the sum over images (\ref{eqn:CHH00}) is a finite sum of $N$ terms, explaining the piecewise behavior of the blue curve. Here we can see the breakdown of the approximation (\ref{eqn:Greg-approximate}) (green) for $N < 3$, and the numerical results based on (\ref{eqn:numerical-subtraction}) (orange) are also shown for reference.}
\label{fig:C00-comparison}
\end{figure}

We will now discuss what happens to the expectation values we have computed as the heavy state increases in mass, crosses the BTZ threshold, and becomes a black hole. We find that the expectation values are nonzero in conical AdS and remain finite at the BTZ threshold, but are exponentially small for large black holes. This story is illustrated by Figure~\ref{fig:C00-comparison}, which shows the behavior of $\avg{\O_{\up{L}}^2}$ as a function of the mass of the heavy state.

\subsection{Pure AdS and Light Defects}
\label{sec:pure-ads-light-defects}

The case of pure AdS is trivial: the propagator $G_{\up{AdS}}(z, \bar{z}) = \abs{1-z}^{-2\Delta}$ has no nontrivial regular part, so all of the OPE coefficients $\cal{C}_{n\ell}^{\up{HH}}$ vanish and no expectation values are turned on. This result is in accordance with the vanishing of all vacuum one-point functions.

The regime of very light defects, for which $N = 1 + \eps$ with $0 < \eps \ll 1$, is of interest because it describes perturbative defects that can form heavy multi-particle bound states without producing a black hole.\footnote{Here it is understood that we take the $c \too \infty$ limit before $\eps$ becomes small, so that even very ``light'' defects are heavy enough to backreact on the geometry.} For these defects one can consider perturbative corrections to the pure AdS spectrum, OPE data, correlators, and expectation values. The LL spectrum and OPE coefficients $\cal{C}_{n\ell}^{\up{LL}}$ are unchanged, while the HL spectrum and the OPE coefficients $\cal{C}_{n\ell}^{\up{HL}}$ receive $O(\eps)$ corrections. The same is also true for the full correlator and its singular part. If we write $G = G_{\up{AdS}} + \eps \d G$ and $G_0 = \abs{1 - z}^{-2\Delta} + \eps \d G_0$, then the difference $G_{\up{reg}} = \eps \big(\d G - \d G_0\big)$ can be used to study the expectation values $\cal{C}_{n\ell}^{\up{HH}}$. The details of this procedure are worked out in Appendix~\ref{sec:appendix-light-defects}. In this regime neither the method of images nor asymptotics are reliable, so one must proceed numerically.

Moving up in mass, the case $N=2$ is somewhat special. The $\up{AdS}_3/\Z_2$ geometry admits only the $k=0$ and $k=1$ images, and the regular part of the propagator is
\begin{align}
G^{N=2}_{\up{reg}}(z, \bar{z}) = \abs{z}^{-\Delta/2} \abs{\f{1 + \sqrt{z}}{2}}^{-2\Delta}\iff \tilde{G}_{\up{reg}}^{N=2}(w, \overline{w}) = \abs{2 - w}^{-2\Delta}.
\end{align}
Thanks to the $\Z_2$ symmetry of the orbifold, all $\cal{C}_{n\ell}^{\up{HH}}$ with odd $n$ (in addition to odd $\ell$) vanish: see Figure~\ref{fig:asymptotics}. The first few nonzero expectation values in this background are
\begin{align}
\cal{C}_{00}^{\up{HH}} = \f{2^{-4\Delta}}{\sqrt{2}}, \qquad \cal{C}_{02}^{\up{HH}} = -\f{2^{-4(\Delta + 1)}}{\sqrt{2}} \sqrt{\f{1 + \Delta}{1 + 2\Delta}}, \qquad \cal{C}_{20}^{\up{HH}} = \f{2^{-4(\Delta + 2)}}{\sqrt{2}} \p{\f{1 + \Delta}{1 + 2\Delta}}.
\end{align}
As expected from the asymptotic analysis of Section~\ref{sec:darboux-results}, these OPE coefficients are exponentially small in $\Delta$ and are even further suppressed as $n$ and $\ell$ increase.

In general, conical defects with $N < 3$ must be studied using the numerical methods described near (\ref{eqn:numerical-subtraction}). It is in this range of moderately small masses that the expectation values exhibit their most interesting behavior. Meanwhile, for defects with $N \geq 3$, the discussion surrounding (\ref{eqn:Greg-approximate}) gives a good approximation analytic in the mass.

\subsection{Critical Behavior At the Threshold}
\label{sec:critical-behavior}

As the defect's mass nears the BTZ threshold, several pieces of evidence point to a phase transition in the limit $N \too \infty$.
In the HL spectrum $\w_{n\ell} = \f{1}{N} \p{\Delta + 2n} + |\ell|$, the gap between neighboring states closes as $1/N$, and for each $|\ell|$ an infinite tower of states with energies $\f{1}{N} \p{\Delta + 2n}$ degenerates. This allows small fluctuations in energy to store much more entropy near the threshold than they can in pure AdS, and as $N \too \infty$ the specific heat of these states diverges: this is typical of a continuous phase transition. The HL OPE coefficients $\cal{C}_{n\ell}^{\up{HL}}$ also exhibit critical behavior, since they scale like
\begin{align}
\p{\cal{C}_{n\ell}^{\up{HL}}}^2 \sim
\begin{cases}
N^{-\Delta}, & \ell \neq 0, \\
N^{1-2\Delta}, & \ell = 0
\end{cases}
\qquad \up{as}\;\, N \too \infty.
\end{align}
The s-wave amplitudes $\cal{C}_{n0}^{\up{HL}}$ decay faster in $N$ than those with $\ell \neq 0$: it is easier to add spinning scalar modes to a heavy defect than it is to add scalar modes. In the bulk, this is because adding any amount of angular momentum to a probe is sufficient to create a long-lived orbiting configuration and prevent it from colliding with the defect. We expect this intuition to fail in BTZ, where there are no stable orbits for massive particles. Consequently, no normal modes of the field can be created in BTZ with positive amplitudes.

The strongest indication of a phase transition comes from the linear response of $\O_{\up{L}}$ as a function of the defect's mass. This was computed in \cite{Berenstein:2022ico} by adding a constant source $J \O_{\up{L}}$ to the boundary Hamiltonian and using the Kubo formula \cite{Kubo:1957mj}. If we define
\begin{align}
\Phi(M) = \f{\d}{\d J} \eval_{J = 0} \mel{\O_{\up{H}}}{\O_{\up{L}}}{\O_{\up{H}}}_J = \int_{0}^{\infty} \dd \t \int_0^{2\pi} \dd\th\, G(\t, \th)
\end{align}
and substitute the mode expansion (\ref{eqn:conical-propagator-modes}), then only the s-wave mode survives the angular integral. The remaining integral over $\t$ can be done explicitly, and gives\footnote{The sum over $n$ in (\ref{eqn:conical-propagator-modes}) can be rewritten as a hypergeometric function, and the integral (\ref{eqn:response-integral}) may be performed by changing variables to $u = e^{-\a\t}$. The integral only converges for $0 < \mathfrak{R}(\Delta) < 1$, but the result is analytic in $\Delta$, so we take it to be the analytic continuation of $\Phi(M)$ to arbitrary $\Delta$. See \cite{Berenstein:2014cia} for a similar discussion in the context of conformal perturbation theory in pure AdS.}
\begin{align}
\label{eqn:response-integral}
\Phi(M) = 2\pi \a^{2\Delta - 1} \int_0^{\infty} \dd\t\, e^{-\a\Delta\t} {}_2 F_1 \p{\Delta, \Delta; 1; e^{-2\a\t}} = \f{\a^{2\Delta - 2} \G(\f{1 - \Delta}{2}) \G(\f{\Delta}{2})}{2^{1 + 2\Delta} \G(\f{1 + \Delta}{2}) \G(1 - \f{\Delta}{2})}.
\end{align}
The factors of $\a$ in the response function reflect the rescaling of the boundary coordinates between pure and conical AdS: the transformation of the fields contributes $\a^{2\Delta}$, while the integral measure contributes the remaining $\a^{-2}$. When we normalize $\Phi(M)$ by the response in pure AdS, the Gamma functions drop out entirely, leaving us with
\begin{align}
\label{eqn:response-normalized}
\f{\Phi(M)}{\Phi(0)} = \a^{2\Delta - 2} = \qty[\f{12}{c} \p{M_* - M}]^{\Delta - 1}, \qquad M < M_* = \f{c}{12}.
\end{align}
The normalized response is not analytic at $M = M_*$, with critical exponent $\Delta - 1$: this signals a continuous phase transition at the BTZ threshold. Since we assume $\Delta \geq 1$, the response dies out as we approach the BTZ threshold from below. Geometrically, this is explained by the emergence of an infinite throat as a horizon develops.

On the other side of the phase transition, the computation of $\Phi(M)$ is complicated by the fact that the s-wave contribution to the BTZ propagator $G(\t,\th)$ involves the unwieldy thermal factor $f_{n0}$ given in (\ref{eqn:fnl-factor}). The dependence of $\cal{C}_{n\ell}^{\up{HL}}$ on $\a$ is the same in BTZ as it is below the threshold, so the normalized response $\Phi(M)/\Phi(0)$ is still proportional to (\ref{eqn:response-normalized}). Thus we have a critical vanishing of the response even when approaching the BTZ threshold from above. However, in BTZ there will also be a $\Delta$-dependent prefactor in $\Phi(M)$ that does not cancel exactly with the one in (\ref{eqn:response-integral}). It would be interesting to understand this factor better, including its potential resonances when $\Delta$ is an integer.

\subsection{The Massless BTZ Geometry}
\label{sec:massless-btz}

Unlike the linear response of $\O_{\up{L}}$, the LL expectation values remain analytic across the BTZ threshold, even as the bulk topology changes abruptly. Figure~\ref{fig:Hawking-Page} shows how a few of these expectation values vary as a function of $M$, and it is natural to examine what happens exactly at the BTZ threshold. The corresponding bulk geometry, the massless BTZ (mBTZ) black hole,\footnote{In our conventions, the massless BTZ black hole is not massless: it has ADM mass $M_* = \f{1}{8G}$. The name is, however, appropriate after accounting for the Casimir energy on the cylinder, $H = L_0 + \bar{L}_0 - \f{c}{12}$. We have chosen not to include this overall shift, instead setting the energy of pure $\up{AdS}_3$ to zero.} can be viewed either as the $N \too \infty$ limit of a conical defect or as a BTZ black hole with vanishing horizon size. An analysis of scalar fields in mBTZ is presented in Appendix~\ref{sec:appendix-mBTZ}: there we solve the wave equation and construct the propagator by summing over field modes. We find that the spectrum becomes continuous, and that the resulting modes are delta-function normalizable and oscillatory near the origin.

It is much more direct to obtain the boundary propagator from the method of images. In the limits $N \too \infty$ and $r_0 \too 0$, the conical and BTZ results (\ref{eqn:images}) both become\footnote{One might worry that a small-angle approximation for $\cos{}_{\!}\big(\a \p{\th + 2\pi k}\big)$ in (\ref{eqn:images}) is poorly controlled because the range of $k$ is unbounded. It is, however, justified to argue that the singular image becomes 
\begin{align}
G_0(t,\th) = \f{\a^{2\Delta}}{\qty[\cos(\a\th) - \cos(\a t)]^{\Delta}} \too \f{1}{\qty[\th^2 - t^2]^{\Delta}} \qquad \up{as}\;\, \a \too 0.
\end{align}
The full propagator is then obtained by summing over images of (\ref{eqn:mBTZ-images1}), which indeed gives (\ref{eqn:mBTZ-images2}).}
\begin{align}
\label{eqn:mBTZ-images1}
G_{\up{mBTZ}}(t,\th) = \sum_{k \in \Z} \qty[\p{\th + 2\pi k}^2 - t^2]^{-\Delta},
\end{align}
in agreement with a sum over geodesic lengths in mBTZ \cite{Martinez:2019nor}. We can also take the limit $\a \too 0$ in (\ref{eqn:propagator-images}) by noting that $z^{\a} \approx 1 + \a \log z$ for small $|\a|$. This gives
\begin{align}
\label{eqn:mBTZ-images2}
G_{\up{mBTZ}}(z, \bar{z}) &= \abs{z}^{-\Delta} \sum_{k \in \Z} \Big|{}_{\!}\log z + 2\pi i k\Big|^{-2\Delta}.
\end{align}
On the CFT side, the semiclassical $t$-channel HHLL blocks (\ref{eqn:$t$-channel-blocks}) simplify considerably in the limit $\a \too 0$. The coordinate $w = 1 - z^{\a} \approx -\a\log z$ vanishes as $\a \too 0$, trivializing the hypergeometric part of the block. All that remains is
\begin{align}
\label{eqn:mBTZ-blocks}
\cal{F}_h^{(\up{t})}(z) \too z^{-h_{\up{L}}} \p{-\log z}^{h - 2h_{\up{L}}} \qquad \up{as}\;\, \a \too 0.
\end{align}

We can use these blocks to get an exact formula for the expectation values of the LL double-twist states at the BTZ threshold. To start, the $t$-channel OPE for $G_{\up{mBTZ}}(z, \bar{z})$ organizes itself into a power series in $\log z$ with coefficients $\cal{C}_{n\ell}^{\up{HH}} \cal{C}_{n\ell}^{\up{LL}}$. We have
\begin{align}
\label{eqn:mBTZ-tchannel-OPE}
\tilde{G}_{\up{reg}}(w, \overline{w}) = \sum_{n\ell} \cal{C}_{n\ell}^{\up{HH}} \cal{C}_{n\ell}^{\up{LL}} \big|{}_{\!}\log z\big|^{2n} \p{\log \bar{z}}^{\ell}.
\end{align}
Meanwhile, the regular part of the sum over images (\ref{eqn:mBTZ-images2}) can be converted into a similar power series by applying the binomial formula to each image. We get
\begin{align}
\label{eqn:mBTZ-image-OPE}
\tilde{G}_{\up{reg}}(w, \overline{w}) &= \sum_{n\ell} \f{2i^{\ell}}{\G(\Delta)^2} \f{\G(\Delta + n + |\ell|) \G(\Delta + n)}{\G(1 + n + |\ell|) \G(1 + n)} \p{2\pi}^{-\Delta_{n\ell}} \zeta(\Delta_{n\ell}) \big|{}_{\!} \log z\big|^{2n} \p{\log z}^{\ell},
\end{align}
where the factor $(2\pi)^{-\Delta_{n\ell}} \zeta(\Delta_{n\ell})$ comes from summing over the factors of $\p{2\pi k}^{-2\Delta_{n\ell}}$ in the binomial expansion of each image. If we had kept only the $k=1$ term, the zeta function would not be there. By matching the coefficients of (\ref{eqn:mBTZ-tchannel-OPE}) and (\ref{eqn:mBTZ-image-OPE}), we read off\footnote{We could also have guessed this result by taking $\a \too 0$ in the asymptotic formula (\ref{eqn:asymptotic-formula}).
}
\begin{align}
\label{eqn:mBTZ-expectation-values}
\cal{C}_{n\ell}^{\up{HH}} = \f{2i^{\ell}}{\cal{C}_{n\ell}^{\up{LL}}} \p{\f{\G(\Delta + n + |\ell|) \G(\Delta + n)}{ \G(\Delta)^2 \G(1 + n + |\ell|) \G(1 + n)}} \p{2\pi}^{-\Delta_{n\ell}} \zeta(\Delta_{n\ell}).
\end{align}

\section{Thermal One-Point Functions in BTZ}
\label{sec:thermal-one-point-functions}

In 2d CFT, conformal invariance forbids any nontrivial primary operator $\O$ from acquiring a nonzero expectation value if the theory lives on a cylinder. On a torus, however, there is no such restriction. Thermal one-point functions like $\avg{\O}_{\b}$ can be nontrivial in a CFT with a spatial a circle, but they must vanish at zero temperature because the thermal circle decompactifies as $\b \too \infty$. It follows from modular invariance that $\avg{\O}_{\b}$ must also rapidly decay to zero at high temperatures. The bulk analog of this argument is that the horizon of a very large BTZ black hole effectively becomes planar. At infinite volume, the bulk geometry is just a solid cylinder, which again forces all one-point functions to vanish.

\subsection{The Cylinder and the Torus}
\label{sec:cylinder-and-torus}

\begin{figure}[t]
\begin{subfigure}{.5\textwidth}
  \centering
  \includegraphics[width=.95\linewidth]{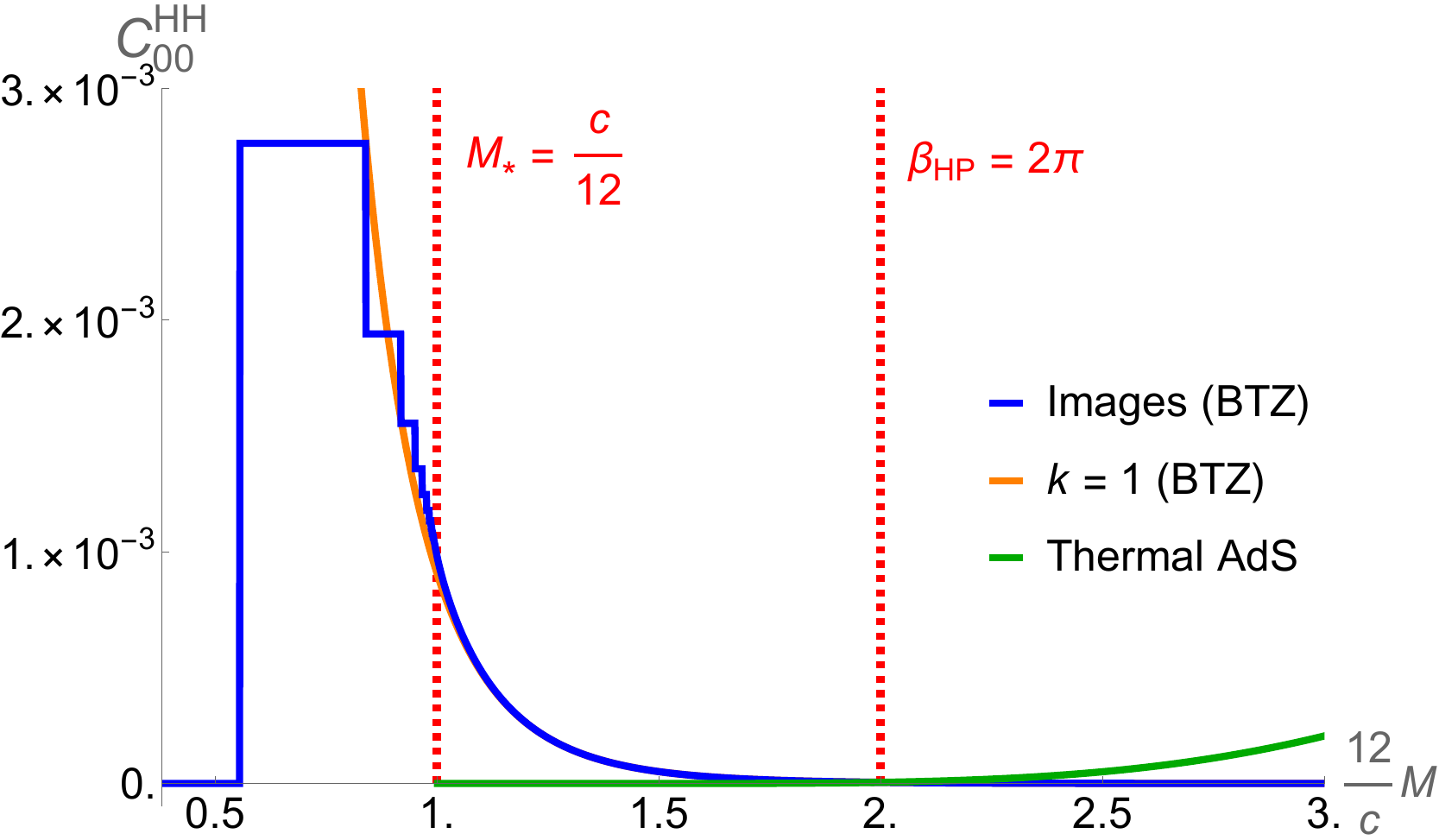}
\end{subfigure}%
\begin{subfigure}{.5\textwidth}
  \centering
  \includegraphics[width=.95\linewidth]{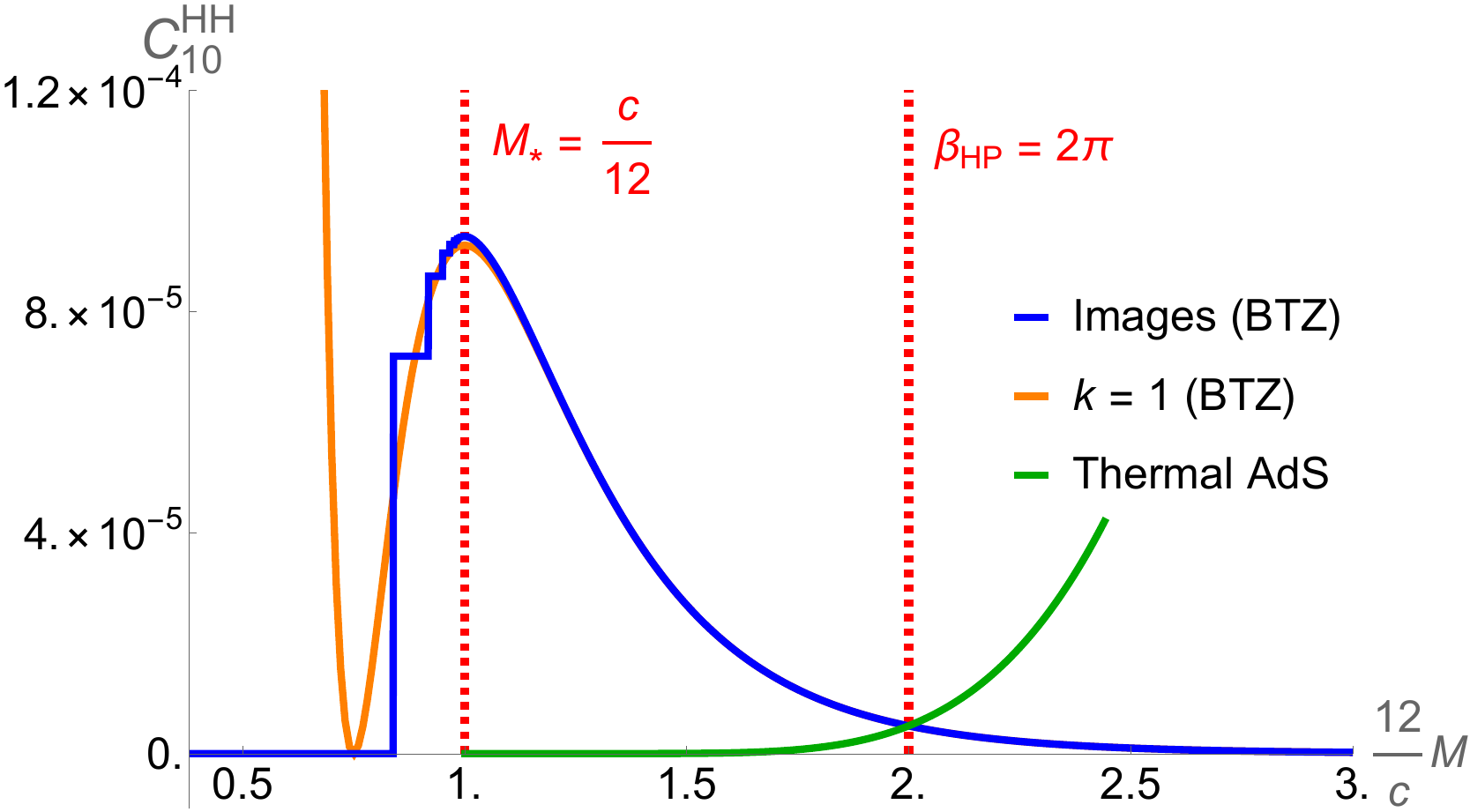}
\end{subfigure} \\[11pt]
\begin{subfigure}{.5\textwidth}
  \centering
  \includegraphics[width=.95\linewidth]{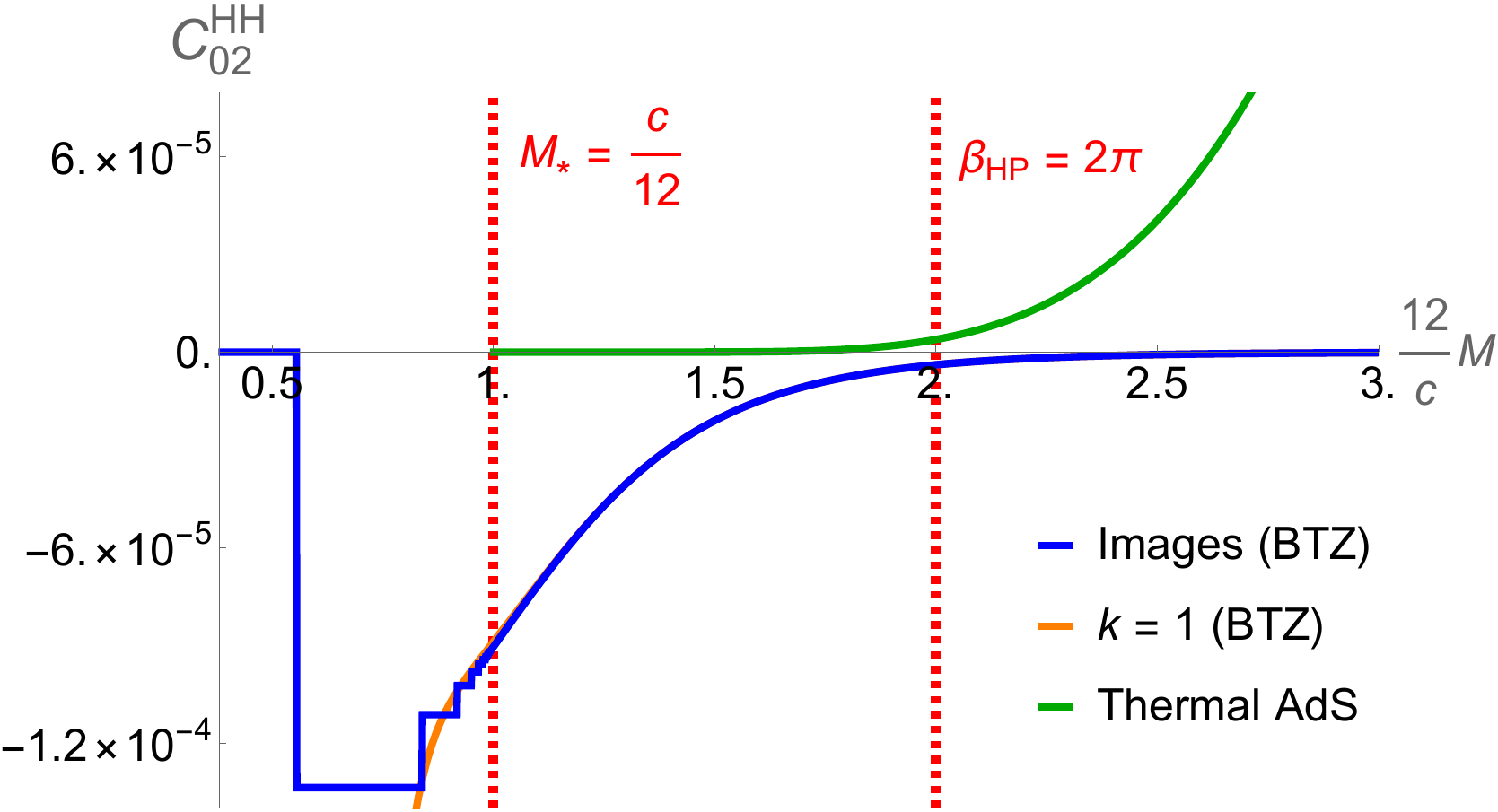}
\end{subfigure}%
\begin{subfigure}{.5\textwidth}
  \centering
  \includegraphics[width=.95\linewidth]{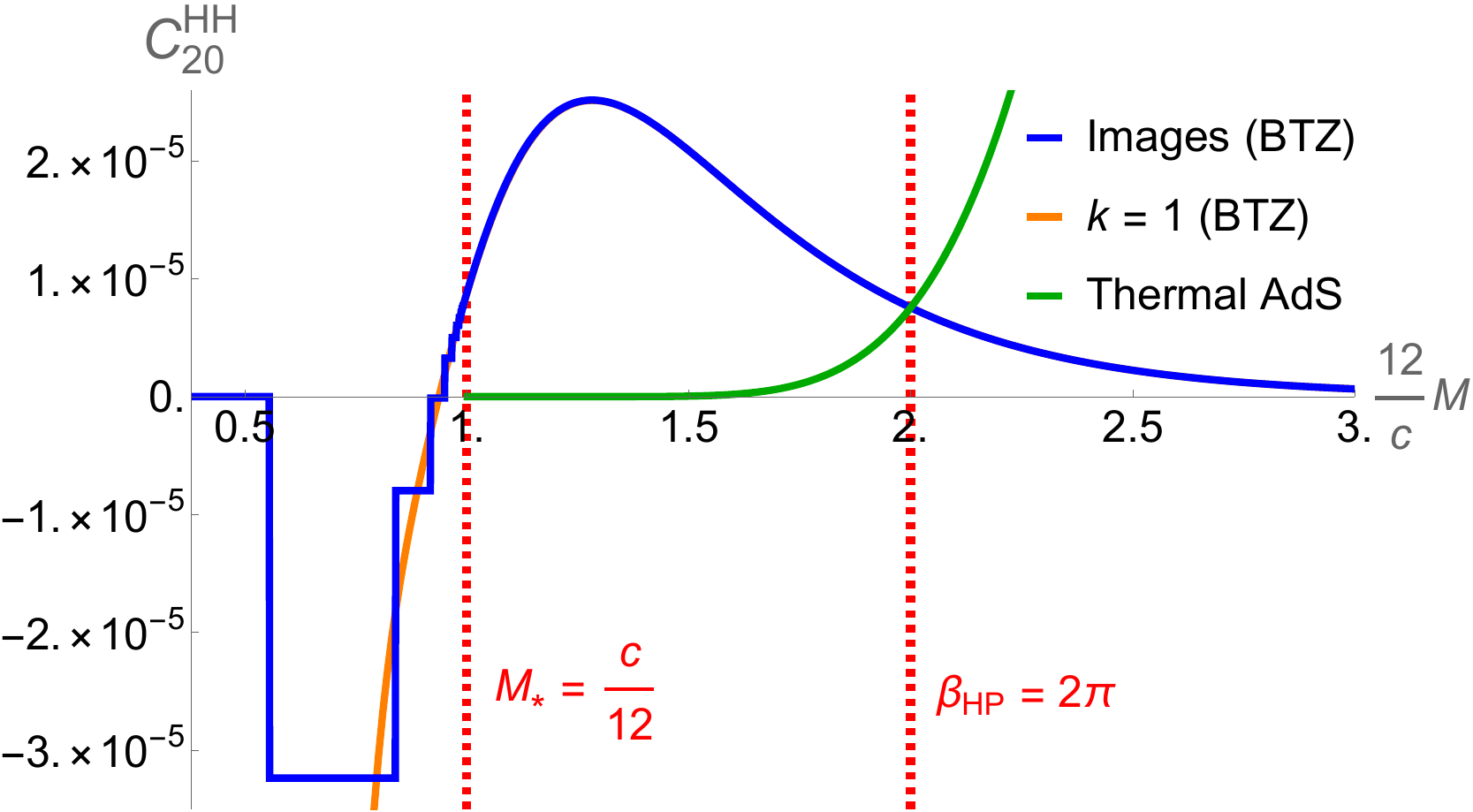}
\end{subfigure}
\caption{The first few low-lying expectation values are plotted as a function of the mass of the heavy state. We take $\Delta = 2$ and focus primarily on black holes. The BTZ expectation values (blue) are dominant for $\b < 2\pi$ (i.e. for $M > \f{c}{6}$), but are subdominant to the corresponding thermal AdS results (green) at low temperatures.}
\label{fig:Hawking-Page}
\end{figure}

We have arrived at essentially the same conclusions from a Lorentzian perspective: instead of working on the torus, we have modeled thermal states by insertions of a heavy CFT primary on the cylinder. In our treatment, the coefficients $\cal{C}_{n\ell}^{\up{HH}} = \mel{\O_{\up{H}}}{[\O_{\up{L}} \O_{\up{L}}]_{n\ell}}{\O_{\up{H}}}$ compute expectation values in highly excited states. But above the BTZ threshold, heavy operators act as effective thermal backgrounds \cite{Fitzpatrick:2015zha}, so we can also view the $\cal{C}_{n\ell}^{\up{HH}}$ as thermal one-point functions $\avg{[\O_{\up{L}} \O_{\up{L}}]_{n\ell}}_{\b}$ at a temperature set by the conformal dimension of $\O_{\up{H}}$. This interpretation is supported by Figure~\ref{fig:Hawking-Page}, which confirms that---as expected for thermal one-point functions---the $\cal{C}_{n\ell}^{\up{HH}}$ quickly decay to zero at high temperature.

At first glance, we encounter a puzzle at low temperatures: thermal one-point functions must vanish at zero temperature, but our expectation values do not. On the contrary, (\ref{eqn:mBTZ-expectation-values}) shows that they are nonzero in massless BTZ, which has zero temperature and a cylinder boundary \cite{Satoh:2002nj}. The resolution of this puzzle relies on the observation that below the Hawking--Page temperature $\b_{\up{HP}} = 2\pi$, BTZ is thermodynamically unstable to thermal AdS \cite{Hawking:1982dh}. Indeed, if scalar perturbations are added to a small BTZ black hole, the system will decay to thermal AdS. As we have seen, these perturbations generate nontrivial one-point functions for light two-particle states. It follows that the true one-point functions at a given temperature must be computed in the dominant geometry,\footnote{We assume that the perturbations are light enough not to backreact on the geometry, so we do not need to worry about their potential effects on the thermodynamic stability of the system.} and these dominant expectation values should vanish at both low and high temperature.

Fortunately, one-point functions in thermal AdS can be obtained from those in BTZ by a modular transformation $\b \too \f{4\pi^2}{\b}$. Their high-temperature decay in BTZ then forces them to vanish in thermal AdS at zero temperature, resolving the puzzle. Our results (\ref{eqn:mBTZ-expectation-values}) represent subdominant one-point functions: they arise only if one holds the bulk topology fixed while varying the temperature, forcing the thermal circle to stay contractible. This is how one obtains massless BTZ, rather than pure AdS, as a zero-temperature limit.

Our choice to work on the cylinder rather than the torus comes with advantages and disadvantages. On one hand, it affords a unified treatment of expectation values in heavy states that analytically continues across the BTZ threshold, and it allows us to describe how thermal behavior arises in black holes without the need to add a thermal circle to the analysis by hand. On the other hand, we have seen that it has the potential to obscure some of the low-temperature physics. Nevertheless, we needed to understand the BTZ one-point functions at all temperatures to get a complete picture of how they arise by analytic continuation from expectation values in pure excited states. 

To be clear, the present discussion is specific to the canonical ensemble. We have been working microcanonically, where there is no Hawking--Page transition and therefore no problem with our expectation values. Our aim here is only to point out that the low-temperature behavior of our expectation values should be interpreted with some care.

\subsection{The Simplest One-Point Function}
\label{sec:illustrative-example}

Let us illustrate the preceding discussion in more concrete terms using the example of the composite operator $\O_{\up{L}}^2$. Above the Hawking--Page temperature ($\b < 2\pi$), large black holes ($r_0 > 1$) dominate, and the thermal one-point function $\avg{\O_{\up{L}}^2}_{\b}$ is given by (\ref{eqn:CHH00}):
\begin{align}
\label{eqn:O2-expectation-BTZ}
\avg{\O_{\up{L}}^2}_{\b}^{\up{BTZ}} = \sqrt{2} \sum_{k = 1}^{\infty} \p{\f{2}{r_0} \sinh\p{\pi k r_0}}^{-2\Delta} = \sqrt{2} \sum_{k = 1}^{\infty} \p{\f{\b}{\pi} \sinh\p{\f{2\pi^2 k}{\b}}}^{-2\Delta}.
\end{align}
At high temperatures ($\b \too 0$ or $r_0 \too \infty$), it decays to zero exponentially:
\begin{align}
\avg{\O_{\up{L}}^2}_{\b}^{\up{BTZ}} \too 
\sqrt{2} r_0^{2\Delta} e^{-2\pi \Delta r_0} =
\sqrt{2}\p{\f{\b}{2\pi}}^{-2\Delta} e^{-\f{4\pi^2 \Delta}{\b}} \quad \up{as}\;\, \b \too 0.
\end{align}
As we pass below the Hawking--Page temperature ($\b > 2\pi$), the BTZ one-point function becomes subdominant and has a nonzero limit at zero temperature (i.e. in mBTZ):
\begin{align}
\avg{\O_{\up{L}}^2}_{\b}^{\up{BTZ}} \too \sqrt{2} \sum_{k = 1}^{\infty} \p{2\pi k}^{-2\Delta} = \sqrt{2} \zeta(2\Delta) \quad \up{as}\;\, \b \too \infty.
\end{align}

To find the dominant low-temperature one-point function, we need to pass to thermal AdS. In Euclidean signature, the boundary tori of BTZ and thermal AdS are related by the modular transformation $\b \too \f{4\pi^2}{\b}$. This transformation also relates their temperatures and exchanges whether the thermal or spatial circle ($\t$ or $\th$) is contractible in the bulk. By using the modular covariance CFT one-point functions on the torus, we can determine the one-point functions in thermal AdS from the corresponding ones in BTZ:
\begin{align}
\label{eqn:modular-transformation}
\avg{\O}_{\f{4\pi^2}{\b}} = i^{\ell} \p{\f{\b}{2\pi}}^{\Delta_{\O}} \avg{\O}_{\b} \implies
\avg{\O}_{\b}^{\up{tAdS}} = i^{\ell} \p{\f{2\pi}{\b}}^{\Delta_{\O}} \avg{\O}_{\f{4\pi^2}{\b}}^{\up{BTZ}}.
\end{align}
(Here $\O$ is a CFT primary with spin $\ell$ and dimension $\Delta_{\O}$.) Applying this to (\ref{eqn:O2-expectation-BTZ}) gives
\begin{align}
\label{eqn:O2-expectation-tAdS}
\avg{\O_{\up{L}}^2}_{\b}^{\up{tAdS}} = \sqrt{2} \sum_{k=1}^{\infty} \p{2 \sinh\p{\f{k \pi}{r_0}}}^{-2\Delta} = \sqrt{2} \sum_{k=1}^{\infty} \p{2 \sinh\p{\f{k \b}{2}}}^{-2\Delta}.
\end{align}
This one-point function dominates below the Hawking--Page temperature, and as we expect, it vanishes as the temperature of thermal AdS goes to zero:
\begin{align}
\avg{\O_{\up{L}}^2}_{\b}^{\up{tAdS}} \too \sqrt{2} e^{-\f{2\pi \Delta}{r_0}} = \sqrt{2} e^{-\Delta \b} \quad \up{as}\;\, \b \too \infty.
\end{align}

So modular invariance tells us that the thermal one-point function $\avg{\O_{\up{L}}^2}_{\b}$ is given by (\ref{eqn:O2-expectation-BTZ}) for $\b < 2\pi$ and by (\ref{eqn:O2-expectation-tAdS}) for $\b > 2\pi$. There is a Hawking--Page transition at $\b = 2\pi$, where the BTZ and thermal AdS results exchange dominance at the critical value
\begin{align}
\avg{\O_{\up{L}}^2}_{2\pi} = \sqrt{2} \sum_{k = 1}^{\infty} \big(2 \sinh\p{k \pi}\big)^{-2\Delta} \approx \sqrt{2} e^{-2\pi \Delta}.
\end{align}
It is amusing to note that the series above can be summed exactly when $\Delta = 1$ (in which case $\O_{\up{L}}^2$ is a massless field) or $\Delta = 2$ (in which case $\O_{\up{L}}$ itself is massless):
\begin{subequations}
\begin{align}
\Delta = 1: \qquad \avg{\O_{\up{L}}^2}_{2\pi} &= \sqrt{2} \sum_{k=1}^{\infty} \f{1}{\big(2 \sinh\p{k \pi}\big)^2} = \f{\sqrt{2}}{8} \p{\f{1}{3} - \f{1}{\pi}} \sim 10^{-3}, \\
\Delta = 2: \qquad \avg{\O_{\up{L}}^2}_{2\pi} &= \sqrt{2} \sum_{k=1}^{\infty} \f{1}{\big(2 \sinh\p{k \pi}\big)^4} = \f{\sqrt{2}}{48} \p{\f{1}{\pi} - \f{11}{30} + \f{\G(1/4)^8}{640 \pi^6}} \sim 10^{-6}.
\end{align}
\end{subequations}

The analysis of the other LL double-twists can be carried out in analogy to what we have done for $\O_{\up{L}}^2$, although the expressions for $\avg{[\O_{\up{L}} \O_{\up{L}}]_{n\ell}}_{\b}$ become more complicated for larger $n$ and $\ell$. Observe because the LL double-twists have even spin, the factor $i^{\ell}$ in (\ref{eqn:modular-transformation}) is $\pm 1$, depending on whether $\ell \equiv 2$ (mod 4) or $\ell \equiv 0$ (mod 4). In the former case, this implies that the expectation values $\avg{[\O_{\up{L}} \O_{\up{L}}]_{n\ell}}_{\b}$ have opposite signs in BTZ and in thermal AdS, and in particular they must vanish at $\b = 2\pi$. It is natural to suspect that a $\Z_2$ symmetry emerges at $\b = 2\pi$ to ensure this: it would be interesting to better understand where this symmetry comes from and why it only affects certain operators.

\section{Discussion and Future Directions}
\label{sec:conclusions-future-work}

\paragraph{Summary of main results.}

In this paper, we discussed holographic correlators in heavy $\up{AdS}_3$ backgrounds and identified an infinite family of composite operators that have nontrivial expectation values in these backgrounds. We derived the propagator for free scalar fields in the bulk by solving the wave equation and summing over modes of the field, as well as by the method of images. On the CFT side, we studied an HHLL four-point function where the heavy insertions create a massive particle or a BTZ black hole in the bulk. By matching its expansion in Virasoro blocks to our bulk results, we obtained the spectrum and OPE data of all states that contribute to the $s$-channel and $t$-channel OPEs.

The OPE coefficients that couple light intermediate states to the heavy insertions represent their expectation values in heavy backgrounds. We determined these expectation values exactly, discussed them in several approximate and asymptotic forms, and described their behavior as the mass of the heavy state varies across the BTZ threshold. We found that they are finite and analytic at the threshold, but become exponentially suppressed for large black holes. In black hole backgrounds, the expectation values are identified with thermal one-point functions on the torus, although below the Hawking--Page temperature they are dominated by the corresponding one-point functions in thermal AdS.

\paragraph{Thermality without temperature.} One broad conclusion of our work is that the physics of conical defects contains hints of thermal behavior, even below the BTZ threshold. From Figure~\ref{fig:propagator}, we see that long-range correlations in conical AdS become suppressed as the mass of the defect increases: this effect resembles the shortening of thermal correlation lengths as the temperature or energy increases. Some other indications of a transition to thermality include the analytic continuation of the metric between conical AdS and BTZ, the phase transition in the linear response function, and the expectation values that analytically continue into BTZ one-point functions.

Much of this discussion is related to our decision to work with individual operator insertions rather than averaged quantities. Our results can be viewed as a realization of the Eigenstate Thermalization Hypothesis (ETH), which is often used to argue that averaged OPE coefficients in thermal states are nearly identical to individual OPE coefficients in pure states with the same energy \cite{Srednicki:1994mfb, Collier:2019weq}. It would be interesting to sharpen what the ETH says about the extent to which we can treat black holes as CFT primaries.

\paragraph{Beyond the vacuum block.} Vacuum dominance is a powerful phenomenon that often allows one to focus exclusively on the Virasoro vacuum block. We have tried to emphasize in this paper that the analytic structure of holographic correlators encodes important physics---present even at leading order in the semiclassical limit---that the vacuum block cannot see. The vacuum contribution correctly predicts the singular structure of four-point functions in the OPE limit, but there are additional regular terms in the OPE that are necessary to preserve crossing symmetry and which have physical interpretations as quantum hair, or nontrivial expectation values of double-trace operators in heavy states. To our knowledge, this careful accounting of non-vacuum $t$-channel exchanges represents the first detailed calculation of double-trace expectation values in AdS/CFT.

\begin{figure}[t]
\centering
\begin{subfigure}{.45\textwidth}
\quad
\begin{tikzpicture}
\draw[thick] (0,0) circle (3); 
\filldraw[] (0,0) circle (.9); 
\draw[very thick, red] (3,0) -- (1.7, 0); 
    \node [red] (a) at (2.3, .3) {$\phi_{\O}$}; 
\draw[very thick, blue] (0,0) circle (1.7); 
\node [blue] (b) at (1, 1.8) {$\phi_{\chi}$}; 
\filldraw[] (1.7,0) circle (2.5 pt); 
\end{tikzpicture}
\caption{The Kraus--Maloney calculation}
\label{fig:Kraus-Maloney}
\end{subfigure}
\begin{subfigure}{.45\textwidth}
\quad
\begin{tikzpicture}
\draw[thick] (0,0) circle (3); 
\filldraw[] (0,0) circle (.9); 
\draw[very thick, blue] 
(3,0) .. controls (1.8,0) and (1.4,1.7) .. (0,1.7)
      .. controls (-.94,1.7) and (-1.7,.94) .. (-1.7,0)
      .. controls (-1.7,-.94) and (-.94,-1.7) .. (0,-1.7)
      .. controls (1.4,-1.7) and (1.8,0) .. (3,0);
\filldraw[] (3,0) circle (2.5 pt); 
\node [blue] (a) at (1, 1.8) {$\phi$}; 
\node [red] (b) at (3.7,0) {$[\up{LL}]_{n\ell}$};
\end{tikzpicture}
\caption{Our calculation}
\label{fig:Grabovsky}
\end{subfigure}
\caption{The one-point function $\avg{\O}_{\b}$ in BTZ is generated by a Witten diagram whose geodesic approximation (left) is shown. In our calculation (right), $\phi_{\O}$ is a weakly bound two-particle state whose CFT dual is an LL double-twist operator $[\O_{\up{L}} \O_{\up{L}}]_{n\ell}$. It splits on the boundary into two copies of $\O_{\up{L}}$, which then propagates in the bulk as a free field $\phi$.}
\label{fig:thermal-1pt}
\end{figure}
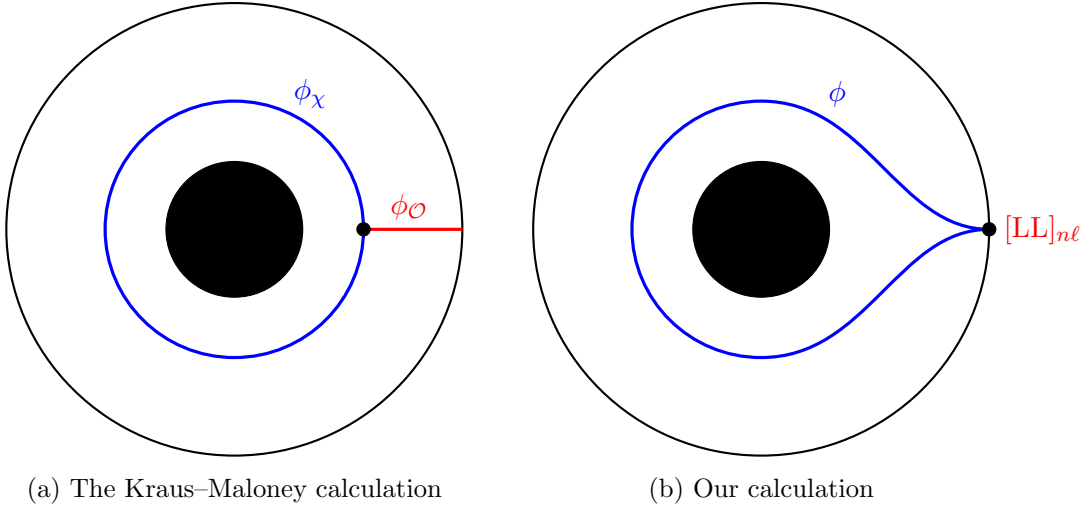

\paragraph{Comparison to Cardyology.} The expectation values we have computed have been studied from a different perspective by Kraus and Maloney \cite{Kraus:2016nwo}. They used modular invariance to derive a universal formula for the averaged three-point coefficients $\overline{\mel{E}{\cal{O}}{E}}$, where $\O$ is a CFT primary and $E$ is the energy of an excited state. They argued that such three-point functions are generated in the bulk by a cubic coupling $\phi_{\chi}^2 \phi_{\O}$ in the action between the fields dual to $\cal{O}$ and the lightest nontrivial primary $\chi$ for which $\mel{\chi}{\O}{\chi} \neq 0$. The corresponding Witten diagram computation, shown in Figure~\ref{fig:Kraus-Maloney}, suggests that these three-point coefficients arise from geodesics (or, more precisely, propagators) that wind nontrivially around the horizon of the black hole when $\O$ splits into two $\chi$ fields.

Our work provides an explicit realization of this story in the case where the composite operators $\O = [\O_{\up{L}} \O_{\up{L}}]_{n\ell}$ acquire expectation values by coupling to the light field $\chi = \O_{\up{L}}$ through the OPE coefficients $\cal{C}_{n\ell}^{\up{LL}} = \mel{\chi}{\O}{\chi}$. Here the cubic coupling arises naturally (because $\cal{C}_{n\ell}^{\up{LL}}$ allows two copies of $\O_{\up{L}}$ to fuse into a composite state), rather than being put in by hand. It should be noted that all of our operator insertions lie on the boundary, so we have no explicit cubic vertices in the bulk, leading to the diagram in Figure~\ref{fig:Grabovsky}. Nevertheless, our results for the expectation values $\cal{C}_{n\ell}^{\up{HL}} = \mel{\O_{\up{H}}}{[\O_{\up{L}} \O_{\up{L}}]_{n\ell}}{\O_{\up{H}}}$ agree with the Kraus--Maloney formula for $\overline{\mel{E}{\cal{O}}{E}}$, and our analysis extends the intuition of winding geodesics to geometries with no horizon. In our case, this bulk interpretation of three-point coefficients doubles as a CFT interpretation of non-minimal bulk geodesics: their role is played on the boundary by the exchange of light double-trace operators.

\paragraph{Multiparticle wave functions.} It is natural to ask whether the wave functions of the HL and LL double-twist primaries can be extended into the bulk. On the boundary, these wave functions are the scaling parts of the Virasoro blocks (\ref{eqn:$s$-channel-blocks}) and (\ref{eqn:$t$-channel-blocks}):\footnote{The full Virasoro blocks contain information about both primaries and descendants. Semiclassically, primaries are responsible for the scaling factor, while descendants contribute the ${}_2 F_1$ part of the blocks.}
\begin{equation}
\begin{aligned}
\phi_{n\ell}^{\up{HL}}(z, \bar{z}) = |z|^{(\a - 1)\Delta} |z|^{2\a n} z^{\ell}, \qquad
\phi_{n\ell}^{\up{LL}}(z, \bar{z}) = |z|^{(\a - 1)\Delta} \abs{\f{1 - z^{\a}}{\a}}^{2n} \p{\f{1 - z^{\a}}{\a}}^{\ell}.
\end{aligned}
\end{equation}
The HL wave functions, which are mode solutions of the wave equation, can be extended into the bulk by including the radial part of the solution. It is less clear how to extend the LL double-twist wave functions into the bulk. Ultimately, one would also want to understand the radial dependence of the expectation values we have computed.

By analogy with our analysis on the boundary, we expect that all of the information necessary to compute wave functions and expectation values in the bulk is contained in the regular part of the bulk-to-bulk propagator. It would be interesting to understand this better, especially with an eye toward similar calculations in higher dimensions. See \cite{Berenstein:2022nlj} and \cite{Dodelson:2022yvn} for progress in higher-dimensional setups, for example.

\paragraph{Connections to the bootstrap.} Several parallels between our work and the conformal bootstrap deserve investigation. It would be instructive to compute the $\cal{C}_{n\ell}^{\up{HH}}$ using the orthogonality of conformal partial waves \cite{Simmons-Duffin:2017nub}, i.e. using Lorentzian inversion. It would also be good to make closer contact with related work on spacetime bananas \cite{Abajian:2023jye}, universal dynamics at finite central charge \cite{Collier:2019weq}, wormholes and ensemble averaging \cite{Chandra:2022bqq}, thermal one-point functions \cite{Grinberg:2020fdj,David:2022nfn}, and the thermal bootstrap program \cite{Iliesiu:2018fao,Alday:2020eua}. 

In higher dimensions, there has been great progress on bootstrapping universal OPE data in the stress tensor sector of HHLL correlators \cite{Fitzpatrick:2019zqz,Fitzpatrick:2019efk,Kulaxizi:2019tkd}. The prospect of extending our work to higher dimensions by using these results is exciting, but also challenging. A closed form for the stress tensor contributions to $G(z,\bar{z})$ is not known; moreover, in the bulk, there are local curvature effects and one generically loses the method of images. So it is not immediately clear how to write down the regular and singular parts of $G(z,\bar{z})$.

\paragraph{Future work.} Our analysis can be extended by adding charge or spin to the light or heavy states: see \cite{li2024spinning} for a study of spinning conical defects. It would also be interesting to compute off-diagonal matrix elements $\mel{E}{\cal{O}}{E'}$ for the LL double-twists and compare the results to \cite{Brehm:2018ipf}. It should then be straightforward to compute generic transition amplitudes of the form $\mel{E, J}{\O_1 \, \O_2}{E', J'}$, where $\ket{E,J}$ represents a heavy state with energy $E$ and angular momentum $J$, and where $\O_1$, $\O_2$ are light operators.

In this work have restricted ourselves to the strict limit of large $c$, but subleading corrections to the Virasoro blocks are well understood \cite{Perlmutter:2015iya}, and the backreaction of free fields in heavy states has also been studied \cite{Belin:2021htw,Emparan:2022ijy}. (In the bulk, backreaction causes a horizon to form, creating a ``quantum BTZ'' geometry.) So in principle, it should be possible to piece together a more refined perturbative understanding of our results.

It may be possible to employ methods similar to ours to study a model of black hole formation. Following the ideas of \cite{Anous:2016kss, Haehl:2023lfo}, one could imagine colliding two light particles or shock waves, sent in at early times, in a two-sided geometry. If they are highly boosted (i.e. highly excited descendants), their center-of-mass energy could exceed the BTZ threshold and a black hole could form \cite{Gott:1990zr}. It would be illuminating to find a CFT interpretation of this process, and to connect it to out-of-time-ordered correlators and quantum chaos.

\acknowledgments

It is a pleasure to thank Xi Dong, Jo\~{a}o Penedones, Tom Hartman, Kwinten Fransen, Adolfo Holguin, Matthew Dodelson, Gabriele di Ubaldo, Sean McBride, Jesse Held, George Hulsey, and Frazier Li for helpful discussions; Adolfo Holguin, Sean McBride, and Jesse Held for comments on the manuscript; and especially David Berenstein for patience, wisdom, and many stimulating conversations. D.G. is grateful to the International Center for Interdisciplinary Science and Education in Quy Nh{\ohorn}n, Vietnam for its hospitality during the 2023 Advanced Summer School in Quantum Field Theory and Quantum Gravity. D.G. was supported in part by the Department of Energy under grant DE-SC 0011702.

\appendix
\section{Wave Equations in Heavy AdS Backgrounds}
\label{sec:appendix-wave-equations}

In this appendix, we carry out a detailed analysis of the wave equation in heavy $\up{AdS}_3$ geometries, supplementary to the story sketched out in Section~\ref{sec:correlators-wave-equation}. After reviewing how the wave equation is solved in pure $\up{AdS}_3$, we generalize to the case of conical AdS and give the analogous derivation in BTZ. In each case, we obtain the momentum-space Green's function. Finally, we close with a discussion of what happens in massless BTZ.

\subsection{Generalities and Setup}

We consider the heavy $\up{AdS}_3$ geometries (\ref{eqn:heavy-metric}), reproduced below for convenience:
\begin{align}
\label{eqn:heavy-AdS-metric}
\dd s^2 &= -f(r) \dd t^2 + \f{\dd r^2}{f(r)} + r^2 \dd\th^2, \qquad
f(r) = \begin{cases}
r^2 + \f{1}{N^2} &\text{below the threshold}, \\
r^2 - r_0^2 &\text{above the threshold}.
\end{cases}
\end{align}
We study the Klein--Gordon equation for a massive scalar field in such backgrounds:
\begin{align}
\label{eqn:Klein-Gordon}
\big(\Box - m^2\big) \phi = \f{1}{\sqrt{-g}} \Big( \pd_{\mu} \p{\sqrt{-g} g^{\mu\nu} \partial_{\nu}} - m^2\Big) \phi = 0.
\end{align}
The mass of the field is related to the conformal dimension of the dual CFT operator by $m^2 = \Delta(\Delta - 2)$, and we assume $\Delta \geq 1$ throughout. We may decompose solutions of (\ref{eqn:Klein-Gordon}) into Fourier modes labeled by their energy $\w \in \R$ and angular momentum $\ell \in \Z$:
\begin{align}
\phi_{\w,\ell}(t,\th,r) = e^{-i\w t} e^{i\ell\th} R_{\w,\ell}(r),
\end{align}
On this ansatz, (\ref{eqn:Klein-Gordon}) reduces to the following equation for the radial modes:
\begin{align}
\label{eqn:radial}
\qty[\f{1}{r} \dv{}{r} \p{r f(r) \dv{}{r}} + \f{\w^2}{f(r)} - \f{\ell^2}{r^2} - \Delta(\Delta - 2)] R_{\w,\ell}(r) = 0.
\end{align}
By a suitable change of variables, (\ref{eqn:radial}) can be brought to the form of the hypergeometric differential equation. We will treat it as an eigenvalue problem for $\w$, and by supplying boundary conditions we will extract a discrete spectrum of (quasi)normal modes. These modes can then be normalized and used to construct the momentum-space propagator.



\subsection{The Propagator in Pure AdS}
\label{sec:appendix-pure-AdS}

In pure $\up{AdS}_3$, the Feynman propagator is built from modes that are regular at the origin and normalizable (i.e. decaying) at infinity. The regular solution to (\ref{eqn:radial}) is\footnote{We assume here that $\Delta \notin \Z$ to avoid special or degenerate cases of the hypergeometric function. However, the final results will be analytic in $\Delta$ and can be extended to integer values of $\Delta$.}
\begin{align}
\label{eqn:AdS-radial-modes}
R_{\w,\ell}(r) = \cal{N}_{\w,\ell} \p{r^2 + 1}^{\w/2} r^{|\ell|} {}_2 F_1 \p{\f{2 - \Delta + \w + |\ell|}{2}, \f{\Delta + \w + |\ell|}{2}; 1 + |\ell|;\, -r^2},
\end{align}
where $\cal{N}_{\w,\ell}$ is a normalization constant to be determined. As for the behavior at infinity, we note that the ${}_2 F_1$ is subject to the following asymptotics:
\begin{align}
{}_2 F_1 \p{a, b; c; z} \sim \p{\f{\G\p{b-a} \G\p{c}}{\G\p{b} \G\p{c-a}}} \p{-z}^{-a} + \p{\f{\G\p{a-b} \G\p{c}}{\G\p{a} \G\p{c-b}}} \p{-z}^{-b} \quad \up{as}\;\, z \too \infty.
\end{align}
It follows that the regular solutions behave near the AdS boundary as a linear combination of normalizable and non-normalizable modes:
\begin{align}
\label{eqn:boundary-behavior}
R_{\w,\ell}(r) \sim \cal{N}_{\w,\ell} \Big( A(\w,\ell) r^{\Delta - 2} + B(\w,\ell) r^{-\Delta}\Big) \quad \up{as}\;\, r \too \infty,
\end{align}
where the coefficients $A(\w,\ell)$ and $B(\w,\ell)$ are given by\footnote{Observe that $B(\w,\ell)$ can be obtained from $A(\w,\ell)$ by sending $\Delta \too 2-\Delta$. Switching $A(\w,\ell)$ and $B(\w,\ell)$ is equivalent to interchanging the solutions $\Delta_{\pm}$ to $m^2 = \Delta(\Delta - 2)$, and corresponds to the alternate quantization of the scalar. This quantization uses $\Delta_-$ in place of $\Delta_+$ and has Neumann boundary conditions at infinity; this exchanges the roles of the $r^{\Delta-2}$ and $r^{-\Delta}$ terms in (\ref{eqn:boundary-behavior}) as the source and response.}
\begin{subequations}
\begin{align}
\label{eqn:A-coeff-AdS}
A(\w,\ell) &= \f{\G(\Delta-1) \G(1+|\ell|)}{\G\p{\f{1}{2}(\Delta + |\ell| - \w)} \G\p{\f{1}{2}(\Delta + |\ell| + \w)}}, \\
\label{eqn:B-coeff-AdS}
B(\w,\ell) &= \f{\G(1-\Delta) \G(1+|\ell|)}{\G\p{\f{1}{2}(2-\Delta + |\ell| - \w)} \G\p{\f{1}{2}(2-\Delta + |\ell| + \w)}}.
\end{align}
\end{subequations}

To ensure that the radial modes are normalizable, we must set $A(\w,\ell) = 0$. This condition, which is satisfied at the poles of the Gamma functions in the denominator of (\ref{eqn:A-coeff-AdS}), quantizes the energies of the normalizable modes in AdS:
\begin{align}
A(\w,\ell) = 0 \implies \w = \w_{n\ell} = \Delta + 2n + |\ell|, \qquad \ell \in \Z, \;\, n \in \N.
\end{align}
The normal modes are therefore $\phi_{n\ell}(t,\th,r) = e^{-i\w_{n\ell} t} e^{i\ell\th} R_{\w_{n\ell},\ell}(r)$. In the dual CFT, the fields $\phi_{n\ell} \sim \Box^n \overset{{}_\leftrightarrow}{\pd}{}^{\ell} \phi$ are descendants of $\phi$ and have energy $\Delta + 2n + |\ell|$ on the cylinder.

It remains to determine the normalization constant $\cal{N}$. The normal modes are required to be orthonormal in the Klein--Gordon inner product, which is defined by
\begin{align}
\label{eqn:KG-product}
\avg{\phi_{n\ell}, \phi_{n'\ell'}} = -i\int_{\Sigma} \dd^2 x \sqrt{g_{\Sigma}}\, \hat{n}^{\mu} \phi_{n\ell}(x) \overset{{}_\leftrightarrow}{\pd}_{\mu} \phi_{n'\ell'}^*(x). 
\end{align}
Here $x = (\th, r)$ are the spatial directions, $\Sigma$ is any hypersurface of constant $t$, $g_{\Sigma}$ is the metric induced on $\Sigma$, and $\hat{n}^{\mu}$ is the future-directed unit normal to $\Sigma$. In pure AdS, we have $\sqrt{g_{\Sigma}} = r/\sqrt{r^2 + 1}$ and $\hat{n}^{\mu} = \d^{\mu,t}/\sqrt{r^2 + 1}$, so the Klein--Gordon product becomes
\begin{align}
\label{eqn:KG-product-AdS}
\avg{\phi_{n\ell}, \phi_{n'\ell'}} = \p{\w + \w'} \int_0^{2\pi} \dd\th \int_0^{\infty} \f{r\, \dd r}{r^2 + 1} R_{\w,\ell}(r) R_{\w',\ell'}(r) e^{i \th(\ell - \ell')}.
\end{align}
The angular integral gives $2\pi \d_{\ell\ell'}$, but for consistency with the normalization of correlators in CFT, we will drop the leading $2\pi$. The radial integral vanishes if $\w \neq \w'$: this can be seen by rewriting the normalizable radial modes in terms of the Jacobi polynomials $P_n^{(\a,\b)}$:\footnote{The Jacobi polynomials $P^{(\a,\b)}_n$ are defined by ${}_2 F_1\p{-n, \a + \b + n + 1; \a + 1; z} = \f{n! \a!}{(n+\a)!} P_n^{(\a,\b)}(1-2z)$. To obtain (\ref{eqn:normalizable-modes-AdS}), we set $\w = \w_{n\ell}$ in (\ref{eqn:AdS-radial-modes}) and used ${}_2 F_1(a,b;c;z) = (1-z)^{c-a-b} {}_2 F_1(c-a,c-b;c;z)$.}
\begin{equation}
\label{eqn:normalizable-modes-AdS}
\begin{aligned}
R_{\w_{n\ell},\ell}(r) &= \cal{N}_{\w_{n\ell},\ell} (r^2 + 1)^{-\w_{n\ell}/2} r^{|\ell|} {}_2 F_1\p{-n, 1-\Delta-n; 1+|\ell|; -r^2} \\
&= \cal{N}_{\w_{n\ell},\ell} \p{\f{\G(1+n) \G(1 + |\ell|)}{\G(1+n+|\ell|)}} (r^2 + 1)^{-\w_{n\ell}/2} r^{|\ell|} P_n^{(|\ell|, -\w_{n\ell})}(1+2r^2).
\end{aligned}
\end{equation}
The orthogonality of the $P_n^{(\a,\b)}$ guarantees that (\ref{eqn:KG-product-AdS}) is proportional to $\d_{nn'}$. Finally, we determine $\cal{N}_{\w_{n\ell},\ell}$ by setting $\ell = \ell'$ and $\w = \w'$ in (\ref{eqn:KG-product-AdS}) and substituting (\ref{eqn:normalizable-modes-AdS}). The integral can be evaluated in \emph{Mathematica} after a change of variables to $z = r^2$:
\begin{align}
\label{eqn:norm-calculation}
\big\Vert \phi_{n\ell} \big\Vert^2 &= \w_{n\ell} \big|\cal{N}_{\w_{n\ell},\ell}\big|^2 \p{\f{\G(1+n) \G(1 + |\ell|)}{\G(1+n+|\ell|)}}^2 \int_0^{\infty} \f{\dd z\, z^{|\ell|}}{\p{1 + z}^{\w_{n\ell} + 1}} \abs{P_n^{(|\ell|, -\w_{n\ell})}(1+2z)}^2 \notag \\
&= \big|\cal{N}_{\w_{n\ell},\ell}\big|^2 \f{\G(\Delta+n)\G(1+n)\G(1+|\ell|)^2}{\G(\Delta+n+|\ell|)\G(1+n+|\ell|)}.
\end{align}
Upon demanding that $\phi_{n\ell}$ have unit norm, we find
\begin{align}
\label{eqn:KG-norm-AdS}
\big\Vert \phi_{n\ell} \big\Vert^2 = \avg{\phi_{n\ell}, \phi_{n\ell}} = 1 \implies \big|\cal{N}_{\w_{n\ell},\ell}\big|^2 = \f{\G(\Delta + n + |\ell|) \G(1 + n + |\ell|)}{\G(\Delta + n) \G(1 + n) \G(1 + |\ell|)^2}.
\end{align}

On the boundary, the operator dual to $\phi_{n\ell}$ is obtained by rescaling the bulk wave function by $r^{\Delta}$ as we take the limit $r \too \infty$ to keep it finite. From (\ref{eqn:boundary-behavior}), we obtain
\begin{align}
\phi_{n\ell}(t,\th) = \lim_{r \to \infty} r^{\Delta} \phi_{n\ell}(t,\th,r) = \big(\cal{N}_{\w_{n\ell},\ell} B(\w,\ell)\big) e^{-i\w_{n\ell}t} e^{i\ell\th}.
\end{align}
Combining (\ref{eqn:B-coeff-AdS}) with (\ref{eqn:KG-norm-AdS}), we obtain the boundary amplitudes
\begin{align}
\p{\cal{C}_{n\ell}^{\up{HL}}}^2 \equiv \big(\cal{N}_{\w_{n\ell},\ell} B(\w,\ell)\big)^2 = \f{\G(\Delta + n) \G(\Delta + n + |\ell|)}{\G(\Delta)^2 \G(1 + n) \G(1 + n + |\ell|)} = \cal{G}(\w_{n\ell}, \ell).
\end{align}
As indicated in the main text, this is the momentum-space Green's function in pure AdS, evaluated at the normal frequencies. As a meromorphic function of $\w$, it reads
\begin{align}
\cal{G}(\w, \ell) = \f{1}{\G(\Delta)^2} \f{\G\p{\f{1}{2}(\Delta + \w + |\ell|)} \G\p{\f{1}{2}(\Delta + \w - |\ell|)}}{\G\p{\f{1}{2}(2-\Delta + \w + |\ell|)} \G\p{\f{1}{2}(2-\Delta + \w - |\ell|)}}.
\end{align}

\subsection{Below the BTZ Threshold}
\label{sec:appendix-below-threshold}

The analysis in conical AdS is nearly identical to the analysis in pure AdS, differing only by certain factors of $N$. See \cite{Berenstein:2022ico} for a more detailed exposition.

The solution to the radial equation (\ref{eqn:radial}) that is regular at $r=0$ is
\begin{align}
\label{eqn:cAdS-radial-modes}
R_{\w,\ell}(r) = \cal{N}_{\w,\ell} \p{r^2 + \f{1}{N^2}}^{N\w/2} r^{N|\ell|} {}_2 F_1 \p{\f{2-\Delta + N(\w + |\ell|)}{2}, \f{\Delta + N(\w + |\ell|)}{2}; 1 + N|\ell|; -N^2 r^2}.
\end{align}
Its asymptotic behavior near infinity is given by (\ref{eqn:boundary-behavior}), this time with
\begin{subequations}
\begin{align}
A(\w,\ell) &= \f{N^{\Delta - 2 - N(|\ell| + \w)} \G(\Delta-1) \G\big(1+N|\ell|\big)}{\G\Big(\f{1}{2}\big(\Delta + N(|\ell| - \w)\big) {}_{\!} \Big) \G\Big(\f{1}{2}\big(\Delta + N(|\ell| + \w)\big) {}_{\!} \Big)}, \\
B(\w,\ell) &= \f{N^{-\Delta - N(|\ell| + \w)} \G(1-\Delta) \G\big(1+N|\ell|\big)}{\G\Big(\f{1}{2}\big(2-\Delta + N(|\ell| - \w)\big) {}_{\!} \Big) \G\Big(\f{1}{2}\big(2-\Delta + N(|\ell| + \w)\big) {}_{\!} \Big)}.
\end{align}
\end{subequations}
The normal mode spectrum resulting from the normalizability condition $A(\w,\ell) = 0$ is
\begin{align}
A(\w,\ell) = 0 \implies \w = \w_{n\ell} = \f{1}{N} \p{\Delta + 2n} + |\ell|, \qquad \ell \in \Z, \;\, n \in \N.
\end{align}
matching (\ref{eqn:normal-frequencies}) in the main text. Note that modes with different angular momenta have integer energy spacing, while the gap between modes with the same $|\ell|$ goes like $1/N$.

The normalizable modes are again expressible in terms of Jacobi polynomials:
\begin{align}
R_{\w_{n\ell},\ell}(r) = \cal{N}_{\w_{n\ell},\ell} \p{\f{\G(1+n) \G(1 + N|\ell|)}{\G(1+n+N|\ell|)}} \p{r^2 + \f{1}{N^2}}^{-N\w_{n\ell}/2} r^{N|\ell|} P_n^{(|\ell|, -N\w_{n\ell})}(1 + 2N^2 r^2).
\end{align}
As before, the modes are orthogonal in the Klein--Gordon inner product (\ref{eqn:KG-product}), where $\sqrt{g_{\Sigma}} = r/\sqrt{r^2 + \f{1}{N^2}}$ and $\hat{n}^{\mu} = \d^{\mu,t}/\sqrt{r^2 + \f{1}{N^2}}$ in conical AdS. Their norms are
\begin{equation}
\begin{aligned}
\big\Vert \phi_{n\ell} \big\Vert^2 &= 2\w_{n\ell} \int_0^{\infty} \f{r\, \dd r}{\p{r^2 + \f{1}{N^2}}} \big|R_{\w_{n\ell},\ell}(r)\big|^2 \\ &= \big|\cal{N}_{\w_{n\ell},\ell} \big|^2 N^{-2\Delta - 4n - 4N|\ell| - 1} \p{\f{\G(\Delta+n)\G(1+n)\G(1+N|\ell|)^2}{\G(\Delta+n+N|\ell|)\G(1+n+N|\ell|)}}
\end{aligned}
\end{equation}
by a calculation similar to (\ref{eqn:norm-calculation}). Thus for the normalization constants we get
\begin{align}
\big|\cal{N}_{\w_{n\ell},\ell} \big|^2 = N^{2\Delta + 4n + 4N|\ell| + 1} \p{\f{\G(\Delta+n+N|\ell|)\G(1+n+N|\ell|)}{\G(\Delta+n)\G(1+n)\G(1+N|\ell|)^2}},
\end{align}
and the corresponding boundary amplitudes (compare (\ref{eqn:HL-coefficients-conical}) in the main text) are
\begin{align}
\p{\cal{C}_{n\ell}^{\up{HL}}}^2 = \big(\cal{N}_{\w_{n\ell},\ell} B(\w_{n\ell},\ell)\big)^2 = \f{N^{1-2\Delta}}{\G(\Delta)^2} \f{\G(\Delta + n + N|\ell|) \G(\Delta + n)}{\G(1 + n + N|\ell|)\, \G(1+n)} = \cal{G}(\w_{n\ell},\ell).
\end{align}
This gives the momentum-space Green's function in conical AdS,
\begin{align}
\label{eqn:cAdS-Green-function}
\cal{G}(\w,\ell) = \f{N^{1-2\Delta}}{\G(\Delta)^2} \f{\G\Big(\f{1}{2} \big(\D + N(\w + \ell)\big){}_{\!} \Big)\, \G\Big(\f{1}{2} \big(\D + N(\w - \ell) \big){}_{\!} \Big)}{\G\Big(\f{1}{2} \big(2-\D + N(\w + \ell) \big){}_{\!} \Big) \G\Big(\f{1}{2} \big(2-\D + N(\w - \ell) \big){}_{\!} \Big)}.
\end{align}
in agreement with (\ref{eqn:momentum-space-G}). We are allowed to remove the absolute values because the Green's function is symmetric under $\ell \too -\ell$, i.e. we have $G(\w, |\ell|) = G(\w, \ell)$.

\subsection{Above the BTZ Threshold}
\label{sec:appendix-above-threshold}

In black hole backgrounds, we are interested in the retarded propagator. Accordingly, we impose ingoing boundary conditions at the horizon in addition to decay at infinity. With these conditions, the solutions to (\ref{eqn:radial}) become rather messy. It is therefore prudent to introduce the ``planar'' coordinate $z = 1 - \p{\f{r_0}{r}}^2$, which puts the horizon at $z=0$ and the AdS boundary at $z=1$. The radial equation transforms into
\begin{align}
\label{eqn:radial-z}
z(1-z) \dv[2]{R_{\w,\ell}}{z} + (1-z) \dv{R_{\w,\ell}}{z} + \f{1}{4} \p{\f{\w^2/r_0^2}{z} - \f{\Delta(\Delta - 2)}{1-z} - \ell^2/r_0^2} R_{\w,\ell} = 0,
\end{align}
and the ingoing boundary conditions (\ref{eqn:ingoing-conditions}) become
\begin{align}
R_{\w\ell}(z) \sim \begin{cases}
\displaystyle z^{-\f{i\w}{2r_0}}, & z \too 0 \quad\; (\text{BTZ horizon}) \\
\displaystyle \p{1-z}^{\Delta/2}, & z \too 1 \quad (\text{AdS boundary})
\end{cases}
\end{align}

The solution to (\ref{eqn:radial-z}) that is purely ingoing at the horizon is\footnote{To lighten the notation, we will take $\ell \geq 0$ throughout. Absolute values around $\ell$ can be restored, where necessary, at the end of the analysis. This is done in the main text.}
\begin{align}
\label{eqn:BTZ-radial-modes}
R_{\w,\ell}(z) &= \cal{N}_{\w,\ell} \p{1 - z}^{\Delta/2}  z^{-\f{i\w}{2r_0}} {}_2 F_1\p{\f{\Delta - \f{i}{r_0}(\ell + \w)}{2}, \f{\Delta - \f{i}{r_0}(\w - \ell)}{2}; 1 - \f{i\w}{r_0}; z}
\end{align}

Near the AdS boundary, the ingoing solutions (\ref{eqn:BTZ-radial-modes}) are once again linear combinations of modes that decay to zero and modes that have unbounded growth at infinity:
\begin{equation}
\label{eqn:boundary-behavior-BTZ}
\begin{aligned}
R_{\w,\ell}(z) &\sim \cal{N}_{\w,\ell} \p{A(\w,\ell) (1-z)^{1 - \Delta/2} + B(\w,\ell) (1-z)^{\Delta/2}} \\ &= \cal{N}_{\w,\ell} \p{ r_0^{2-\Delta} A(\w,\ell) r^{\Delta - 2} + r_0^{\Delta} B(\w,\ell) r^{-\Delta}},
\end{aligned}
\end{equation}
where the coefficients $A(\w,\ell)$ and $B(\w,\ell)$ are, including their scaling with $r_0$,
\begin{subequations}
\label{eqn:source-response}
\begin{align}
r_0^{2-\Delta} A(\w,\ell) &= \f{r_0^{2-\Delta} \G(\Delta-1) \G\p{1 - \f{i\w}{r_0}}}{\G\p{\f{1}{2} \p{\Delta - \f{i}{r_0} \p{\w + \ell}}} \G\p{\f{1}{2} \p{\Delta - \f{i}{r_0} \p{\w - \ell}}}}, \\
r_0^{\Delta} B(\w,\ell) &= \f{r_0^{\Delta} \G(1-\Delta) \G\p{1 - \f{i\w}{r_0}}}{\G\p{\f{1}{2} \p{2-\Delta - \f{i}{r_0} \p{\w + \ell}}} \G\p{\f{1}{2} \p{2-\Delta - \f{i}{r_0} \p{\w - \ell}}}}.
\end{align}
\end{subequations}

To respect the boundary condition at infinity, we must take $A(\w,\ell) = 0$. This gives us the spectrum of quasinormal modes (QNMs), matching (\ref{eqn:quasinormal-frequencies}) in the main text:
\begin{align}
A(\w,\ell) = 0 \implies \w = \tilde{\w}_{n\ell} = -\ell - ir_0 \p{\Delta + 2n}, \qquad \ell \in \Z, \;\, n \in \N.
\end{align}
The modes themselves can once again be expressed in terms of Jacobi polynomials:
\begin{equation}
\begin{aligned}
R_{\tilde{\w}_{n\ell},\ell}(z) &= \cal{N}_{\tilde{\w}_{n\ell},\ell} \p{1-z}^{\Delta/2} z^{-\f{i\tilde{\w}_{n\ell}}{2r_0}} {}_2 F_1 \p{-n, -n + \f{i\ell}{r_0}; 1 - \Delta - 2n + \f{i\ell}{r_0}; z} \\ &= \cal{N}_{\tilde{\w}_{n\ell},\ell} \p{\f{\G(1+n) \G\big(1 - \f{i\tilde{\w}_{n\ell}}{r_0}\big)}{\G\big(1+n- \f{i\tilde{\w}_{n\ell}}{r_0}\big)}} \p{1-z}^{\Delta/2} z^{-\f{i\tilde{\w}_{n\ell}}{2r_0}} P^{\big(-\f{i\tilde{\w}_{n\ell}}{r_0},\, \Delta-1\big)}_n(1 - 2z).
\end{aligned}
\end{equation}
These modes are not normalizable because they can fall through the horizon: thus we cannot determine $\cal{N}_{\tilde{\w}_{n\ell},\ell}$ from a Klein--Gordon norm, nor can we obtain the propagator by summing over modes. Instead, we write down the momentum-space Green's function by taking the ratio of the response $r_0^{\Delta} B(\w,\ell)$ to the source $r_0^{2-\Delta} A(\w,\ell)$ in (\ref{eqn:source-response}):
\begin{align}
\label{eqn:BTZ-Green-function}
\cal{G}_{\up{R}}(\w,\ell) \propto \f{r_0^{2\Delta - 2} \G(1-\Delta) \G\p{\f{1}{2} \p{\Delta - \f{i}{r_0} \p{\w + \ell}}} \G\p{\f{1}{2} \p{\Delta - \f{i}{r_0} \p{\w - \ell}}}}{\G(\Delta-1) \G\p{\f{1}{2} \p{2-\Delta - \f{i}{r_0} \p{\w + \ell}}} \G\p{\f{1}{2} \p{2-\Delta - \f{i}{r_0} \p{\w - \ell}}}}.
\end{align}
This agrees with (\ref{eqn:momentum-space-G}), although the overall normalization here is not quite the same as in (\ref{eqn:cAdS-Green-function}). We shall explain the origin of this apparent mismatch shortly.

As described in Section~\ref{sec:correlators-above-threshold}, the position-space retarded propagator is the Fourier inverse of $\cal{G}_{\up{R}}(\w,\ell)$. Its calculation involves a contour integral over $\w$, closed in the lower half plane, which picks up the residues of $\cal{G}_{\up{R}}(\w,\ell)$ at the quasinormal frequencies:
\begin{align}
G_{\up{R}}(t,\th) = \sum_{n\ell} \big(\tilde{\cal{C}}_{n\ell}^{\up{HL}}\big)^2 e^{-i\tilde{\w}_{n\ell} t} e^{i\ell\th}, \qquad \textbf{}\big(\tilde{\cal{C}}_{n\ell}^{\up{HL}}\big)^2 \equiv i\, \up{Res}\big[\cal{G}_{\up{R}}(\w,\ell),\, \tilde{\w}_{n\ell}\big].
\end{align}
These residues can be computed directly from (\ref{eqn:BTZ-Green-function}). We obtain\footnote{$\cal{G}_{\up{R}}(\w,\ell)$ has simple poles for $\ell \neq 0$ and double poles for $\ell = 0$. Both sets of poles have nonzero residues, but the s-wave residues must be computed separately because the $\ell \neq 0$ expression diverges at $\ell = 0$. The result (\ref{eqn:thermal-factor-again}) shows that the s-wave QNMs give extra $n$-dependence relative to the $\ell \neq 0$ QNMs.}
\begin{equation}
\begin{aligned}
\label{eqn:thermal-factor-again}
\big(\tilde{\cal{C}}_{n\ell}^{\up{HL}}\big)^2 &= \f{2\p{\Delta-1}r_0^{2\Delta - 1}}{\G(\Delta)^2} \f{\G(\Delta + n -i\ell/r_0) \G(\Delta + n)}{\G(1 + n -i\ell/r_0) \G(1+n)} f_{n\ell}, \\
f_{n\ell} &=\begin{cases}
\displaystyle \f{\sinh\p{\pi\ell/r_0 + i\pi\Delta}}{\sinh\p{\pi\ell/r_0}}, & \ell \neq 0, \\
\displaystyle 2\cos\p{\pi \Delta} - \f{2}{\pi} \sin\p{\pi\Delta} \big(\psi(1+n) - \psi(\Delta + n)\big), & \ell = 0.
\end{cases}
\end{aligned}
\end{equation}
This agrees with (\ref{eqn:HL-coefficients-BTZ}) up to a constant prefactor of $\Delta_+ - \Delta_- = 2\Delta - 2$. Fortunately, this is precisely the factor obtained in \cite{Berenstein:2014cia} relating the bulk and boundary normalizations of the Green's function. This extra factor needs to be taken into account in (\ref{eqn:BTZ-Green-function}) to restore consistency between the conical AdS and BTZ normalizations.

\subsection{The Case of Massless BTZ}
\label{sec:appendix-mBTZ}

The massless BTZ (mBTZ) geometry is a somewhat singular limit of both conical AdS ($N \too \infty$) and BTZ ($r_0 \too 0$). The hypergeometric solutions to the wave equation, both above and below the BTZ threshold, become ill-defined in mBTZ, and it is not immediately obvious which boundary conditions one should impose at the origin. The metric is
\begin{align}
\dd s^2 = -r^2 \dd t^2 + \f{\dd r^2}{r^2} + r^2\dd\th^2,
\end{align}
and the radial equation (\ref{eqn:radial}) in mBTZ takes the form
\begin{align}
\label{eqn:radial-mBTZ}
\qty[\f{1}{r} \dv{}{r} \p{r^3 \dv{}{r}} + \f{\W^2}{r^2} - \Delta(\Delta - 2)] R_{\w,\ell}(r) = 0, \qquad \W^2 \equiv \w^2 - \ell^2.
\end{align}
By a change of variables, this differential equation can be brought to the form of the Bessel equation. The Bessel equation is a degenerate limit of the hypergeometric equation: it has one regular singular point (at the AdS boundary) and one irregular singular point (at the origin), as opposed to the three regular singularities of the hypergeometric equation. This structure suggests that we should still be able to demand algebraic decay of the solutions at infinity, but that near $r=0$ they could behave rather erratically.

The general solution to (\ref{eqn:radial-mBTZ}) is given by
\begin{align}
R_{\w,\ell}(r) = \f{\W}{2r} \qty[A(\w,\ell) J_{1-\Delta}\p{\f{\W}{r}} + B(\w,\ell) J_{\Delta - 1}\p{\f{\W}{r}}],
\end{align}
where $J_{\nu}(x)$ is a Bessel function of the first kind, and $A(\w,\ell)$ and $B(\w,\ell)$ are constants analogous to those in (\ref{eqn:boundary-behavior}) and (\ref{eqn:boundary-behavior-BTZ}). Since $J_{\nu}(z) \sim \f{1}{\G(\nu + 1)} \p{\f{z}{2}}^{\nu}$ as $z \too 0$, we have
\begin{align}
\label{eqn:boundary-behavior-mBTZ}
R_{\w,\ell}(r) \sim \f{A(\w,\ell)}{\G(2-\Delta)} \p{\f{2r}{\W}}^{\Delta - 2} + \f{B(\w,\ell)}{\G(\Delta)} \p{\f{2r}{\W}}^{-\Delta} \quad \up{as}\;\, r \too \infty.
\end{align}
In order to ensure that $R_{\w,\ell}(r) \sim r^{-\Delta}$ at infinity, we must set $A(\w,\ell) = 0$. As for the behavior near the origin, the large-argument asymptotics of $J_{\nu}(z)$ give
\begin{align}
\label{eqn:pathological-behavior}
R_{\w,\ell}(r) \sim \f{B(\w,\ell)}{2} e^{-\Delta\pi i/2} \sqrt{\f{i\W}{2\pi r}} \qty[e^{i\W/r} - ie^{\Delta \pi i} e^{-i\W/r}] \quad \up{as}\;\, r \too 0.
\end{align}
From this behavior, we see that there is no way to make the modes regular, purely ingoing, or purely outgoing at the origin. The best we can do is to admit the boundary condition (\ref{eqn:pathological-behavior}), conceding that all normalizable solutions to (\ref{eqn:radial-mBTZ}) will behave pathologically near the singularity. Since there is no constraint on $\w$, the spectrum will be continuous. We will, however, demand that $\w^2 > \ell^2$, so that $\W$ is real and positive. This will at least force the solutions to oscillate near the origin instead of blowing up exponentially.\footnote{The case $\w = \ell$ leads to the normalizable modes $R_{\ell}(r) = B(\ell) r^{-\Delta}$. These behave worse at the origin than the $\w^2 > \ell^2$ modes, which go like $r^{-1/2}$, but better than the exponentially divergent $\w^2 < \ell^2$ modes.}

Despite the problems at $r=0$, it is nevertheless possible to normalize the modes
\begin{align}
\phi_{\w,\ell}(t,\th,r) = B(\w,\ell) \f{\W}{2r} J_{\Delta-1}\p{\f{\W}{r}} e^{-i\w t} e^{i\ell\th}
\end{align}
in the Klein--Gordon inner product (\ref{eqn:KG-product}). We have $\sqrt{g_{\Sigma}} = 1$ and $\hat{n}^{\mu} = \f{1}{r} \d^{\mu,t}$; as before, the $\th$ integral gives $2\pi \d_{\ell\ell'}$, but we drop the $2\pi$. The inner product reduces to
\begin{align}
\avg{\phi_{\w,\ell}, \phi_{\w',\ell'}} = \d_{\ell\ell'} \p{\w + \w'} \f{\W\W'}{4} B(\w,\ell) B^*(\w',\ell) \int_0^{\infty} \f{\dd r}{r^3} J_{\Delta-1}\p{\f{\W}{r}} J_{\Delta-1}^*\p{\f{\W'}{r}},
\end{align}
where $\W^2 = \w^2 - \ell^2$ and $\W'^2 = \w'^2 - \ell^2$. To evaluate the integral, we change variables to $z = \f{1}{r}$ and use the orthogonality relation for Bessel functions, which reads
\begin{align}
\int_0^{\infty} \dd z\, z J_{\Delta-1}(\W z) J_{\Delta - 1}(\W'z) = \f{\d(\W - \W')}{\W}.
\end{align}
It follows that our modes are delta-function normalizable, with amplitudes given by
\begin{align}
\avg{\phi_{\w,\ell}, \phi_{\w',\ell'}} = \f{\big| B(\w,\ell)\big|^2}{2} \w \sqrt{\w^2 - \ell^2} \d_{\ell\ell'} \d(\w - \w') \implies
\big|B(\w,\ell)\big|^2 = \f{2}{\w \sqrt{\w^2 - \ell^2}}.
\end{align}

The bulk-to-bulk propagator is then constructed by integrating over modes:
\begin{align}
G(x,x') = \sum_{\ell \in \Z} \int_{\ell}^{\infty} \dd\w\, R_{\w,\ell}(r) R_{\w,\ell}(r') e^{-i\w t} e^{i\ell\th}.
\end{align}
Finally, we pass to the AdS boundary using (\ref{eqn:boundary-behavior-mBTZ}) and obtain the propagator:
\begin{align}
\lim_{r \to \infty} r^{\Delta} R_{\w,\ell}(r) = \f{B(\w,\ell)}{\G(\Delta)} \p{\f{\W}{2}}^{\Delta} \implies
G(t,\th) = \f{1}{\G(\Delta)^2} \sum_{\ell \in \Z} \int_{\ell}^{\infty} \f{\dd\w}{\w} \p{\f{\W}{2}}^{2\Delta - 1} e^{-i\w t} e^{i\ell\th}.
\end{align}

\section{Very Light Conical Defects}
\label{sec:appendix-light-defects}

The spectrum of the HL double-twist modes shifts by
\begin{align}
\label{eqn:spectrum-shift}
\w_{n\ell} = \a(\Delta + 2n) + |\ell| \too \w_{n\ell}^{(\up{AdS})} - \eps\p{\Delta + 2n},
\end{align}
where $\w_{n\ell}^{(\up{AdS})} = \Delta + 2n + |\ell|$ are the energies of the modes $\phi_{n\ell}$ in pure AdS.

The corresponding HL OPE coefficients (\ref{eqn:HL-coefficients}) become
\begin{equation}
\label{eqn:OPE-data-shift}
\begin{aligned}
\p{\cal{C}_{n\ell}^{\up{HL}}}^2 &\too \f{\G(\Delta + n + |\ell|) \G(\Delta + n)}{\G(\Delta)^2 \G(1 + n + |\ell|) \G(1 + n)} \qty[1 + \eps \d\p{\cal{C}_{n\ell}^{\up{HL}}}^2], \\
\d\p{\cal{C}_{n\ell}^{\up{HL}}}^2 &=
1 - 2\Delta + |\ell| \big(\psi(\Delta + n + |\ell|) - \psi(1 + n + |\ell|) \big),
\end{aligned}
\end{equation}
where $\psi(z) = \f{\G'(z)}{\G(z)}$ is the digamma function. Using (\ref{eqn:spectrum-shift}) and (\ref{eqn:OPE-data-shift}) in the $s$-channel OPE, one can work out the first-order correction to the pure AdS correlator for light defects. Writing the correlator in the form $G(z, \bar{z}) = G_{\up{AdS}}(z, \bar{z}) + \eps \d G(z, \bar{z})$, we obtain
\begin{align}
\label{eqn:correlator-correction}
\d G(z, \bar{z}) &= \abs{z}^{(\a - 1)\Delta} \sum_{n\ell} \f{\G(\Delta + n + |\ell|) \G(\Delta + n)}{\G(\Delta)^2 \G(1 + n + |\ell|) \G(1 + n)} 
\p{\d \p{\cal{C}_{n\ell}^{\up{HL}}}^2 + 2n \log |z|} \abs{z}^{2n} z^{\ell}.
\end{align}
Meanwhile, the vacuum block expands as $G_0(z, \bar{z}) = \abs{1-z}^{-2\Delta} + \eps \d G_0(z, \bar{z})$, with
\begin{align}
\label{eqn:vacuum-correction}
\d G_0(z, \bar{z}) = 2\Delta \abs{z}^{(\a - 1)\Delta} \abs{1-z}^{-2\Delta} \mathfrak{R} \p{1 + \f{z \log z}{1-z}}.
\end{align}

Interestingly, (\ref{eqn:vacuum-correction}) behaves near $z, \bar{z} = 1$ like $\abs{1 - z}^{-2\Delta - 1}$, a stronger singularity than in the original vacuum block. Accordingly, the asymptotics of the coefficients of (\ref{eqn:correlator-correction}) must be such that $\d G(z, \bar{z})$ exhibits the same behavior, in the sense that the difference
\begin{align}
\label{eqn:Greg-correction}
G_{\up{reg}}(z, \bar{z}) = \eps \big(\d G(z, \bar{z}) - \d G_0(z, \bar{z})\big) + O(\eps^2)
\end{align}
remains finite at $z, \bar{z} = 1$. One can then study the expectation values $\cal{C}_{n\ell}^{\up{HH}}$ at first order in $\eps$ by evaluating (\ref{eqn:Greg-correction}) and its derivatives numerically.

\bibliographystyle{jhep}
\bibliography{refs} 

\end{document}